\documentclass [11pt]{article}

\usepackage{jheppub}

\usepackage[utf8]{inputenc}

\usepackage{amsfonts}

\usepackage{amsmath,amssymb}

\usepackage{mathtools}

\usepackage{graphicx,subfigure}
\usepackage[space]{grffile}
\usepackage{mathtools}

\usepackage{simplewick}

\usepackage{array}

\usepackage{float}

\usepackage{color}

\usepackage{mathrsfs}

\usepackage{ytableau}

\usepackage{tikz}
\usepackage{tikz-feynman}
\usepackage{contour}
\usepackage{slashed}
\def\nn{\nonumber}

\usetikzlibrary{external, calc} 
\usepackage{iftex}
\ifluatex
	\RequirePackage{shellesc}
\else
\fi

\newcommand{\cO}{\mathcal{O}}

\newcommand{\<}{\langle}
\renewcommand{\>}{\rangle}
\newcommand{\p}{\partial}
\newcommand{\be}{\begin{eqnarray}\displaystyle}
\newcommand{\ee}{\end{eqnarray}}
\newcommand{\f}{\frac}

\newcommand{\ZZ}{\mathbb{Z}_2}

\newcommand{\myRho}{{\mathfrak{r}}}

 \definecolor{verde}{rgb}{0,0.7,0.2}

\newcommand{\myP}{\boldsymbol{p}}

\title{\boldmath Bootstrapping the $a$-anomaly in $4d$ QFTs}

\author[a]{Denis Karateev,}
\author[b]{Jan Marucha,}
\author[b]{Jo\~ao Penedones}
\author[b]{and Biswajit Sahoo}

\affiliation[a]{
	D\'epartment de Physique Th\'eorique, Universit\'e de Gen\`eve,\\
	24 quai Ernest-Ansermet, 1211 Gen\`eve 4, Switzerland}

\affiliation[b]{Fields and Strings Laboratory, Institute of Physics\\ École Polytechnique Fédéral de Lausanne (EPFL)
	\\ Route de la Sorge, CH-1015 Lausanne, Switzerland}

\abstract{
	We study gapped 4d quantum field theories (QFTs) obtained from a relevant deformation of a UV conformal field theory (CFT). For simplicity, we assume the existence of a $\mathbb{Z}_2$ symmetry   and  a single $\mathbb{Z}_2$-odd stable particle and no $\mathbb{Z}_2$-even particles at low energies. Using unitarity, crossing and the assumption of maximal analyticity we compute numerically a lower bound on the value of the $a$-anomaly of the UV CFT as a function of various non-perturbative parameters describing the two-to-two scattering amplitude of the particle.
}

\vspace{4cm}

\dedicated{
\vspace{3cm}We condemn the current war in Ukraine and dedicate this paper to its people.\\
We also express gratitude  to the independent journalists who are reporting the truth in Russia \\
and to all the people who are trying to stop this war.}

\begin{document}
\maketitle

\section{Introduction and summary}
\label{introduction}

The dream of a \emph{bootstrapper} is to map out the space of consistent Quantum Field Theories (QFTs).
Given the vastness of this space, in practice, one can only plot its projection onto a few cleverly chosen coordinates in theory space. Ideally, these coordinates correspond to physical observables that characterize the QFT across all length scales, from the UV Conformal Field Theory (CFT) to the IR, which we assume to be gapped.
On the one hand, the S-matrix bootstrap focuses on IR observables like ratios of masses of stable particles and coupling constants defined in terms of scattering amplitudes, for recent progress see \cite{Paulos:2016fap, Paulos:2016but, Paulos:2017fhb,Doroud:2018szp, He:2018uxa, Cordova:2018uop, Guerrieri:2018uew, Paulos:2018fym, Homrich:2019cbt, EliasMiro:2019kyf, Cordova:2019lot, Bercini:2019vme, Gabai:2019ryw,Bose:2020shm,Bose:2020cod,Correia:2020xtr,Kruczenski:2020ujw, Guerrieri:2020bto,Hebbar:2020ukp,Karateev:2019ymz,Karateev:2020axc,Guerrieri:2020kcs,Tourkine:2021fqh,Guerrieri:2021ivu,He:2021eqn,EliasMiro:2021nul,Guerrieri:2021tak,Chen:2021pgx,Cordova:2022pbl,Albert:2022oes,Sinha:2020win,Chowdhury:2021ynh,Chen:2022nym,Miro:2022cbk}.\footnote{For an overview of recent results and discussion of some future directions, see \cite{Kruczenski:2022lot}.}
On the other hand, the conformal bootstrap focuses on UV data like  scaling dimensions and  Operator Product Expansion (OPE) coefficients of the UV CFT, for a review see \cite{Poland:2018epd}.\footnote{For an overview of the most recent progress and discussion of some further directions see also \cite{Poland:2022qrs}.}
In this work, we extend the S-matrix bootstrap  method in four spacetime dimensions to gain access to the $a$-anomaly of the UV CFT, which is a precise measure of   its degrees of freedom.
This is similar to the spirit of \cite{Karateev:2019ymz} in two spacetime dimensions, that incorporated the central charge of the UV CFT into the S-matrix bootstrap framework.

Our strategy is to follow Komargodski and Schwimmer \cite{Komargodski:2011vj} and probe the QFT with an external massless scalar field $\varphi(x)$ which creates a massless particle $B$ from the vacuum, usually called the \emph{dilaton}.\footnote{The dilaton $B$ in this paper should not be confused with the Nambu–Goldstone boson of spontaneously broken conformal symmetry which is also called the dilaton. As explained in section \ref{KS_setup} the better name for the former dilaton would be the compensator particle. However in this paper we keep the commonly used terminology.}
As we review in section \ref{S:review}, this construction does not affect the scattering amplitudes of the original QFT but it generates non-trivial scattering between dilaton particles. In particular, the $a$-anomaly $a^\text{UV}$ of the UV CFT can be read off from the low energy behavior
\begin{equation}  
 \widetilde{\mathcal{T}}_{BB\rightarrow BB}  =  a^\text{UV}  \,(s^2+t^2+u^2) + \dots \,,
\end{equation}
where $s,t,u$ are the standard Mandelstam invariants for two to two scattering.

For simplicity, we study QFTs with a $\ZZ$ symmetry and a single stable scalar particle of mass $m$ and $\ZZ$ odd, which we shall denote by $A$.  The combined system QFT + \emph{dilaton}
 has two asymptotic states: the original scalar particle $A$ and the new massless particle describing the dilaton $B$. We   study the complete set of 2 to 2 scattering amplitudes in this system, see figure \ref{fig:set_amplitudes}. 
 Then we write  all  crossing equations and unitarity conditions such a system must satisfy. All the details of this setup are given in section \ref{sec:S-matrix_setup}. In addition, we impose the universal soft behavior of $\widetilde{\mathcal{T}}_{AB \rightarrow  AB}$ derived in section \ref{S:EFT_constraints}.
 In order to obtain concrete results we employ the numerical approach of \cite{Paulos:2017fhb,Homrich:2019cbt}. 
 Let us now briefly summarize our main findings.

\begin{center}
	\begin{figure}[h!]
%

	\begin{tikzpicture}[baseline=(a)]
		\begin{feynman}
			\vertex [blob, minimum size=1.8cm] (a) {};
			\vertex [above left = of a, label=left:$A$ ] (i1);
			\vertex [below left = of a, label=left:$A$ ] (i2);
			\vertex [above right = of a, label=right:$A$] (o1);
			\vertex [below right = of a, label=right:$A$] (o2);
			\diagram* {
				(i1)--(a)--(i2), 
				(o1)--(a)--(o2),
			};
		\end{feynman} 
	\end{tikzpicture}\qquad \qquad \qquad
	\begin{tikzpicture}[baseline=(a)]
		\begin{feynman}
			\vertex [blob, minimum size=1.8cm] (a) {};
			\vertex [above left = of a, label=left:$A$ ] (i1);
			\vertex [below left = of a , label=left:$A$] (i2);
			\vertex [above right = of a, label=right:$B$] (o1);
			\vertex [below right = of a, label=right:$B$] (o2);
			\diagram* {
				(i1)--(a)--(i2), 
				(o1)--[dashed](a)--[dashed](o2),
			};
		\end{feynman} 
	\end{tikzpicture}\qquad \qquad \qquad
	\begin{tikzpicture}[baseline=(a)]
		\begin{feynman}
			\vertex [blob, minimum size=1.8cm] (a) {};
			\vertex [above left = of a, label=left:$B$ ] (i1);
			\vertex [below left = of a, label=left:$B$ ] (i2);
			\vertex [above right = of a, label=right:$B$] (o1);
			\vertex [below right = of a, label=right:$B$] (o2);
			\diagram* {
				(i1)--[dashed](a)--[dashed](i2), 
				(o1)--[dashed](a)--[dashed](o2),
			};
		\end{feynman} 
	\end{tikzpicture}
	\caption{The complete system of scattering amplitudes of the $\ZZ$ odd particle $A$ with mass $m$ and the massless dilaton $B$.}
	\label{fig:set_amplitudes}
\end{figure}
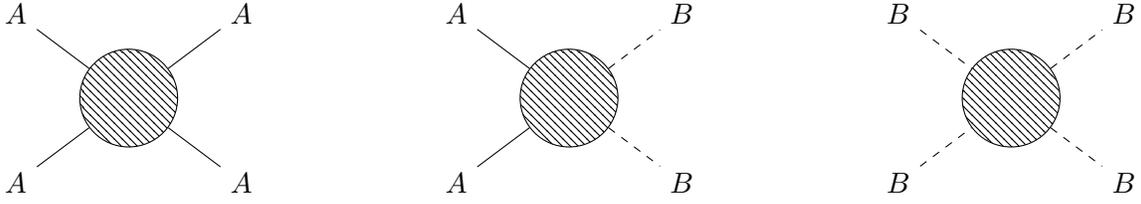
\end{center}
We define {\bf non-perturbative couplings}   in terms of the physical scattering amplitude.
In this work we   focus  on 
\begin{equation}
	\mathcal{T}_{AA\rightarrow AA} (s_0,\, t_0,\, u_0),\qquad
	\partial_s^2\mathcal{T}_{AA\rightarrow AA} (s_0,\, t_0,\, u_0),
\end{equation}
where $(s_0, t_0, u_0)$ is a point inside the Mandelstam triangle defined by   $0\leq s_0, t_0, u_0\leq 4m^2$.
We will consider two choices. The first choice is the crossing symmetric point $s_0=t_0=u_0=4m^2/3$ which leads to the definition of the parameters $\lambda_0$ and $\lambda_2$, namely
\begin{equation}
	\label{eq:lambdas}
	\begin{aligned}
		\lambda_0 &\equiv \frac{1}{32 \pi }\mathcal{T}_{AA\rightarrow AA} (4m^2/3,\, 4m^2/3,\, 4m^2/3),\\
		\lambda_2 &\equiv \frac{1}{32 \pi }m^4\partial_s^2\mathcal{T}_{AA\rightarrow AA} (4m^2/3,\, 4m^2/3,\, 4m^2/3).
	\end{aligned}
\end{equation}
The second choice is the ``forward'' point  $s_0=u_0=2m^2$ and $t=0$ which leads to the definition of the parameters  $\Lambda_0$ and $\Lambda_2$, namely
\begin{equation}
	\label{eq:Lambdas}
	\Lambda_0\equiv \frac{1}{32 \pi }\mathcal{T}_{AA\rightarrow AA} (2m^2,\, 0,\, 2m^2),\qquad
	\Lambda_2\equiv \frac{1}{32 \pi }m^4\partial_s^2\mathcal{T}_{AA\rightarrow AA} (2m^2,\, 0,\, 2m^2).
\end{equation}

\begin{figure}
	\centering
	\includegraphics[width=0.45\textwidth]{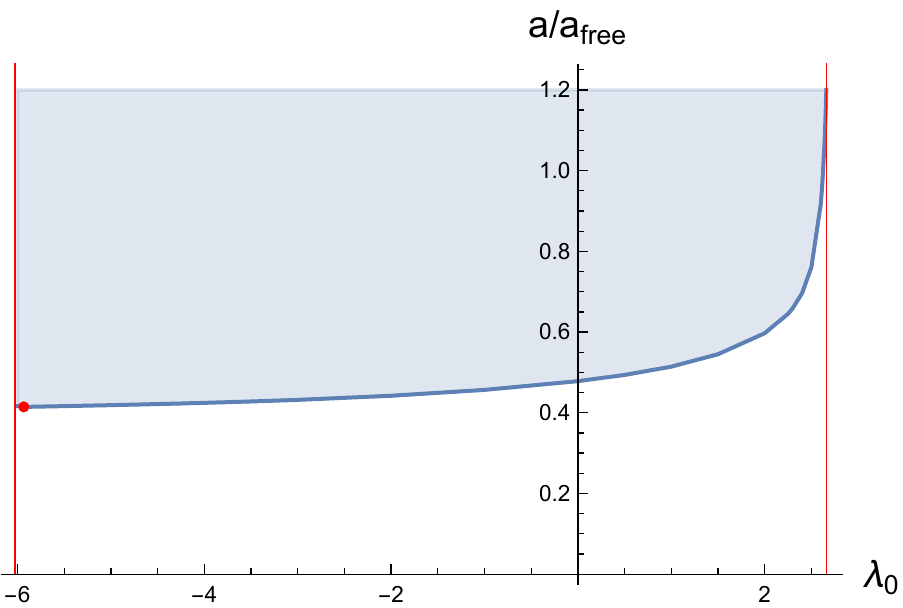}
	\includegraphics[width=0.45\textwidth]{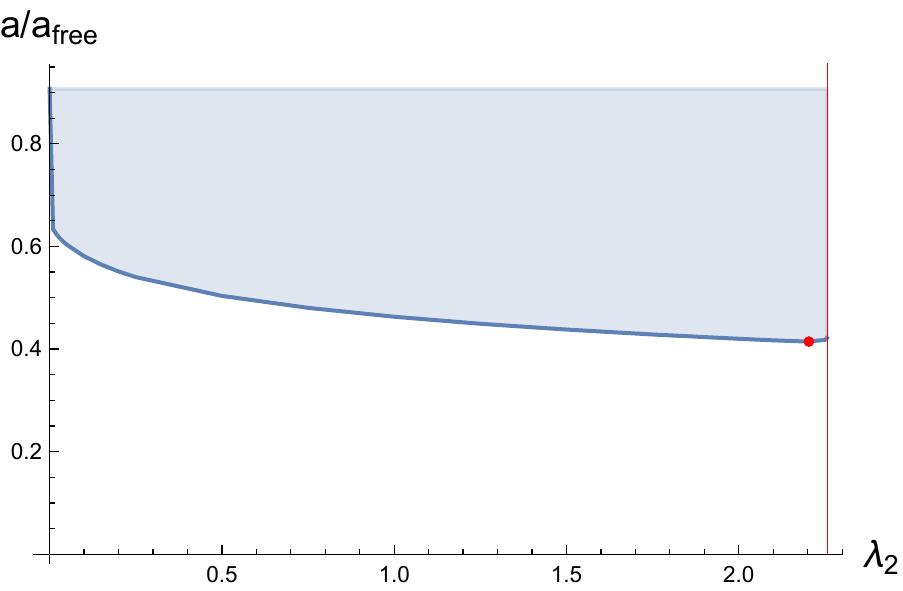}
	\caption{Minimum of the $a$-anomaly of the UV CFT as a function of the parameters $\lambda_0$ and $\lambda_2$ defined in \eqref{eq:lambdas}.
	The red dot marks the absolute minimum. The red vertical lines indicate the boundaries of the allowed regions for $\lambda_0$ and $\lambda_2$.
	}
	\label{fig:plot_bounds_lambda}
\end{figure}

\begin{figure}
	\centering
	\includegraphics[width=0.45\textwidth]{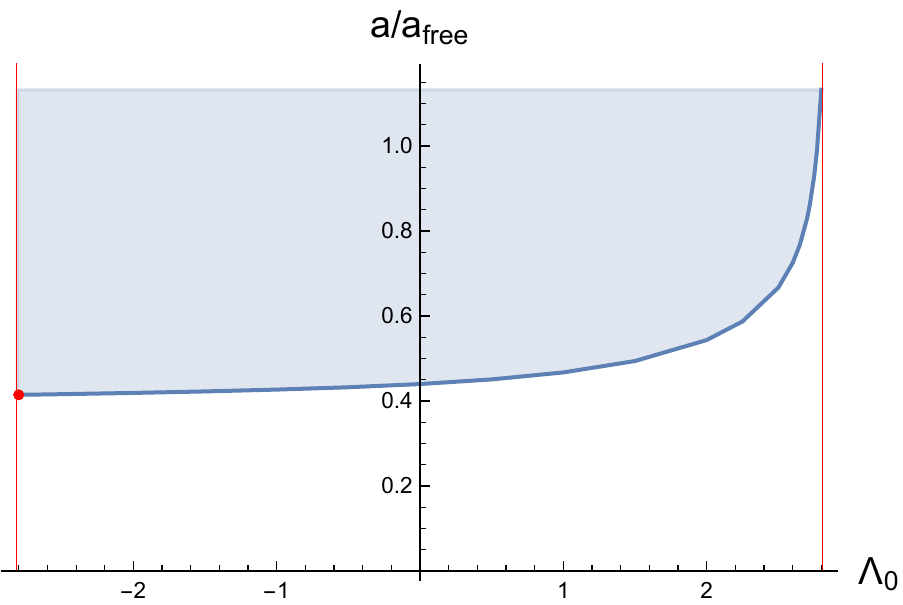}
	\includegraphics[width=0.45\textwidth]{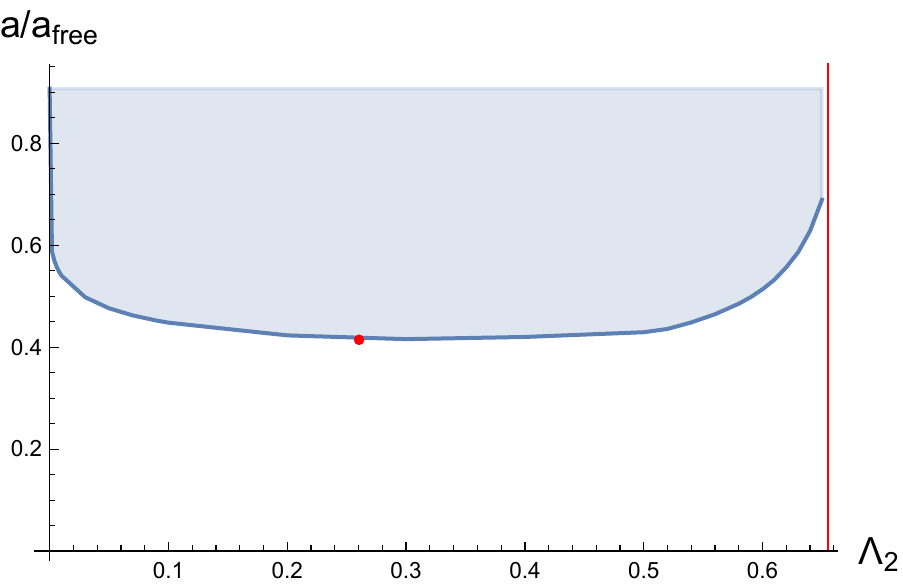}
	\caption{Minimum of the $a$-anomaly of the UV CFT as a function of the parameters $\Lambda_0$ and $\Lambda_2$ defined in \eqref{eq:Lambdas}.
	The red dot marks the absolute minimum. The red vertical lines indicate the boundaries of the allowed regions for $\Lambda_0$ and $\Lambda_2$.}
	\label{fig:plot_bounds_Lambda}
\end{figure}

Crossing, unitarity and analyticity put strong bounds on the above parameters. For instance we found that\footnote{These numerical bounds are a rough estimate based on our numerical results described in section \ref{sec:num}. The exception is the upper bound on $\lambda_0$ which can be determined quite precisely  \cite{Paulos:2017fhb}. For the recent more detailed study of these observables see \cite{Chen:2022nym}.}
\begin{align}
	-6 \lesssim&\,\lambda_0\lesssim + 2.6613,\qquad
	&&0\leq \lambda_2\lesssim +2 ,\\
	-3 \lesssim&\,\Lambda_0 \lesssim + 3 ,\qquad
	&&0\leq \Lambda_2\lesssim +0.7 .
\end{align}
The minimum of the $a$-anomaly as a function of the above parameters is given in figures \ref{fig:plot_bounds_lambda} and \ref{fig:plot_bounds_Lambda}.
 All consistent QFTs must live in the allowed region which we shaded in blue. We mark the absolute minimum of the a-anomaly with a red dot in these figures.  
Our best numerical estimate is
\begin{equation}
\label{amin}
	a/a_\text{free} \gtrsim 0.3\,,
\end{equation}
with $a_\text{free}$ the $a$-anomaly of a free scalar field.
We refer the reader to section \ref{sec:num} for a detailed discussion of the numerical uncertainties of these results.
We do not know if there is any 4d QFT that saturates the lower bound \eqref{amin}. In fact, we do not know of any theory with an $a$-anomaly smaller than  $a_\text{free}$.
We conclude in section \ref{sec:conc} with a discussion of open questions and future work.

\section{Review of classic results}
\label{S:review}
In this paper we work in $4d$ Minkowski flat space with  metric: 
\begin{equation}
	\label{eq:flat_metric}
\eta_{\mu\nu} = \eta^{\mu\nu} = \text{diagonal}\{ -1,+1,+1,+ 1 \}.
\end{equation}
All the quantum field theories have a very special operator called  the stress-tensor $T^{\mu\nu}(x)$. It is symmetric in its two Lorentz indices and obeys conservation law, namely
\begin{equation}
T^{\mu\nu}(x) = T^{\nu\mu}(x),\qquad
\partial_\mu T^{\mu\nu}(x) =0.
\end{equation}

\subsection{Stress tensor and trace anomaly in CFTs }
\label{S:trace_anomaly_from_TTT_correlator}
Let us start the discussion by considering conformally invariant quantum field theory. The conformal symmetry puts severe constraints on the form of correlation functions. In \cite{Osborn:1993cr} it was shown that the most general two- and three-point functions of the stress tensor in CFTs have the following form
\begin{align}
\label{eq:TT}
\< 0|T^{\mu\nu}(x_1) T^{\rho\sigma}(x_2)|0\> &= \f{C_T}{x_{12}^8} \times \mathbf{T}^{\mu\nu;\rho\sigma}_0,\\
\label{eq:TTT}
\< 0|T^{\mu\nu}(x_1) T^{\rho\sigma}(x_2)T^{\alpha\beta}(x_3)|0\> &=
\f{1}{x_{12}^{4}x_{23}^{4}x_{31}^{4}}\Big(\mathbb{A} \mathbf{T}^{\mu\nu;\rho\sigma;\alpha\beta}_1+\mathbb{B} \mathbf{T}^{\mu\nu;\rho\sigma;\alpha\beta}_2 +\mathbb{C} \mathbf{T}^{\mu\nu;\rho\sigma;\alpha\beta}_3\Big).
\end{align}
Here the objects $\mathbf{T}_0$, $\mathbf{T}_1$, $\mathbf{T}_2$ and $\mathbf{T}_3$ take care of the correct behaviour of the correlation functions under conformal transformations. They are called tensor structures. The standard basis for these tensor structures is defined in appendix \ref{app:tensor_structures}.
The coefficient $C_T\geq 0$ is usually referred to as the central charge. The coefficients $\mathbb{A}$, $\mathbb{B}$ and $\mathbb{C}$ are called the OPE coefficients since they appear in the OPE expansion of the stress-tensor with itself. All of them are real quantities. Due to the conformal Ward identities\footnote{All the generators of the conformal transformation can be written as certain integral of the stress-tensor, see for example \cite{Osborn:1993cr}. By performing appropriate integrals over one stress-tensor in \eqref{eq:TTT} and using the properties of the generators we effectively obtain the two-point function \eqref{eq:TT}.} the following relation holds
\begin{equation}
C_{T}\ =\ \f{\pi^2}{3}\Big(14\mathbb{A}-2\mathbb{B}-5\mathbb{C}\Big).
\end{equation}
Summarizing, there are three independent parameters describing the 2- and the 3-point function of the stress-tensor. One can choose these three parameters to be for example $\{C_T,\;\mathbb{A},\;\mathbb{B}\}$. It is standard to also define the following quantities
\begin{equation}
\label{eq:ac-coefficients}
a \equiv
\f{\pi^4}{64\times 90}\Big(9\mathbb{A}-2\mathbb{B}-10\mathbb{C}\Big),\qquad
c \equiv \f{\pi^4}{64\times 30}\Big(14\mathbb{A}-2\mathbb{B}-5\mathbb{C}\Big) 
=\f{\pi^2}{64\times 10}C_T.
\end{equation}
Finally we recall that in CFTs the trace of the stress-tensor vanishes, namely
\begin{equation}
\label{eq:trace_cft}
T^\mu{}_\mu(x) =0.
\end{equation}

Let us now discuss CFTs on the curved background which is described by the metric $g_{\mu\nu}(x)$. Conformal invariance on a curved background is achieved by requiring $\text{diff}\times \text{Weyl}$ invariance. We recall that the Weyl transformation is defined as
\begin{equation}
\label{eq:weyl_transformation}
g_{\mu\nu}(x) \rightarrow e^{2\sigma(x)}g_{\mu\nu}(x),\qquad
\cO(x)\rightarrow e^{-\Delta_\cO \sigma(x)}\cO(x),
\end{equation}
where $\sigma(x)$ is an arbitrary scalar function, $\cO(x)$ is a local scalar operator  and $\Delta_\cO$ is the scaling dimension of the operator $\cO(x)$.
Contrary to the flat space-time where \eqref{eq:trace_cft} holds, for CFTs on the curved background we instead have
\begin{equation}
\label{eq:trace_anomay}
\<0| T^\mu{}_\mu(x) |0\>_g =\ -a \times E_{4} + c \times W^2 ,
\end{equation}
where $E_4$ is the Euler density defined in \eqref{eq:euler} and $W^2$ is the square of Weyl tensor defined in \eqref{eq:weyl}. The subscript $g$ in the left-hand side of \eqref{eq:trace_anomay} indicates that the CFT is on the curved background rather than on the flat one. The coefficients $a$ and $c$ are exactly the ones introduced in 
\eqref{eq:ac-coefficients}. They are called the Weyl anomalies as well as trace anomalies. 
The name Weyl anomaly is appropriate because  exact Weyl invariance implies $\<0| T^\mu{}_\mu(x) |0\>_g=0$.

\subsection{Compensator field and the dilaton particle}
\label{KS_setup}
Let us define a generic quantum field theory as the renormalization group flow from the UV to the IR fixed points which are described by the UV and the IR conformal field theories. Such a theory in curved background can be described by the action
\begin{equation}
\label{eq:action}
A(g,M_i) \equiv A_\text{UV\;CFT}(g) + A_\text{deformation}(g,M_i),
\end{equation}
where the deformation of the UV CFT has the form
\begin{equation}
\label{eq:deformation_T}
A_\text{deformation}(g,M_i) =  \sum_{i}\int d^4 x \sqrt{-g}   \left( \lambda_i M_i^{4-\Delta_i}\cO_i(x)\right).
\end{equation}
Here $\cO_i(x)$ are relevant scalar UV CFT operators (operators obeying $\Delta_i<4$), $\lambda_i$ are dimensionless coefficients and $M_i$ are the mass scales which control when the deformation due to a particular operator becomes important. The explicit dependence on $g_{\mu\nu}(x)$ indicates that we work on a generic curved background. The QFT in flat space-time is recovered by setting  $g_{\mu\nu}(x)$ to the flat metric \eqref{eq:flat_metric}.
The determinant of the metric is defined as follows
\begin{equation}
g\equiv \text{det}\ g_{\mu\nu}(x).
\end{equation}
The action \eqref{eq:deformation_T} is diff invariant by construction.

In curved background the stress-tensor is defined as
\begin{equation}
T^{\mu\nu}(x) = \frac{2}{\sqrt{-g}}\,\frac{\delta A(g,M_i)}{\delta g_{\mu\nu}(x)}.\label{Tmunu}
\end{equation}
Under the Weyl transformation \eqref{eq:weyl_transformation}, the trace of the stress-tensor can be defined as a variation of the action with respect to the infinitesimal Weyl transformation parameter $\sigma$ in the following way,
\begin{equation}
T^{\mu}_{\mu}(x) \equiv  \frac{1}{\sqrt{-g}}\,\frac{\delta_{W} A(g,M_i)}{\delta \sigma(x)}.\label{trT}
\end{equation}
Performing the Weyl transformation \eqref{eq:weyl_transformation} in \eqref{eq:action} and focusing on   flat space-time we obtain the trace of the stress-tensor using the above definition, which reads
\begin{equation}
\label{eq:trace_defomration}
g_{\mu\nu}=\eta_{\mu\nu}:\qquad
T^\mu{}_\mu(x)=
 \sum_{i} \lambda_i (4-\Delta_i) M_i^{4-\Delta_i} \cO_i(x).
\end{equation}
This is the standard result in QFT, namely the trace is proportional to the deforming operators.

The correlation functions of the stress-tensor both in the UV and IR are described by \eqref{eq:TT} and \eqref{eq:TTT} where the coefficients $C_T$, $\mathbb{A}$, $\mathbb{B}$ and $\mathbb{C}$ have an additional label UV and IR respectively. 
Out of all the above coefficients the $a$ trace anomaly is the most interesting. In \cite{Cardy:1988cwa} it was conjectured that
\begin{equation}
\label{eq:a-theorem}
a^\text{UV} - a^\text{IR} \geq 0,
\end{equation}
where the equality can hold only if there is no flow and the theory remains conformal, in other words if $A_\text{deformation}=0$ in \eqref{eq:action}. The inequality \eqref{eq:a-theorem} is known as the $a$-theorem. It was shown to hold in perturbation theory in \cite{Osborn:1989td,Jack:1990eb}. It was proven non-perturbatively in \cite{Komargodski:2011vj}, for further discussion see also \cite{Komargodski:2011xv,Luty:2012ww,Komargodski:2015grt}. The proof of \cite{Komargodski:2011vj} gives also the prescription on how to probe/compute the difference $(a^\text{UV} - a^\text{IR})$ in a given QFT. One of the main ingredients of this proof is the compensator field and the associated particle which we call the dilaton. In the rest of this section we will define the compensator field and the dilaton particle.

Let us work with the action \eqref{eq:action} on a curved background. It is diff invariant but not Weyl invariant. There are two sources which break the Weyl symmetry, namely the trace anomaly of the UV and IR CFTs given by \eqref{eq:trace_anomay} and the deformation part of the action $A_\text{deformation}$ in \eqref{eq:action} which explicitly depends on the scale.
The latter breaking can be compensated for in a modified theory with the following action
\begin{equation}
\label{eq:action_modified}
A^{\prime}(g,M_i, \Omega) \equiv  A(g,M_i \Omega(x))
+A_\text{dynamics}(g, \Omega),
\end{equation} 
where $\Omega(x)$ is a real scalar field called the compensator field and the action $A$ was defined in \eqref{eq:action}. Both the metric $g_{\mu\nu}(x)$ and the compensator field  $\Omega(x)$ are  non-dynamical fields. We can however promote them to dynamical probe fields by adding a kinetic term $A_\text{dynamics}(g, \Omega)$. We will discuss possible convenient choices for this term in the end of this subsection.  The compensator field $\Omega(x)$ can be represented in the following two ways
\begin{equation}
\label{eq:dilaton}
\Omega(x) = e^{-\tau(x)}= 1-\frac{\varphi(x)}{\sqrt{2}f}.
\end{equation}
We refer to the real scalar fields $\tau(x)$ and $\varphi(x)$ also as the compensator or the dilaton fields interchangeably. 
Here $f$ is a new parameter with mass dimension one. The following relation holds
\begin{equation}
\tau(x) = \frac{\varphi(x)}{\sqrt{2}f} + O\left(\frac{1}{f^2}\right).
\end{equation}
Let us now emphasize that the action \eqref{eq:action_modified} can be made invariant under the Weyl transformation \eqref{eq:weyl_transformation}
given that the dilaton transforms in the following way
\begin{equation}
\tau(x) \rightarrow \tau(x) + \sigma(x)
\end{equation}
and that the term $A_\text{dynamics}$ is chosen appropriately.
The particle created by the compensator (or the dilaton) field $\varphi(x)$ from the vacuum is called the dilaton particle. It will be denoted by $B$ throughout this paper.

The simplest choice for $A_\text{dynamics}(g, \Omega)$, already used in \cite{Komargodski:2011vj}, reads as
\begin{equation}
	\label{eq:dynamics}
	A_\text{dynamics}(g, \Omega) =  \frac{1}{6}\, f^2 \int d^4x\ \sqrt{-\widehat{g}}\, R(\widehat{g}), 
\end{equation}
where we have defined
\begin{equation}
	\label{eq:ghat}
	\widehat{g}_{\mu\nu} \equiv e^{-2\tau}g_{\mu\nu}.
\end{equation}
The action \eqref{eq:dynamics} is Weyl invariant at the classical level but not at the quantum level. One can simply see this for instance by taking the $f\rightarrow\infty$ limit and focusing on the flat space. The action \eqref{eq:dynamics} then reduces to the standard kinetic term describing free massless scalar. Free massless scalar gives a particular example of a CFT with trace anomalies $a$ and $c$ reported in \eqref{eq:ac_free_scalar}. The latter break Weyl invariance via \eqref{eq:trace_anomay}. There are many other possible choices of $A_\text{dynamics}(g, \Omega)$. Let us stress, that these choices do not have to be Weyl invariant even classically.

Let us now focus on  flat space-time $g_{\mu\nu}=\eta_{\mu\nu}$.
Using \eqref{eq:deformation_T}, \eqref{eq:trace_defomration}, \eqref{eq:dilaton} and \eqref{eq:dynamics} we can rewrite the modified action \eqref{eq:action_modified} in the following equivalent way\footnote{If in the UV CFT there exists more than one relevant operator which can be used to deform the theory (i.e. for $i>1$), the neglected terms starting from order $O(f^{-2})$ in equation \eqref{eq:action_modified_equivalent} can not be expressed in terms of the trace of the stress-tensor in general. Indeed,  the order $O(f^{-2})$ contribution in $A'(M_i)$ turns out to be $\f{1}{4f^2}\sum\limits_{i} (4-\Delta_i)(3-\Delta_i)\int d^4x \ \varphi^2(x)\left( \lambda_i M_i^{4-\Delta_i}\cO_i(x)\right)$, which is not expressible in terms of the trace of the stress tensor given in equation \eqref{eq:trace_defomration} for $i>1$. }
\begin{equation}
\label{eq:action_modified_equivalent}
A'(M_i,\varphi) \equiv  A(M_i) - \frac{1}{\sqrt{2}f}\int d^4 x\; T^\mu{}_\mu(x) \varphi(x)-
 \int d^4x \left(\frac{1}{2}\partial_\mu\varphi(x) \partial^\mu\varphi(x)\right)+ O\left(\frac{1}{f^2}\right).
\end{equation} 
From \eqref{eq:action_modified_equivalent} it becomes obvious that in the limit $f\rightarrow \infty$ the interaction between the dilaton and the rest of the system disappears and the action \eqref{eq:action_modified} simply becomes the original one plus the freely propagating dilaton field.
The dilaton field $\varphi(x)$ should be seen as a probe for a given QFT which does not disturb it in the limit $f\rightarrow \infty$.

\subsection{$a$-theorem}\label{S:KS_a_theorem_review}
Let us now take the UV and IR limits of the action  \eqref{eq:action_modified}. They can be written as\footnote{Let us emphasize that we have made here a very non-trivial statement that the IR dilaton EFT action completely decouples from the IR CFT even though dilaton self interaction is present. One can argue for this at least in the limit of flat space-time: by construction \eqref{eq:action_modified}, dilaton couples only to mass parameters, IR CFT instead does not have dimensionful parameters.}
\begin{equation}
\label{eq:IR_action}
\begin{aligned}
A'_\text{UV}(g,\Omega) &\equiv A_\text{UV CFT}(g) + A_\text{dynamics}(g, \Omega),\\
A'_\text{IR}(g,\Omega)  &\equiv  A_\text{IR CFT}(g) + A_\text{dilaton EFT}(g, \Omega) + A_\text{dynamics}(g, \Omega).
\end{aligned}
\end{equation}
Here $A_\text{dilaton EFT}$ is the effective field theory (EFT) action describing the dilaton interaction at low energy.
In order to obtain it in some explicit QFT model one needs to integrate out all the ``massive'' degrees of freedom throughout the RG flow which is almost impossible in practice. Luckily there is a model independent way to compute $A_\text{dilaton EFT}$ which we will now review.\

Consider the action \eqref{eq:action_modified}. It breaks Weyl invariance in a very special way. The Weyl symmetry breaking is coming only from the UV and IR fixed points \eqref{eq:IR_action}. Taking into account \eqref{eq:trace_anomay} we get
\begin{equation}
	\label{eq:transformations_1}
	\begin{aligned}
		\delta_W A^{\prime}_\text{UV}(g,\Omega) &= \int d^4 x \sqrt{-g} \sigma(x)\,\Big(-a^\text{UV} \times E_{4} + c^\text{UV} \times W^2\Big) + \delta_W A_\text{dynamics}(g, \Omega),\\
		\delta_W A^{\prime}_\text{IR}(g,\Omega) &= \int d^4 x \sqrt{-g} \sigma(x)\,\Big(-a^\text{IR} \times E_{4} + c^\text{IR} \times W^2\Big)\\
		&+\delta_WA_\text{dilaton EFT}(g, \Omega)+ \delta_W A_\text{dynamics}(g, \Omega).
	\end{aligned}
\end{equation}
Here $\delta_W$ stands for the infinitesimal Weyl variation.

Let us now assume that the Weyl anomaly in the UV matches the Weyl anomaly in the IR, in other words
\begin{equation}
\label{eq:matching_condition}
\delta_WA^{\prime}_\text{UV}(g,\Omega)=
\delta_WA^{\prime}_\text{IR}(g,\Omega).
\end{equation}
Notice that contrary to the 't Hooft anomaly matching, there is no proof for the Weyl anomaly matching \eqref{eq:matching_condition} and it might not be true in general.\footnote{In fact the authors of \cite{Niarchos:2020nxk}, see also \cite{Andriolo:2022lcb}, found an apparent mismatch of the $c$-anomaly on the Higgs branch of $\mathcal{N}=2$ super-conformal field theory where conformal symmetry is spontaneously broken.} For further discussion on Weyl anomaly matching and its consequences see \cite{karateev:2023}.
Plugging  \eqref{eq:transformations_1} in \eqref{eq:matching_condition} we obtain the following variational equation
\begin{equation}
	\delta_WA_\text{dilaton EFT}(g, \Omega) = \int d^4 x \sqrt{-g} \sigma(x)\,\Big(-\left(a^\text{UV} -a^\text{IR} \right)\times E_{4} + \left(c^\text{UV} -c^\text{IR}\right) \times W^2\Big).
\end{equation}
The most general solution for this equation can be written in the following form
\begin{equation}
\label{eq:dilaton_effective}
A_\text{dilaton EFT}(g, \Omega) =  -(a^\text{UV}-a^\text{IR})\times A_a(g,\Omega) + (c^\text{UV}-c^\text{IR}) \times A_c(g,\Omega)+\ A_{\text{invariant}}(g,\Omega),
\end{equation}
where the two newly introduced terms $A_a(g,\Omega)$ and $A_c(g,\Omega)$  behave in the following way under the infinitesimal Weyl transformation
\begin{equation}
\label{eq:transformations_2}
\delta_W A_a(g,\Omega) = \int d^4 x \sqrt{-g} \sigma(x)\, E_{4},\qquad
\delta_W A_c(g,\Omega) = \int d^4 x \sqrt{-g} \sigma(x)\, W^2.
\end{equation}
The term $A_{\text{invariant}}$ instead remains completely invariant.
The solution to the above requirement was found in \cite{Schwimmer:2010za, Komargodski:2011vj}, it reads
	\begin{align}
		\nn
		A_a(g,\Omega) &= \int d^4 x \sqrt{-g}\left( \tau E_{4}+4\left(R^{\mu\nu}-\frac{1}{2}g^{\mu\nu}R\right)(\partial_\mu\tau)(\partial_\nu\tau)
		-4(\partial\tau)^2(\partial^2\tau)+2(\partial\tau)^4
		\right),\\ 
		A_c(g,\Omega) &= \int d^4 x \sqrt{-g} \tau(x)\, W^2.
		 \label{eq:KS_solution}
	\end{align}
This solution is not easy to obtain but it is easy to check that it satisfies \eqref{eq:transformations_2}. It is also important to stress that even though \eqref{eq:matching_condition} might not hold for every QFT, the weaker Wess-Zumino consistency condition exists, see  \cite{karateev:2023}, which implies that at the very least the first line in \eqref{eq:KS_solution} always holds true.

The most general Weyl invariant action can be parametrized as follows
\begin{equation}
	A_\text{invariant}(g,\Omega) =\int d^4x\ \sqrt{-\widehat{g}}\Bigg(M^4\lambda+M^2r_0\widehat R\ +\ r_1 \widehat R^{2}+r_2 \widehat W^{2}+r_3 \widehat E_4+\ldots\Bigg). \label{eq:Action_invariant}
\end{equation}
Here the   Ricci scalar, Weyl tensor and the Euler density are built out of the metric \eqref{eq:ghat}. The real dimensionless parameters $\lambda$, $r_0$, $r_1$, $r_2$ and $r_3$ depend on a particular QFT. The EFT cut-off scale $M$ can be chosen to be the lowest deformation energy scale of the UV CFT. In spontaneously broken QFTs $\lambda=0$, but in generic QFTs $\lambda\neq 0$. However, when defining the action \eqref{eq:action_modified}, if needed, one can fine tune the counterterms in such a way that $\lambda=0$.

In flat space the solution \eqref{eq:dilaton_effective} together with  \eqref{eq:KS_solution} and \eqref{eq:Action_invariant} simply leads to
\begin{multline}
	\label{eq:IR_action_interacting}
	A_\text{dilaton EFT}(\varphi) =\f{M^4 \lambda}{4f^4}\int d^4x \big(\varphi(x))^4\ +\f{6M^2 r_0}{f^2}\int d^4 x\Big(-\f{1}{2}\p_\mu\varphi(x)\p^\mu\varphi(x)\Big) \\
	+36r_1\int d^4 x\Big(\f{1}{2f^2}+\f{\varphi(x)}{\sqrt{2}f^3}+\f{3\varphi(x)^2}{4f^4}\Big)(\p^2\varphi(x))^2\\
	+ \frac{a^\text{UV}-a^\text{IR}}{2f^4}\times \int d^4x \ \big(\partial \varphi(x)\big)^4 +\ O\big( f^{-5}\p^4\varphi^5\ ,\  f^{-4}\p^6\varphi^4\big).
\end{multline}
The term proportional to $r_0$ gives an $O(f^{-2})$ correction to the dilaton kinetic term coming from $A_\text{dynamics}(g, \Omega)$. In the limit $f\rightarrow \infty$ it should be neglected.
The interacting part of the dilaton scattering process $B(p_1)B(p_2) \rightarrow B(p_3)B(p_4)$ at low energy is described by the effective action  \eqref{eq:IR_action_interacting} and has the following form
\begin{equation}
\label{eq:sum_rule}
\lim_{f\rightarrow \infty} f^4 \mathcal{T}_{BB\rightarrow BB}(s,t,u) =6M^4 \lambda + (a^\text{UV} - a^\text{IR})\times(s^2+t^2+u^2) + O(s^3).
\end{equation}
where $s=-(p_1+p_2)^2\ ,\ t=-(p_1-p_3)^2\ ,\ u=-(p_1-p_4)^2$ with $s+t+u=0$. Note, that the term proportional to $r_1$ in \eqref{eq:IR_action_interacting} does not contribute to this scattering amplitude, since it vanishes under the dilaton equation of motion. Using the standard approach one can write the following dispersion relation in the $f\rightarrow\infty$ limit
\be
a^\text{UV}-a^\text{IR}\ &=&\ \f{f^{4}}{2}\ \f{1}{2\pi i}\oint_{0}\f{ds}{s^3} \ \mathcal{T}_{BB \rightarrow BB}(s,0,-s)\nn\\
&=&\ \f{f^4}{\pi}\ \int_{0}^\infty\f{ds}{s^3}\ \text{Im}\mathcal{T}_{BB \rightarrow BB}(s,0,-s).
\label{eq:dispersion_relation_general}
\ee
Since $\text{Im}\mathcal{T}_{BB\rightarrow BB}(s,0,-s)= s\ \sigma(s) \geq 0$  where $\sigma(s)$ is the total cross section for the scattering of $BB\rightarrow \text{anything}$, $(a^\text{UV}-a^\text{IR})$ is non-negative. This proves the $a$-theorem.

\paragraph{Application in free scalar theory}
As an application of the $a$-theorem consider the UV CFT which is generated by the free massless field $\Phi(x)$, namely we have the action
\begin{equation}
A_\text{UV\;CFT} = - \int d^4 x \left(\frac{1}{2}\partial_\mu\Phi(x) \partial^\mu\Phi(x)\right).
\end{equation}
It is straightforward to compute then the two- and tree-point correlation functions of the stress-tensor with itself. One obtains \eqref{eq:TT} and \eqref{eq:TTT} with
\begin{equation}
\label{eq:constants_UV}
C_T^\text{UV} = \frac{1}{3\pi^4},\qquad
\mathbb{A}^\text{UV}=\f{1}{27\pi^6}\ ,\qquad \mathbb{B}^\text{UV}=-\f{4}{27\pi^6}\ ,\qquad \mathbb{C}^\text{UV}=-\f{1}{27\pi^6}.
\end{equation}
Let us now add the following deformation
\begin{equation}
A_\text{deformation}(m) = -\frac{1}{2} m^2 \Phi(x)^2, \label{mass_deformation}
\end{equation}
where $m$ becomes the mass of the field $\Phi$. This triggers the flow to an empty IR fixed point, thus in the deep IR we simply have
\begin{equation}
\label{eq:constants_IR}
C_T^\text{IR} = 0,\qquad
\mathbb{A}^\text{IR} =0,\qquad
\mathbb{B}^\text{IR} = 0,\qquad
\mathbb{C}^\text{IR} = 0.
\end{equation}
As a result according to \eqref{eq:ac-coefficients} we get the following UV an IR a-anomaly
\begin{equation}
\label{eq:a_free}
a^\text{UV} =  \frac{1}{5760\pi^2},\qquad
c^\text{UV} =  \frac{1}{1920\pi^2},\qquad
a^\text{IR} =  0,\qquad
c^\text{IR}  = 0.
\end{equation}

Using the the modified action \eqref{eq:action_modified_equivalent} we can also compute the dilaton scattering at low energies. We get
\begin{equation}
\label{eq:dilaton_example}
\lim_{f\rightarrow \infty} f^4 \mathcal{T}_{BB \rightarrow BB}(s,t,u) = \frac{1}{5760\pi^2}\times(s^2+t^2+u^2) + O(s^3).
\end{equation}
The details of the computation are given in appendix \ref{sec:example}. The result \eqref{eq:dilaton_example} is in a perfect agreement with the $a$-theorem sum-rule \eqref{eq:sum_rule} and \eqref{eq:a_free}.

\section{Non-perturbative S-matrix bootstrap setup}
\label{sec:S-matrix_setup}

In this section we explain our framework in details. We define scattering and partial amplitudes of particles $A$ and $B$ in section \ref{sec:amplitudes}. We write down all the crossing equations for the scattering amplitudes in section \ref{sec:crossing}. We construct unitarity conditions on the partial amplitudes in section \ref{sec:unitarity} assuming that $B$ is a generic massless particle. In section \ref{sec:dilaton} we discuss further restrictions on the scattering amplitudes and modifications of the unitarity conditions in the case when $B$ is the dilaton in the $f\rightarrow \infty$ limit.

\subsection{Scattering and partial amplitudes}
\label{sec:amplitudes}
Let us consider a QFT which contains two different scalar particles $A$ and $B$ with the masses
\begin{equation}
	\label{eq:masses}
	m_A=m,\qquad
	m_B=0.
\end{equation}
Here and in the next two section we will not impose that $B$ is the dilaton particle, it will be enough to treat it as a generic massless scalar particle. The requirement that $B$ is the dilaton will be imposed only in section \ref{sec:dilaton}. The particles $A$ and $B$ are described by the following asymptotic states
\begin{equation}
|A\>_{in} \equiv |m_A,\vec p\,\>_{in},\quad
|A\>_{out}\equiv |m_A,\vec p\,\>_{out},\quad
|B\>_{in} \equiv |m_B,\vec p\,\>_{in},\quad
|B\>_{out}\equiv |m_B,\vec p\,\>_{out}.
\end{equation}

For simplicity we assume the presence of the $\ZZ$ symmetry. We require that the particle of type $A$ is odd and the particle of type $B$ is even under this symmetry. In other words
\begin{equation}
\ZZ:\qquad
|A\>_{in}\rightarrow - |A\>_{in},\quad
|B\>_{in}\rightarrow + |B\>_{in}.
\end{equation}
The same transformation properties hold for the out states.

In this work we will be interested in the following scattering amplitudes
\begin{equation}
\label{eq:processes}
AA \rightarrow AA,\qquad
AA \rightarrow BB,\qquad
AB \rightarrow AB,\qquad
BB \rightarrow BB.
\end{equation}
The presence of the $\ZZ$ symmetry required above forbids all the two to two scattering process with an odd number of particles of type $A$. We define the scattering amplitudes describing the processes \eqref{eq:processes} as
\begin{equation}
\label{eq:amplitudes}
\begin{aligned}
\mathcal{S}_{AA \rightarrow AA}(s,t,u)\times(2\pi)^4\delta^4(p_1+p_2-p_3-p_4) &\equiv
{}_{out}\<A_3,A_4|A_1,A_2\>_{in},\\
\mathcal{S}_{AA \rightarrow BB}(s,t,u)\times(2\pi)^4\delta^4(p_1+p_2-p_3-p_4) &\equiv
{}_{out}\<B_3,B_4|A_1,A_2\>_{in},\\
\mathcal{S}_{AB \rightarrow AB}(s,t,u)\times(2\pi)^4\delta^4(p_1+p_2-p_3-p_4) &\equiv
{}_{out}\<A_3,B_4|A_1,B_2\>_{in},\\
\mathcal{S}_{BB \rightarrow BB}(s,t,u)\times(2\pi)^4\delta^4(p_1+p_2-p_3-p_4) &\equiv
{}_{out}\<B_3,B_4|B_1,B_2\>_{in}.
\end{aligned}
\end{equation}
Here we used the following short-hand notation for describing the two-particle in and out asymptotic states
\begin{equation}
\label{eq:2PS}
\begin{aligned}
|A_1,A_2\>_{in} &\equiv \frac{1}{\sqrt{2}}\Big(
|m_A,\vec p_1\,\>_{in}\otimes |m_A,\vec p_2\,\>_{in}+
|m_A,\vec p_2\,\>_{in}\otimes |m_A,\vec p_1\,\>_{in}
\Big),\\
|A_1,B_2\>_{in} &\equiv 
|m_A,\vec p_1\,\>_{in}\otimes |m_B,\vec p_2\,\>_{in},\\
|B_1,B_2\>_{in} &\equiv \frac{1}{\sqrt{2}}\Big(
|m_B,\vec p_1\,\>_{in}\otimes |m_B,\vec p_2\,\>_{in}+
|m_B,\vec p_2\,\>_{in}\otimes |m_B,\vec p_1\,\>_{in}
\Big),
\end{aligned}
\end{equation}
where $\otimes$ denotes the ordered tensor product.
Notice that the two-particle states which consist of particles of type $A$ or $B$ only are totally symmetric. The factor $\sqrt{2}$ is introduced to comply with the standard normalization conditions which will be provided below. The Mandelstam variables are defined as
\begin{equation}
\label{eq:mandelstam}
s\equiv -(p_1+p_2)^2,\quad
t\equiv -(p_1-p_3)^2,\quad
u\equiv -(p_1-p_4)^2.
\end{equation}
As usual these are not linearly independent due to the following relations
\begin{equation}
\label{eq:stu_constraint}
s+t+u=m_1^2+m_2^2+m_3^2+m_4^2,
\end{equation} 
where $m_i$ are the masses of particles participating in the two to two scattering. This constraint is thus different for each process in \eqref{eq:amplitudes}. Let us write here this constraint precisely for each process separately
\begin{equation}
\begin{aligned}
AA\rightarrow AA:&\quad s+t+u=4m^2,\\
AA\rightarrow BB:&\quad s+t+u=2m^2,\\
AB\rightarrow AB:&\quad s+t+u=2m^2,\\
BB\rightarrow BB:&\quad s+t+u=0.
\end{aligned}
\end{equation}

It is useful to write the four-momentum $p^\mu$ in the spherical coordinates. One has
\begin{equation}
p^\mu = \{p^0, \vec p\,\},\qquad
\vec p = \{\myP\cos\phi\sin\theta,\, \myP\sin \phi\sin \theta,\, \myP\cos \theta\},\qquad
\myP\equiv |\vec p\,|.
\end{equation}
Let us discuss the normalization of states. The same normalizations holds for both  in and out states. As a result below we write explicitly only the normalization of in states.
The one-particle states are normalized in the following way
\begin{equation}
\begin{aligned}
{}_{in}\<m_A,\vec p_1|m_A,\vec p_2\>_{in} &= 2\,\sqrt{m^2+\myP_1^2}
\times(2\pi)^3\delta^{(3)}(\vec p_1-\vec p_2),\\
{}_{in}\<m_A,\vec p_1|m_B,\vec p_2\>_{in} &= 0,\\
{}_{in}\<m_B,\vec p_1|m_B,\vec p_2\>_{in} &= 2\myP_1
\times(2\pi)^3\delta^{(3)}(\vec p_1-\vec p_2).
\end{aligned}
\end{equation}
The normalization of two-particle states \eqref{eq:2PS} follows immediately and read
\begin{align}
\nn
{}_{in}\<A_3,A_4|A_1,A_2\>_{in} &= 4\,(2\pi)^6\,\sqrt{m^2+\myP_1^2}\sqrt{m^2+\myP_2^2}\\
\nn
&\times\Big(\delta^{(3)}(\vec p_1-\vec p_3)\delta^{(3)}(\vec p_2-\vec p_4)
+\delta^{(3)}(\vec p_1-\vec p_4)\delta^{(3)}(\vec p_2-\vec p_3)\Big),\\
\nn
{}_{in}\<B_3,B_4|A_1,A_2\>_{in} &= 0,\\
\label{eq:normalization_2PS}
{}_{in}\<A_3,B_4|A_1,B_2\>_{in} &= 4\,(2\pi)^6\,\myP_2\,\sqrt{m^2+\myP_1^2}
\times\delta^{(3)}(\vec p_1-\vec p_3)\delta^{(3)}(\vec p_2-\vec p_4),\\
{}_{in}\<B_3,B_4|B_1,B_2\>_{in} &= 4\,(2\pi)^6\,\myP_1\myP_2
\times\Big(\delta^{(3)}(\vec p_1-\vec p_3)\delta^{(3)}(\vec p_2-\vec p_4)
+\delta^{(3)}(\vec p_1-\vec p_4)\delta^{(3)}(\vec p_2-\vec p_3)\Big).
\nn
\end{align}

We can finally define the interacting part of the scattering amplitudes. This is done by subtracting the trivial normalization terms (which describe particles propagating without interacting), more precisely
\begin{equation}
\label{eq:interacting_amplitudes}
\begin{aligned}
i\mathcal{T}_{AA \rightarrow AA}(s,t,u)\times(2\pi)^4\delta^4(p) &\equiv
{}_{out}\<A_3,A_4|A_1,A_2\>_{in}-{}_{in}\<A_3,A_4|A_1,A_2\>_{in},\\
i\mathcal{T}_{AA \rightarrow BB}(s,t,u)\times(2\pi)^4\delta^4(p) &\equiv
{}_{out}\<B_3,B_4|A_1,A_2\>_{in}-{}_{in}\<B_3,B_4|A_1,A_2\>_{in},\\
i\mathcal{T}_{AB \rightarrow AB}(s,t,u)\times(2\pi)^4\delta^4(p) &\equiv
{}_{out}\<A_3,B_4|A_1,B_2\>_{in}-{}_{in}\<A_3,B_4|A_1,B_2\>_{in},\\
i\mathcal{T}_{BB \rightarrow BB}(s,t,u)\times(2\pi)^4\delta^4(p) &\equiv
{}_{out}\<B_3,B_4|B_1,B_2\>_{in}-{}_{in}\<B_3,B_4|B_1,B_2\>_{in},
\end{aligned}
\end{equation}
where in order to make the formulas more compact we have introduced the following short-hand notation
\begin{equation}
\delta^4(p) \equiv
\delta^4(p_1+p_2-p_3-p_4).
\end{equation}
Notice the presence of the imaginary unit in the left-hand side which is introduced in order to match the standard conventions.

\subsection{Crossing equations}
\label{sec:crossing}
The crossing equations for the process $AA\rightarrow AA$ and $BB\rightarrow BB$ are extremely simple.  They require that the associated amplitudes are fully symmetric under any permutations of the Mandelstam variables, namely
\begin{align}
\label{eq:crossin_1}
\mathcal{T}_{AA \rightarrow AA}(s,t,u) &= \mathcal{T}_{AA \rightarrow AA}(t,s,u) \,= \mathcal{T}_{AA \rightarrow AA}(u,t,s),\\
\label{eq:crossin_2}
\mathcal{T}_{BB \rightarrow BB}(s,t,u) &= \mathcal{T}_{BB \rightarrow BB}(t,s,u) = \mathcal{T}_{BB \rightarrow BB}(u,t,s).
\end{align}

Things are more complicated for the processes $AA\rightarrow BB$ and $AB\rightarrow AB$. They are related by crossing.  Recall that the exchange of particles $14$ and $23$ lead to the s-t crossing equations, instead the exchange of particles $13$ and $24$ lead to the s-u crossing equations.  In other words
\begin{align}
\label{eq:crossin_3}
\mathcal{T}_{AB \rightarrow AB}(s,t,u) &= \mathcal{T}_{AA \rightarrow BB}(t,s,u) = \mathcal{T}_{BB \rightarrow AA}(t,s,u),\\
\label{eq:crossin_4}
\mathcal{T}_{AB \rightarrow AB}(s,t,u) &= \mathcal{T}_{AB \rightarrow AB}(u,t,s) .
\end{align} 
From the relation \eqref{eq:crossin_3} it follows immediately that
\begin{equation}
\mathcal{T}_{BB \rightarrow AA}(s,t,u) = \mathcal{T}_{AA \rightarrow BB}(s,t,u).
\end{equation}

\subsection{Unitarity}
\label{sec:unitarity}
The two-particle states \eqref{eq:2PS} do not transform in the irreducible representation of the Little group $SO(3)$. We can however decompose them into irreducible representations, see for example \cite{Karateev:2019ymz,Hebbar:2020ukp}.

To begin with let us evaluate the states \eqref{eq:2PS} in the center of mass frame,  namely when the directions of particles are aligned and opposite to each other
\begin{equation}
\vec p_1  = +\vec p,\qquad
\vec p_2 = - \vec p.
\end{equation}
In spherical coordinates the momentum $\vec p$ reads as
\begin{equation}
\label{eq:negative_spherical}
+\vec p =(\myP,\,   \theta,\,\phi),\quad
-\vec p =(\myP,\, \pi-\theta,\,\pi+\phi),\quad
\theta\in[0,\pi], \quad
\phi\in[0,2\pi].
\end{equation}
Here $\myP\equiv |\vec p\,|$.
We denote the states \eqref{eq:2PS} in this particular frame by
\begin{equation}
\label{eq:2PS_com}
|A_1,A_2\>_{in}^{com},\qquad
|A_1,B_2\>_{in}^{com},\qquad
|B_1,B_2\>_{in}^{com}.
\end{equation}
These states depend on the masses $m_A$,  $m_B$ and the spherical coordinates $\myP$,  $\theta$ and $\phi$.
The same notation holds for the out states.
According to equations (2.29) and (2.30) in \cite{Hebbar:2020ukp} we can write
\begin{equation}
\label{eq:2PS_decomposition}
\begin{aligned}
|A_1,A_2\>_{in}^{com} &= \sqrt{2}\sum_{\ell,\lambda}
C_\ell^{AA}(\myP) e^{-i \lambda\phi} d^{(\ell)}_{\lambda 0}(\theta) |\sqrt s, \vec 0; \ell, \lambda\rangle_{in}^{AA},\\
|A_1,B_2\>_{in}^{com} &= \quad\;\,\sum_{\ell,\lambda} C_\ell^{AB}(\myP) \,
e^{-i\lambda\phi}
d^{(\ell)}_{\lambda 0}(\theta) |\sqrt s, \vec 0; \ell, \lambda\rangle_{in}^{AB},\\
|B_1,B_2\>_{in}^{com} &= \sqrt{2}\sum_{\ell,\lambda}
C_\ell^{BB}(\myP) e^{-i\lambda\phi} d^{(\ell)}_{\lambda 0}(\theta) |\sqrt s, \vec 0; \ell, \lambda\rangle_{in}^{BB},
\end{aligned}
\end{equation}
where $d_{\lambda_1\lambda_2}^{(\ell)}(\theta)$ is the small Wigner d matrices and $|\sqrt s, \vec 0; \ell, \lambda\rangle$ are states in the irreducible representation of the $SO(3)$ Little group.
Here we have also introduced the object $C_\ell$ which is defined as
\begin{align}
\label{eq:coefficient_Cj}
C_\ell(\myP)^2    \equiv 4 \pi\, (2\ell+1) \times \frac{\sqrt s}{\myP},
\end{align}
where $\sqrt s$ is the total energy of the two-particle state which is
\begin{equation}
\begin{aligned}
AA:\quad \sqrt{s} &= 2\sqrt{m^2+\myP^2},\\
AB:\quad \sqrt{s} &=\sqrt{m^2+\myP^2} + \myP,\\
BB:\quad \sqrt{s} &= 2\myP.
\end{aligned}
\end{equation}
The small Wigner $d$-matrices are defined for example in equation (2.9) of \cite{Hebbar:2020ukp}.
The states in the right-hand side of \eqref{eq:2PS_decomposition} transform in the irreducible representation of the Lorentz group.  They have the Little group spin $\ell$ and helicity $\lambda=-\ell,\ldots,+\ell$.  They also have the zero total spatial momentum $\vec 0$.  Their normalization is completely fixed by \eqref{eq:2PS_decomposition} and \eqref{eq:normalization_2PS}.

By using the orthogonality of the small Wigner d-matrices we can invert the decompositions \eqref{eq:2PS_decomposition} and write
\begin{equation}
\label{eq:2PS_decomposition_inverted}
\begin{aligned}
|\sqrt s, \vec 0; \ell, \lambda\rangle_{in}^{AA} &= \Pi_\ell^{AA} |A_1,A_2\>_{in}^{com} ,\\
|\sqrt s, \vec 0; \ell, \lambda\rangle_{in}^{AB} &= \Pi_\ell^{AB} |A_1,B_2\>_{in}^{com} ,\\
|\sqrt s, \vec 0; \ell, \lambda\rangle_{in}^{BB} &= \Pi_\ell^{BB} |B_1,B_2\>_{in}^{com} ,
\end{aligned}
\end{equation}
where $\Pi_\ell$ are the projectors to the definite Little group spin. Their explicit form reads
\begin{equation}
\label{eq:projectors}
\begin{aligned}
\Pi_\ell^{AA}&\equiv \frac{2\ell +1}{4 \pi \sqrt{2} C^{AA}_\ell(\myP)}
\int_0^{2\pi}d\phi\, e^{i\lambda\phi} \int_{-1}^{+1}d\cos\theta\;
d^{( \ell)}_{\lambda 0} (\theta),\\
\Pi_\ell^{AB}&\equiv \;\;\frac{2\ell +1}{4 \pi C^{AB}_\ell(\myP)}\;\;\;
\int_0^{2\pi}d\phi\, e^{i\lambda\phi} \int_{-1}^{+1}d\cos\theta\;
d^{( \ell)}_{\lambda 0} (\theta),\\
\Pi_\ell^{BB}&\equiv \frac{2\ell +1}{4 \pi \sqrt{2} C^{BB}_\ell(\myP)}
\int_0^{2\pi}d\phi\, e^{i\lambda\phi} \int_{-1}^{+1}d\cos\theta\;
d^{( \ell)}_{\lambda 0} (\theta).
\end{aligned}
\end{equation}

Let us define the partial amplitudes $\mathcal{S}_\ell(s)$ as the following matrix elements
\begin{equation}
\label{eq:partial_amplitudes}
\begin{aligned}
\mathcal{S}_{AA\rightarrow AA}^\ell(s) \times \delta_{\ell'\ell}\delta_{\lambda'\lambda}(2\pi)^4 \delta(\sqrt s' - \sqrt s)\delta^3(\vec 0) &\equiv
{}_{out}^{AA}\<\sqrt s', \vec 0; \ell', \lambda' |\sqrt s, \vec 0; \ell, \lambda\rangle_{in}^{AA},\\
\mathcal{S}_{AA\rightarrow BB}^\ell(s) \times \delta_{\ell'\ell}\delta_{\lambda'\lambda}(2\pi)^4 \delta(\sqrt s' - \sqrt s)\delta^3(\vec 0) &\equiv
{}_{out}^{BB}\<\sqrt s', \vec 0; \ell', \lambda' |\sqrt s, \vec 0; \ell, \lambda\rangle_{in}^{AA},\\
\mathcal{S}_{AB\rightarrow AB}^\ell(s) \times \delta_{\ell'\ell}\delta_{\lambda'\lambda}(2\pi)^4 \delta(\sqrt s' - \sqrt s)\delta^3(\vec 0) &\equiv
{}_{out}^{AB}\<\sqrt s', \vec 0; \ell', \lambda' |\sqrt s, \vec 0; \ell, \lambda\rangle_{in}^{AB},\\
\mathcal{S}_{BB\rightarrow BB}^\ell(s) \times \delta_{\ell'\ell}\delta_{\lambda'\lambda}(2\pi)^4 \delta(\sqrt s' - \sqrt s)\delta^3(\vec 0) &\equiv
{}_{out}^{BB}\<\sqrt s', \vec 0; \ell', \lambda' |\sqrt s, \vec 0; \ell, \lambda\rangle_{in}^{BB}.
\end{aligned}
\end{equation}
Using \eqref{eq:2PS_decomposition_inverted} and \eqref{eq:projectors} we obtain the explicit relations between the partial amplitudes and the scattering amplitudes. The detailed steps in this derivation are explained in section 2.5 in \cite{Hebbar:2020ukp}. They read
\begin{equation}
\label{eq:partial_vs_scattering amplitudes}
\begin{aligned}
\mathcal{S}_{AA\rightarrow AA}^\ell(s)  &= \frac{1}{32\pi}\left(1-4m^2/s\right)^{1/2}\int_{-1}^{+1}d\cos\theta P_\ell(\cos\theta)\mathcal{S}_{AA\rightarrow AA}(s,t,u),\\
\mathcal{S}_{AA\rightarrow BB}^\ell(s)  &= \frac{1}{32\pi}\left(1-4m^2/s\right)^{1/4}\int_{-1}^{+1}d\cos\theta P_\ell(\cos\theta)\mathcal{S}_{AA\rightarrow BB}(s,t,u),\\
\mathcal{S}_{AB\rightarrow AB}^\ell(s)  &= \frac{1}{16\pi}(1-m^2/s)\int_{-1}^{+1}d\cos\theta P_\ell(\cos\theta)\mathcal{S}_{AB\rightarrow AB}(s,t,u),\\
\mathcal{S}_{BB\rightarrow BB}^\ell(s)  &= \frac{1}{32\pi}\int_{-1}^{+1}d\cos\theta P_\ell(\cos\theta)\mathcal{S}_{BB\rightarrow BB}(s,t,u),
\end{aligned}
\end{equation}
where $P_\ell(\cos\theta)$ are the Legendre polynomials. They are related to the small Wigner d matrices as  $P_\ell(\cos\theta)=d_{00}^{(\ell)}(\theta)$.
To complete these equations it is important to specify the relation between the Mandelstam variables and the scattering angle for each process.  One has
\begin{align}
\nn
AA\rightarrow AA:&\quad t=\frac{1}{2}(4m^2-s)(1-\cos\theta),\quad u=\frac{1}{2}(4m^2-s)(1+\cos\theta),\\
\nn
AA\rightarrow BB:&\quad t=m^2-\frac{s}{2}+\frac{1}{2}\sqrt{s(s-4m^2)}\cos\theta,\quad u=m^2-\frac{s}{2}-\frac{1}{2}\sqrt{s(s-4m^2)}\cos\theta,\\
\nn
AB\rightarrow AB:&\quad t=\frac{(s-m^2)^2(\cos\theta-1)}{2s},\quad u=\frac{2m^4-(s-m^2)^2(\cos\theta+1)}{2s},\\
BB\rightarrow BB:&\quad t=-\frac{s}{2}\,(1-\cos\theta),\quad u=-\frac{s}{2}\,(1+\cos\theta).
\label{eq:definitionsTU}
\end{align} 
In \eqref{eq:partial_vs_scattering amplitudes} we found the relations between the partial amplitudes and the full scattering amplitudes.  We can further rewrite them by splitting the scattering amplitudes into their trivial and their interacting parts given by \eqref{eq:normalization_2PS} and \eqref{eq:interacting_amplitudes}. Performing the explicit integral over the trivial part we obtain the final expressions
\begin{equation}
\label{eq:partial_vs_interacting_scattering amplitudes}
\begin{aligned}
\mathcal{S}_{AA\rightarrow AA}^\ell(s)  &= 1+i\mathcal{T}_{AA\rightarrow AA}^\ell(s),\\
\mathcal{S}_{AA\rightarrow BB}^\ell(s)  &= 0+i\mathcal{T}_{AA\rightarrow BB}^\ell(s),\\
\mathcal{S}_{AB\rightarrow AB}^\ell(s)  &= 1+i\mathcal{T}_{AB\rightarrow AB}^\ell(s),\\
\mathcal{S}_{BB\rightarrow BB}^\ell(s)  &= 1+i\mathcal{T}_{BB\rightarrow BB}^\ell(s),
\end{aligned}
\end{equation}
where we have defined
\begin{equation}
\label{eq:Tell}
\begin{aligned}
\mathcal{T}_{AA\rightarrow AA}^\ell(s)  &\equiv \frac{1}{32\pi}\left(1-4m^2/s\right)^{1/2}\int_{-1}^{+1}d\cos\theta P_\ell(\cos\theta)\mathcal{T}_{AA\rightarrow AA}(s,t,u),\\
\mathcal{T}_{AA\rightarrow BB}^\ell(s)  &\equiv \frac{1}{32\pi}\left(1-4m^2/s\right)^{1/4}\int_{-1}^{+1}d\cos\theta P_\ell(\cos\theta)\mathcal{T}_{AA\rightarrow BB}(s,t,u),\\
\mathcal{T}_{AB\rightarrow AB}^\ell(s)  &\equiv \frac{1}{16\pi}(1-m^2/s)\int_{-1}^{+1}d\cos\theta P_\ell(\cos\theta)\mathcal{T}_{AB\rightarrow AB}(s,t,u),\\
\mathcal{T}_{BB\rightarrow BB}^\ell(s)  &\equiv \frac{1}{32\pi}\int_{-1}^{+1}d\cos\theta P_\ell(\cos\theta)\mathcal{T}_{BB\rightarrow BB}(s,t,u),
\end{aligned}
\end{equation}

We are now in position to address the unitarity constraints.  Let us first consider the $\ZZ$ odd two-particle in and out states projected to the definite spin,  namely
\begin{equation}
|\sqrt s, \vec 0; \ell, \lambda\rangle_{in}^{AB}, \qquad
|\sqrt s, \vec 0; \ell, \lambda\rangle_{out}^{AB} .
\end{equation}
We take all possible inner products of these states.  Unitarity requires that such a matrix is semipositive definite.  Using the definition \eqref{eq:partial_amplitudes} we conclude that
\begin{equation}
\label{eq:unitarity_1}
\forall \ell\geq0,\quad
\forall s\in[m^2,\infty):\qquad
\begin{pmatrix}
1 & \mathcal{S}_{AB\rightarrow AB}^{*\ell}(s) \\
\mathcal{S}_{AB\rightarrow AB}^\ell(s) & 1
\end{pmatrix}\succeq 0.
\end{equation}
Let us consider now the $\ZZ$ even two-particles in and out states projected to the definite spin, namely
\begin{equation}
|\sqrt s, \vec 0; \ell, \lambda\rangle_{in}^{AA}, \qquad
|\sqrt s, \vec 0; \ell, \lambda\rangle_{in}^{BB}, \qquad
|\sqrt s, \vec 0; \ell, \lambda\rangle_{out}^{AA} ,\qquad
|\sqrt s, \vec 0; \ell, \lambda\rangle_{out}^{BB} .
\end{equation}
Analogously we obtain the following unitarity constraint
\begin{equation}
\label{eq:unitarity_2}
\begin{aligned}
\forall \ell=0,2,4,\ldots\\
\forall s\in[4m^2,\infty)
\end{aligned}
:\qquad
\begin{pmatrix}
1 & 0 &\mathcal{S}_{AA\rightarrow AA}^{*\ell}(s)  & \mathcal{S}_{AA\rightarrow BB}^{*\ell}(s) \\
0 & 1 & \mathcal{S}_{BB\rightarrow AA}^{*\ell}(s) & \mathcal{S}_{BB\rightarrow BB}^{*\ell}(s)   \\
\mathcal{S}_{AA\rightarrow AA}^{\ell}(s) & \mathcal{S}_{BB\rightarrow AA}^{\ell}(s) & 1 & 0  \\
\mathcal{S}_{AA\rightarrow BB}^{\ell}(s)  & \mathcal{S}_{BB\rightarrow BB}^{\ell}(s) & 0 & 1
\end{pmatrix} \succeq 0.
\end{equation}
For the energy range $s\in[0,4m^2)$ the two particle state $AA$, do not exist,  however the two particle state $BB$ do.  Considering the states
\begin{equation}
|\sqrt s, \vec 0; \ell, \lambda\rangle_{in}^{BB}, \qquad
|\sqrt s, \vec 0; \ell, \lambda\rangle_{out}^{BB} ,
\end{equation}
we conclude that the following unitarity constraint also holds
\begin{equation}
\label{eq:unitarity_3}
\forall \ell=0,2,4,\ldots,\quad
\forall s\in[0,4m^2):\qquad
\begin{pmatrix}
1 & \mathcal{S}_{BB\rightarrow BB}^{*\ell}(s) \\
\mathcal{S}_{BB\rightarrow BB}^\ell(s) & 1
\end{pmatrix}\succeq 0.
\end{equation}

\subsection{Dilaton scattering}
\label{sec:dilaton}
So far the particle $B$ was simply a generic massless scalar particle.  We would like now to identify it with the dilaton. First, let us define rescaled amplitudes that remain finite in the probe limit $f\to \infty$,
\begin{equation}
\label{eq:new_amplitudes}
\begin{aligned}
 \widetilde{\mathcal{T}}_{AA \rightarrow BB}(s,t,u) = \lim_{f\to \infty} f^2\, \mathcal{T}_{AA \rightarrow BB}(s,t,u)\,,\\ \widetilde{\mathcal{T}}_{AB \rightarrow AB}(s,t,u) = \lim_{f\to \infty} f^2 \,\mathcal{T}_{AB \rightarrow AB}(s,t,u) \,,\\
   \widetilde{\mathcal{T}}_{BB \rightarrow BB}(s,t,u)= \lim_{f\to \infty} f^4 \, \mathcal{T}_{BB \rightarrow BB}(s,t,u) \,.
\end{aligned}
\end{equation}
 
The main result of \cite{Komargodski:2011vj} which we use in this paper is the relation between the $a$-anomaly in the UV with the dilaton field scattering.   More precisely
\begin{equation}
\label{eq:soft_BBBB}
\lim_{f\rightarrow \infty} f^4 \mathcal{T}_{BB \rightarrow BB}(s,t,u) = a^\text{UV} \times(s^2+t^2+u^2) + O(s^3).
\end{equation}
The limit $f\rightarrow \infty$ corresponds to the full decoupling of the dilaton from the system.  Nevertheless it still carries some non-trivial information about the system.

The goal of this section is to formulate the bootstrap setup in terms of amplitudes \eqref{eq:new_amplitudes} and take explicitly the limit $f\rightarrow \infty$. 

\paragraph{Analyticity and soft behavior}
We start by noticing that dilatons in the limit $f\rightarrow \infty$ cannot contribute to the scattering amplitudes as intermediate states. In practice this means that in all the scattering amplitudes there are no branch cuts $s\in[0,\infty]$ due to $n$-dilaton particle states, where $n=1,2,3,4,\ldots$. In other words
\begin{equation}
\label{eq:disc1}
s\in[4m^2,\infty]:\qquad
\begin{aligned}
&\text{disc}_s\mathcal{T}_{AA \rightarrow AA}(s,t,u)\neq 0,\\
&\text{disc}_s\widetilde{\mathcal{T}}_{AA \rightarrow BB}(s,t,u)\neq 0,\\
&\text{disc}_s\widetilde{\mathcal{T}}_{BB \rightarrow BB}(s,t,u)\neq 0,
\end{aligned}
\end{equation}
where the non-zero discontinuity is due to the $n$-particle states of type A, where $n\geq 2$ is only allowed to be even due to the $\ZZ$ symmetry. Analogously we conclude that
\begin{equation}
\label{eq:disc2}
s\in[9m^2,\infty]:\qquad
\text{disc}_s\widetilde{\mathcal{T}}_{AB \rightarrow AB}(s,t,u)\neq 0,
\end{equation}
where the discontinuity appears due to the $n$-particle states of type A, where $n\geq 3$ is only allowed to be odd due to the $\ZZ$ symmetry. It is important also to discuss the contribution of the $n=1$ particle state of type A to the scattering amplitudes describing the processes $AB\rightarrow AB$ and $AA\rightarrow BB$. It appears as an s-channel pole in the amplitude, namely
\begin{equation}
\label{eq:pole}
\begin{aligned}
\widetilde{\mathcal{T}}_{AB \rightarrow AB}(s,t,u) &= -\frac{h^2}{s-m^2}+\ldots,\\
\widetilde{\mathcal{T}}_{AA \rightarrow BB}(s,t,u) &= \qquad 0 \quad\;\,+\ldots,
\end{aligned}
\end{equation}
where $h$ is the real number which describes the interaction strength between the particle $A$ and the dilaton.
The second amplitude in \eqref{eq:pole} does not have an s-channel pole due to the $\ZZ$ symmetry.
Imposing the crossing symmetry given by equations  \eqref{eq:crossin_3} and \eqref{eq:crossin_4} we conclude that
\begin{equation}
\label{eq:poles}
\begin{aligned}
\widetilde{\mathcal{T}}_{AB \rightarrow AB}(s,t,u) &= -\frac{h^2}{s-m^2}-\frac{h^2}{u-m^2} + g(s,t,u)\\
\widetilde{\mathcal{T}}_{AA \rightarrow BB}(s,t,u) &= -\frac{h^2}{t-m^2}-\frac{h^2}{u-m^2} + g(t,s,u),
\end{aligned}
\end{equation}
where the function $g$ describes the finite part of the amplitudes.
No poles in the other amplitudes are allowed either due to the $\ZZ$ symmetry or due to the limit $f\rightarrow \infty$. In appendix \ref{app:poles} we give a non-perturbative argument that the residue $h$ has the following value
\begin{equation}
\label{eq:residue}
h^2 = 2m^4.
\end{equation}
In the next section we will obtain   a stronger result which not only fixes the residue of the poles but also the constant piece of the $AB\rightarrow AB$ and $AA\rightarrow BB$ amplitudes in the soft limit. The result reads
\be
\widetilde{\mathcal{T}}_{AA\rightarrow BB}(s,t,u)\ = - \f{2m^4}{t-m^2}-\f{2m^4}{u-m^2} -m^2 +\ O(u-m^2,t-m^2). \label{eq:soft_AABB}
\ee

\paragraph{Unitarity constraints}
The final step is to write the unitarity constraints \eqref{eq:unitarity_1}, \eqref{eq:unitarity_2} and \eqref{eq:unitarity_3} in the limit $f\rightarrow \infty$.  Let us start with the constraint \eqref{eq:unitarity_1}. We use the relations \eqref{eq:new_amplitudes}, compute the matrix eigenvalues and expand them in the inverse-powers of $f$ to the sub-leading order.  Semipositive definiteness requires that all the eigenvalues are non-negative. Explicitly we get
\begin{equation}
\begin{aligned}
0+f^{-2}\text{Im}\widetilde{\mathcal{T}}^\ell_{AB \rightarrow AB}(s)+O(f^{-4})\geq 0,\\
2-f^{-2}\text{Im}\widetilde{\mathcal{T}}^\ell_{AB \rightarrow AB}(s)+O(f^{-4})\geq 0.
\end{aligned}
\end{equation}
In the limit $f\rightarrow \infty$ the second condition is satisfied automatically since the second term gives a negligible contribution. The leading term in the first condition vanishes however and we are forced to study the sub-leading contribution. Clearly in the limit $f\rightarrow \infty $ the condition \eqref{eq:unitarity_1} simply reduces to
\begin{equation}
\label{eq:final_condition_1}
\forall \ell\geq0,\quad
\forall s\in[9m^2,\infty):\qquad
\text{Im}\widetilde{\mathcal{T}}^\ell_{AB \rightarrow AB}(s)\geq 0.
\end{equation}
Here we have also used \eqref{eq:disc2}.
Analogous reasoning holds for the condition \eqref{eq:unitarity_3}. However due to the conditions \eqref{eq:disc1} the imaginary part of $\widetilde{\mathcal{T}}^\ell_{BB \rightarrow BB}(s)$ vanishes in the interval $s\in [0,4m^2]$.  As a consequence the condition \eqref{eq:unitarity_3} is trivially satisfied.

The best way to analyse the condition \eqref{eq:unitarity_2} is to cast it into a different form where the limit $f\rightarrow\infty$ is obvious. Most of the manipulations bellow were already presented in appendix B in \cite{Homrich:2019cbt}. First, due to \eqref{eq:partial_vs_interacting_scattering amplitudes} we can write \eqref{eq:unitarity_2} as
\begin{equation}
\label{eq:step_1}
\begin{pmatrix}
\mathbb{I} & \mathbb{I}-i\mathbb{T}^\dagger(s) \\
\mathbb{I}+i\mathbb{T}(s) & \mathbb{I}
\end{pmatrix} \succeq 0,
\end{equation}
where we have defined
\begin{equation}
\label{eq:defs}
\mathbb{I}\equiv\begin{pmatrix}
1 & 0  \\
0 & 1
\end{pmatrix},\qquad
\mathbb{T}(s)\equiv
\begin{pmatrix}
\mathcal{T}_{AA\rightarrow AA}^{\ell}(s) & \mathcal{T}_{BB\rightarrow AA}^{\ell}(s) \\
\mathcal{T}_{AA\rightarrow BB}^{\ell}(s)  & \mathcal{T}_{BB\rightarrow BB}^{\ell}(s)
\end{pmatrix}.
\end{equation}
Second, we notice that the condition \eqref{eq:step_1} is equivalent to
\begin{equation}
\label{eq:step_2}
2\text{Im} \mathbb{T}- \mathbb{T}^\dagger(s) \mathbb{T}(s) \succeq 0.
\end{equation}
Third, we rewrite this condition in another equivalent form
\begin{equation}
\label{eq:step_3}
\begin{pmatrix}
\mathbb{I} & \mathbb{T}^\dagger(s) \\
\mathbb{T}(s) & 2\text{Im} \mathbb{T}
\end{pmatrix} \succeq 0.
\end{equation}
The matrix appearing in the left-hand side of \eqref{eq:step_3} is 4x4. Let us write it explicitly using \eqref{eq:defs} and \eqref{eq:new_amplitudes}. We get
\begin{equation}
\label{eq:step_4}
\begin{pmatrix}
1 & 0 &\mathcal{T}_{AA\rightarrow AA}^{*\ell}(s)  & \frac{1}{f^2}\widetilde{\mathcal{T}}_{AA\rightarrow BB}^{*\ell}(s) \\
0 & 1 & \frac{1}{f^2}\widetilde{\mathcal{T}}_{BB\rightarrow AA}^{*\ell}(s) & \frac{1}{f^4}\widetilde{\mathcal{T}}_{BB\rightarrow BB}^{*\ell}(s)   \\
\mathcal{T}_{AA\rightarrow AA}^{\ell}(s) & \frac{1}{f^2}\widetilde{\mathcal{T}}_{BB\rightarrow AA}^{\ell}(s) & 2\text{Im}\mathcal{T}_{AA\rightarrow AA}^{\ell}(s) & \frac{2}{f^2}\text{Im}\widetilde{\mathcal{T}}_{BB\rightarrow AA}^{\ell}(s)   \\
\frac{1}{f^2}\widetilde{\mathcal{T}}_{AA\rightarrow BB}^{\ell}(s)  & \frac{1}{f^4}\widetilde{\mathcal{T}}_{BB\rightarrow BB}^{\ell}(s) & \frac{2}{f^2}\text{Im}\widetilde{\mathcal{T}}_{AA\rightarrow BB}^{\ell}(s)  & \frac{2}{f^4}\text{Im}\widetilde{\mathcal{T}}_{BB\rightarrow BB}^{\ell}(s)
\end{pmatrix} \succeq 0.
\end{equation}
We notice that multiplying the last row and subsequently the last column of the matrix in the left-hand side of \eqref{eq:step_4} by $f^2$ does not change its semidefinite positive property.\footnote{This can be proven in several different ways. For instance one can multiply the condition \eqref{eq:step_4} by the diagonal matrix $\{1,1,1,f^2\}$ to the left and to the right.  Alternatively one can use the Sylvester's criterion. Considering all the principal minors of the original matrix and the one with the last row and column multiplied by $f^2$ we see that their non-negativity conditions are equivalent as long as $f^2>0$.}

Taking the limit $f\rightarrow \infty$ we obtain the unitarity condition \eqref{eq:unitarity_2} in its final form
\begin{equation}
\label{eq:unitarity_2_final}
\begin{aligned}
\forall \ell=0,2,4,\ldots\\
\forall s\in[4m^2,\infty)
\end{aligned}
:\qquad
\begin{pmatrix}
1 &\mathcal{T}_{AA\rightarrow AA}^{*\ell}(s)  & \widetilde{\mathcal{T}}_{AA\rightarrow BB}^{*\ell}(s) \\
\mathcal{T}_{AA\rightarrow AA}^{\ell}(s) &  2\text{Im}\mathcal{T}_{AA\rightarrow AA}^{\ell}(s) & 2\text{Im}\widetilde{\mathcal{T}}_{BB\rightarrow AA}^{\ell}(s)   \\
\widetilde{\mathcal{T}}_{AA\rightarrow BB}^{\ell}(s)  & 2\text{Im}\widetilde{\mathcal{T}}_{AA\rightarrow BB}^{\ell}(s)  & 2\text{Im}\widetilde{\mathcal{T}}_{BB\rightarrow BB}^{\ell}(s)
\end{pmatrix} \succeq 0.
\end{equation}
Notice that in going from \eqref{eq:step_4} to \eqref{eq:unitarity_2_final} after taking the limit $f\rightarrow \infty$ we dropped the second row and the second column which were simply $(0,1,0,0)$, did not affect the semidefinite positiveness. 

Summarizing, the unitarity conditions which must be satisfied for the system of a particle and the dilaton are \eqref{eq:final_condition_1} and \eqref{eq:unitarity_2_final}. We can use the Sylvester's criterion in order to rewrite the latter condition as a set of inequalities. As a result the conditions \eqref{eq:final_condition_1} and \eqref{eq:unitarity_2_final} can be written as
\begin{gather}
\nn
\text{Im}\widetilde{\mathcal{T}}^\ell_{AB \rightarrow AB}(s)\geq 0,\\
\label{eq:conditions}
2\text{Im}\mathcal{T}_{AA\rightarrow AA}^{\ell}(s)\geq \left|\mathcal{T}_{AA\rightarrow AA}^{\ell}(s)\right|^2\geq 0,\qquad
2\text{Im}\widetilde{\mathcal{T}}_{BB\rightarrow BB}^{\ell}(s)\geq 
\left|\widetilde{\mathcal{T}}_{AA\rightarrow BB}^{\ell}(s)\right|^2\geq 0,\\
\nn
\text{Im}\mathcal{T}_{AA\rightarrow AA}^{\ell}(s)\times
\text{Im}\widetilde{\mathcal{T}}_{BB\rightarrow BB}^{\ell}(s)\geq
\left(\text{Im}\widetilde{\mathcal{T}}_{AA\rightarrow BB}^{\ell}(s)\right)^2
\end{gather}
together with
\begin{multline}
2\text{Im}\widetilde{\mathcal{T}}_{BB\rightarrow BB}^{\ell}(s)\times
\left(2\text{Im}\mathcal{T}_{AA\rightarrow AA}^{\ell}(s)-
\left|\mathcal{T}_{AA\rightarrow AA}^{\ell}(s)\right|^2\right)
+\\
2\text{Im}\widetilde{\mathcal{T}}_{AA\rightarrow BB}^{\ell}(s) \times
\left(
\mathcal{T}_{AA\rightarrow AA}^{\ell}(s)\widetilde{\mathcal{T}}_{AA\rightarrow BB}^{*\ell}(s)+
\mathcal{T}_{AA\rightarrow AA}^{*\ell}(s)\widetilde{\mathcal{T}}_{AA\rightarrow BB}^{\ell}(s)
\right)\geq\\
2\text{Im}\mathcal{T}_{AA\rightarrow AA}^{\ell}(s)\times
\left|\widetilde{\mathcal{T}}_{AA\rightarrow BB}^{\ell}(s)\right|^2 +4\left(\text{Im}\widetilde{\mathcal{T}}_{AA\rightarrow BB}^{\ell}(s)\right)^2.
\end{multline}
The last condition comes from the determinant of the matrix \eqref{eq:final_condition_1}.

\paragraph{Sum-rules for the $a$-anomaly}
Let us start by recalling the formula \eqref{eq:dispersion_relation_general}. In the notation of this section it reads
\begin{equation}
	\label{eq:sum_rule_amplitude}
	a^\text{UV} = \f{1}{\pi}\ \int_{4m^2}^\infty\f{ds}{s^3}\ \text{Im}\widetilde{\mathcal{T}}_{BB\rightarrow BB}(s,0,-s).
\end{equation}
Inverting the last equation in \eqref{eq:Tell} and plugging it to the above sum-rule we obtain
\begin{align}
		\label{eq:sum_rule_pamplitude}
a^\text{UV} =
16\sum_{\ell=0,2,4,\ldots} (2\ell+1)\ \int_{4m^2}^{\infty}\f{ds}{s^3}\ \text{Im}\widetilde{\mathcal{T}}^\ell_{BB\rightarrow BB}(s).
\end{align}
Using this formula and the conditions \eqref{eq:conditions} we conclude that
\begin{equation}
\label{eq:sum_rule_a}
a^\text{UV} \geq 8\sum_{\ell=0,2,4,\ldots} (2\ell+1)\ \int_{4m^2}^{\infty}\f{ds}{s^3}\ \left|\widetilde{\mathcal{T}}_{AA\rightarrow BB}^{\ell}(s)\right|^2.
\end{equation}
Equivalently using \eqref{eq:Tell}, the completeness relation of Legendre polynomials and changing integration variables from $(s,\cos\theta)$ to $(t,u)$ the above equation can be written as 
\begin{equation}
	\label{eq:sum_rule_a_better}
	a^\text{UV} \geq  \frac{1}{32\pi^2} \int_{t u > m^4 \atop t,u<0 }\f{dt du}{(2m^2-t-u)^4}\ \left|\widetilde{\mathcal{T}}_{AA\rightarrow BB}(s,t,u)\right|^2.
\end{equation}

In appendix \ref{sec:example}, in particular see equation \eqref{TAABB}, we will show that in the case when particle $A$ is a free boson, the matter - dilaton scattering amplitude has the following simple form 
\begin{equation}
	\label{eq:matter_compensator_free}
	\widetilde{\mathcal{T}}^{\text{free}}_{AA\rightarrow BB}(s,t,u)=
	-m^2-\f{2m^4}{t-m^2}-\f{2m^4}{u-m^2}.
\end{equation}
Plugging it into \eqref{eq:sum_rule_a_better} we obtain $a^\text{UV} \ge a_\text{free} $ and conclude that the inequality \eqref{eq:sum_rule_a_better} is saturated in this case. Equivalently plugging \eqref{eq:matter_compensator_free} into \eqref{eq:sum_rule_a} we obtain
\begin{equation}
	a/a_\text{free} = 0.83864 + 0.14908 + 0.01048 + 0.00140 + \ldots = 0.99960 + \ldots,
\end{equation}
where the numbers indicated correspond to $\ell=0$, $2$, $4$ and $6$ and $\ldots$ represent all the higher spin contributions to the $a$-anomaly of a free scalar field.

\section{Matter - dilaton scattering at low energy}
\label{S:EFT_constraints}
In \cite{Komargodski:2011vj} the authors derived the most general low energy effective action of the dilaton field in a curved background. As reviewed in section \ref{S:KS_a_theorem_review} this action is given in \eqref{eq:dilaton_effective}. It is found by solving the 't Hooft anomaly matching conditions which can be written as a system of differential equations \eqref{eq:transformations_2}. In flat space this action reduces to \eqref{eq:IR_action_interacting} and leads to the explicit expression of the low energy $BB\rightarrow BB$ amplitude given by equation \eqref{eq:sum_rule}.

Here we perform a similar analysis for the $AB\rightarrow AB$ amplitude. We start in section \ref{sec:ABtoAB_EFT} by writing the most general low energy effective action in curved background which describes this process. In section \ref{secABtoAB_amplitude} we evaluate it in flat space and derive the resulting scattering amplitude at low energy. The final result turns out to be universal (independent of a particular model) and is given by equation \eqref{eq:soft_ABtoAB_final}.

\subsection{The most general effective action}
\label{sec:ABtoAB_EFT}
The assumption of this paper is the presence of a single $\ZZ$ asymptotic particle $A$.  Let us denote by $\Phi(x)$ the effective field associated to this particle.
We would like to find the most general low energy effective action $A_\text{eff}[\Phi,\varphi]$, using which we can compute the S-matrix for the scattering process $AB\rightarrow AB$.\footnote{One can think of the effective action  $A_\text{eff}[\Phi,\varphi]$ as the Legendre transform of the generating function $W[\zeta,\varphi]$ where $\zeta(x)$ is the source that couples to a local operator $\mathcal{O}_\Delta(x)$ that can create particle $A$ from the vacuum.} We remind that the filed $\varphi(x)$ creates the dilaton particle $B$ from the vacuum.

Following the discussion of section \ref{KS_setup} we work with the diff and Weyl invariant action \eqref{eq:action_modified}. We would like to rewrite this action in terms of low energy degrees of freedom $\Phi(x)$ and $\varphi(x)$ using only diff and Weyl symmetry. The main ingredients for doing this are the Weyl invariant metric $\widehat{g}_{\mu\nu}(x)$ and  scalar $\widehat \Phi(x)$, defined in terms of background metric $g_{\mu\nu}(x)$ and scalar field $\Phi(x)$ in the following way
\begin{equation}
	\widehat{g}_{\mu\nu}(x)=e^{-2\tau(x)}\ g_{\mu\nu}(x),\qquad
	\widehat\Phi(x)= e^{\Delta \tau(x)}\Phi(x),
\end{equation}
where $\tau(x)$ is the dilaton field related to $\varphi(x) $ according to \eqref{eq:dilaton} and $\Delta$ is some effective scaling dimension of the field $\Phi(x)$. Now we want to write the most generic form of the effective action $A^{g}_\text{eff}[\Phi,\varphi]$ in the curved background, which should be general coordinate invariant in the metric $\widehat{g}_{\mu\nu}(x)$ and should contain the scalar $\widehat \Phi(x)$. Finally we need to substitute $g_{\mu\nu}=\eta_{\mu\nu}$ in $A^{g}_\text{eff}[\Phi,\varphi]$ to get the flat space effective action  $A_\text{eff}[\Phi,\varphi]$ which describes the scattering process $AB\rightarrow AB$. For the construction of $A^{g}_\text{eff}[\Phi,\varphi]$ we adapt a variation of the covariantization prescription originally developed in \cite{Sen:2017nim,Chakrabarti:2017ltl} for the soft gravitational background.

Let us start from the tangent space with locally flat metric $\eta_{ab}$. The connection between the curved and tangent space is provided by the objects $e_\mu^a(x)$ and $E_a^\mu(x)$ called the vierbein and inverse vierbein respectively. They are defined via the relations
\begin{equation}
	\widehat{g}_{\mu\nu}(x) = e_\mu^a(x)e_\nu^b(x) \eta_{ab},\qquad
    \eta_{ab} = E_a^\mu(x)E_b^\nu(x) \widehat{g}_{\mu\nu},
\end{equation} 
where $\eta_{ab}$ is the flat Minkowski metric. We start with the quadratic part of the tangent space 1PI effective action for scalar field $\Phi(x)$, 
\begin{equation}
	\label{eq:efective_action_tangent}
	A^\text{tangent} = \f{1}{2}\int d^4x \ \Phi(x)K(\partial_a)\Phi(x)\ ,
\end{equation}
where we dropped all the terms with cubic and higher powers of $\Phi$ since they will not contribute to the scattering process $AB\rightarrow AB$ after covariantization. The most general form of the kinetic operator $ K$ can be written as\footnote{The form of the kinetic operator as Taylor series expansion in derivatives is always possible for a EFT where massless particles are absent. }
\begin{equation}
 K(\partial_a) \equiv \sum_{n=0}^\infty c_n^{a_1 a_2\ldots a_n} \partial_{a_1}\partial_{a_2}\ldots \partial_{a_n}, \label{eq:kinetic_operator_x}
\end{equation}
where the coefficients $c_n$ are built out of all possible combinations of $\eta^{ab}$ with arbitrary numerical coefficients and $c_n^{a_1\ldots a_n}$ is symmetric under the exchange of tangent space indices.  
Now we would like to first covariantize the tangent space action \eqref{eq:efective_action_tangent}. This requires to make the following replacements: $d^4x \rightarrow d^4 x\sqrt{-\widehat{g}}$ together with
\begin{align}
	K(\partial_a)  \rightarrow
	 \sum_{n=0}^\infty c_n^{a_1a_2\ldots a_n}E_{a_1}^{\mu_1}E_{a_2}^{\mu_2}\cdots E_{a_n}^{\mu_n}D_{\mu_1}D_{\mu_2}\cdots D_{\mu_n}\ \equiv F.
\end{align}
Above $F$ is obtained only by minimally covariantizing the kinetic operator. In principle $F$ could contain non-minimal terms involving one or more Riemann tensors and derivatives on them. Since those non-minimal terms contain two or more derivatives of the dilaton field, they  start contributing at higher orders in low dilaton momentum expansion of $AB\rightarrow AB$ scattering amplitude. Since our interest is to constrain first two terms in low dilaton momentum expansion of the scattering amplitude $AB\rightarrow AB$, we are not including those terms under our covariantization process. Now on top of the covariantization we also need to make the action Weyl invariant which is achieved by replacing $\Phi(x) \rightarrow \widehat\Phi(x)$. Also in $A^{g}_\text{eff}[\Phi,\varphi]$ we should include general coordinate invariant scalar term purely constructed out of metric $\widehat{g}_{\mu\nu}(x)$ given in \eqref{eq:Action_invariant} as well as the classically Weyl invariant action \eqref{eq:dynamics} to make the dilaton dynamical. As a result of all these, our curved space effective action becomes
\begin{equation}
		A^{g}_\text{eff}[\Phi,\varphi] =\f{1}{2}\int d^4x \sqrt{-\widehat{g}}\ \widehat\Phi(x)F\widehat\Phi(x)\ +  A_\text{dynamics}(g, \Omega)+A_\text{invariant}(g, \Omega)+\ \cdots \label{eq:curved_space_eff_action}
\end{equation}
We denote by $\ldots $ all the possible non-minimal terms appearing in the covariantization procedure as described earlier. From here on to reduce complexity we ignore $A_\text{invariant}(g, \Omega)$ part of the above action as it always contribute terms at higher power in $f^{-1}$ compare to similar terms coming from $A_\text{dynamics}(g, \Omega)$ and won't affect our result later on.

Now to read off the flat space effective action from $A_\text{eff}[\Phi,\varphi]$ from \eqref{eq:curved_space_eff_action} we need to set $g_{\mu\nu}(x)=\eta_{\mu\nu}$ which in turn requires the following substitutions in \eqref{eq:curved_space_eff_action}
\begin{eqnarray}
	\widehat{g}_{\mu\nu}(x)&=&e^{-2\tau(x)}\ \eta_{\mu\nu}=\Big(1-\f{\varphi}{\sqrt{2}f}\Big)^2\ \eta_{\mu\nu},\nonumber\\
   e_\mu^a(x) &=& e^{-\tau(x)}\delta_\mu^a =\Big(1-\f{\varphi}{\sqrt{2}f}\Big)\ \delta_\mu^a,\nonumber\\
   E_a^\mu(x) &=& e^{\tau(x)}\delta_a^\mu=\Big(1-\f{\varphi}{\sqrt{2}f}\Big)^{-1}\delta _a^\mu.
\end{eqnarray}
With these substitutions the flat space effective action becomes
\begin{eqnarray}	
	A_\text{eff}[\Phi,\varphi]& =& \f{1}{2}\int d^4x\ \Big(1-\f{\varphi(x)}{\sqrt{2f}}\Big)^{4-\Delta}\Phi(x)\sum_{n=0}^\infty c_n^{a_1a_2\ldots a_n}\Big(1-\f{\varphi(x)}{\sqrt{2f}}\Big)^{-n}\delta_{a_1}^{\mu_1}\delta_{a_2}^{\mu_2}\cdots \delta_{a_n}^{\mu_n}\nonumber \\
	&&\times\ D_{\mu_1}D_{\mu_2}\cdots D_{\mu_n}\Big(1-\f{\varphi(x)}{\sqrt{2f}}\Big)^{-\Delta}\Phi(x)\nonumber\\
	&& +\int d^4x \left(-\frac{1}{2}\partial_\mu\varphi(x) \partial^\mu\varphi(x) + \f{1}{f}O(\p^4\varphi^3)\right)   \ +\cdots \label{eq:efective_action}
\end{eqnarray}
 Now we expand the effective action $A_\text{eff}$ in power of $f$ to the second order around $f=\infty$. The result can be written in the following form
\begin{equation}
	A_\text{eff} =A^0_\text{eff}+ \ A^1_\text{eff} +  \ A^2_\text{eff}+\ldots.\label{eq:A_eff_expansion}
\end{equation} 
where the superscript $n$ in $A^n_\text{eff}$ indicates the order in  $f^{-n}$ expansion of $A_\text{eff}$ around $f\rightarrow \infty$. The first term in \eqref{eq:A_eff_expansion} reads as
\begin{equation}
	A^0_\text{eff} = \frac{1}{2}\int d^4x\sum_{n=0}^\infty c_n^{a_1 a_2\ldots a_n}\Big(\Phi(x)\p_{a_1}\p_{a_2}\cdots \p_{a_n}\Phi(x) \Big)-\frac{1}{2}\int d^4x \partial_\mu\varphi(x) \partial^\mu\varphi(x).\label{eq:A0_eff_x}
\end{equation}
In the second term $A^1_\text{eff}$ we grouped all the contributions proportional to a single dilaton field $\varphi$ up to one derivative acting on it and also schematically kept the order of three dilaton field interaction term. The result reads
\begin{equation}
	\label{eq:A1_eff_x}
	\begin{aligned}
		A^1_\text{eff} = \frac{1}{2\sqrt{2}f}\int d^4x \Bigg(&\sum_{n=0}^\infty c_n^{a_1 a_2\ldots a_n}\Big\lbrace 
		n-(4-2\Delta)\Big\rbrace\varphi(x)\Phi(x)\p_{a_1}\p_{a_2}\cdots \p_{a_n}\Phi(x) \\
		+& \sum_{n=2}^{\infty}c_n^{a_1 \ldots a_n}\sum_{\substack{i, j=1\\ i<j}}^{n}\Big\lbrace \delta_{a_i}^\nu \p_{a_j}\varphi(x)+\delta_{a_j}^\nu \p_{a_i}\varphi(x)-\eta_{a_i a_j}\p^\nu \varphi(x)\Big\rbrace\\
		& \qquad\;\;\,\times\Phi(x)\p_{a_1}\cdots \p_{a_{i-1}}\p_{a_{i+1}}\cdots \p_{a_{j-1}}\p_{a_{j+1}}\cdots \p_{a_n}\p_\nu\Phi(x)\\
		+&\sum_{n=1}^{\infty}c_n^{a_1\ldots a_n}\sum_{i=1}^{n}\Delta\p_{a_i}\varphi(x)\ \Phi(x)\p_{a_1}\cdots \p_{a_{i-1}}\p_{a_{i+1}}\cdots \p_{a_n}\Phi(x)\\
		& +O(\Phi^2\p^2 \varphi,\;\p^4 \varphi^3)\Bigg)\ .
	\end{aligned}
\end{equation}
In the above expression the first term is linear in $\varphi(x)$. It is obtained by replacing all the covariant derivatives in \eqref{eq:efective_action} by ordinary derivatives and commuting $\Big(1-\f{\varphi(x)}{\sqrt{2}f}\Big)^{-\Delta}$ through the derivatives (in other words neglecting terms containing derivatives operating on $\varphi(x)$). Expansion of the resulting expression at linear order in $\varphi(x)$ generates the first term. In the second and third terms in \eqref{eq:A1_eff_x} one derivative operates on $\varphi(x)$. In presence of two covariant derivatives on $\Phi(x)$ in the first term in \eqref{eq:efective_action}, we need to substitute $D_{\mu_i}D_{\mu_j}\Phi(x)=\p_{\mu_i}\p_{\mu_j}\Phi(x)-\Gamma_{\mu_i \mu_j}^\nu \p_{\nu}\Phi(x)$. Then writing down the Christoffel connection up to linear order in $\varphi(x)$ for any pair of such covariant derivatives and setting $\varphi=0$ in all other places we get the second term above. On the other hand when any one of the ordinary derivative from the set of covariant derivatives operates on $\Big(1-\f{\varphi(x)}{\sqrt{2}f}\Big)^{-\Delta}$ in the first term  in \eqref{eq:efective_action}, we get the third term above at the linear order in $\varphi(x)$ from the expansion of the resulting expression. First non-vanishing contribution to three dilaton interaction appears at four derivative order as schematically written as $O(\p^4 \varphi^3)$ in the above expression.

In the third term of \eqref{eq:A_eff_expansion}, we grouped all the contributions involving $\varphi(x)^2$ and no derivative on it, and the result reads
\begin{eqnarray}
	A^2_\text{eff} &=& \f{1}{2}\int d^4x\ \Big(\f{\varphi(x)}{\sqrt{2}f}\Big)^2\ \Phi(x)\sum_{n=0}^{\infty}\Bigg(\f{(4-2\Delta)(3-2\Delta)}{2} -(4-2\Delta)\ n+\f{n(n+1)}{2}\Bigg)\nonumber\\
	&&\times \ c_n^{a_1 a_2\cdots a_n}\p_{a_1}\p_{a_2}\cdots \p_{a_n}\Phi(x) \ +\ \f{1}{f^2}O(\Phi^2 \varphi\p\varphi).\label{eq:A2_eff_x}
\end{eqnarray}
To get the above contribution we first replace all the covariant derivatives in \eqref{eq:efective_action} by ordinary derivatives and commute $\Big(1-\f{\varphi(x)}{\sqrt{2}f}\Big)^{-\Delta}$ through the derivatives neglecting terms containing derivatives operating on $\varphi(x)$. Then we expand the resulting contribution and collect terms at quadratic order in $\varphi(x)$. We do not need to compute $A_{\text{eff}}^{n}$ for $n\geq 3$ as these terms of the effective action do not contribute to $AB\rightarrow AB$ scattering amplitude. Another important point is that once the expansion is done in power of $f$ about $f\rightarrow \infty$ in \eqref{eq:A_eff_expansion}, in the expressions of $A_{\text{eff}}^{n}$ both tangent space indices $a_1, a_2,\cdots$ and curved space indices $\mu_1,\mu_2,\cdots$ can be treated as just flat space Lorentz indices.

\paragraph{Momentum space} It is cleaner to present further discussion in momentum space. We would like to rewrite $A_{\text{eff}}^0$, $A_{\text{eff}}^1$ and $A_{\text{eff}}^2$ as integrals in momenta variables. 
Let us start by Fourier transforming the object in \eqref{eq:kinetic_operator_x} in momentum variable $q$ 
\begin{equation}
		\mathcal{K}(\partial_a) \longrightarrow \mathcal{K}(q)\equiv\sum_{n=0}^\infty (i)^n\ c_n^{a_1 a_2\ldots a_n}  q_{a_1}q_{a_2}\ldots q_{a_n} . \label{eq:K_fourier_transform}
\end{equation}
In appendix \ref{S:Fourier_transforms} we have derived some other important identities under Fourier transformations which we use below. 

Using the definition of \eqref{eq:K_fourier_transform} the momenta space expression of \eqref{eq:A0_eff_x} becomes
\begin{eqnarray}
	A_\text{eff}^0 &=&\f{1}{2}\int \f{d^4 q_1}{(2\pi)^4}\f{d^4 q_2}{(2\pi)^4}\ (2\pi)^4\delta^{(4)}(q_1+q_2)\ \Phi(q_1)\mathcal{K}(q_2)\Phi(q_2) \nonumber\\ 
	&&-\f{1}{2}\int\f{d^4k_1}{(2\pi)^4}\f{d^4 k_2}{(2\pi)^4}\ (2\pi)^4\delta^{(4)}(k_1+k_2)\varphi(k_1)k_2^2 \varphi(k_2)\label{eq:low_energy_action_0}.
\end{eqnarray}
The kinetic operator $\mathcal{K}(q)$ should vanish on-shell as follows from the equation of motion. Using the definition \eqref{eq:K_fourier_transform} and equalities from appendix  \ref{S:Fourier_transforms} in momentum space the expression \eqref{eq:A1_eff_x} can be written as
\begin{eqnarray}
	A_{\text{eff}}^{1} &=& \f{1}{2\sqrt{2}f}\int \f{d^4 q_1}{(2\pi)^4}\f{d^4 q_2}{(2\pi)^4}\f{d^4 k}{(2\pi)^4}\ (2\pi)^4 \delta^{(4)}(q_1 +q_2+k)\ \Phi(q_1)\Phi(q_2)\ \varphi(k)\nonumber \\
	&&\Bigg(
	\Big\lbrace -(4-2\Delta)\mathcal{K}(q_2)+q_{2}^{\mu}\f{\p \mathcal{K}(q_2)}{\p q_2^\mu}\Big\rbrace
	+\f{1}{2}\big\lbrace \delta_\mu^\nu k_\rho +\delta_\rho^\nu k_\mu -\eta_{\mu\rho}k^\nu\big\rbrace \nonumber\\
	&&\times q_{2\nu} \f{\p^2 \mathcal{K}(q_2)}{\p q_{2\mu}\p q_{2\rho}}
	+\Delta\ k^{\mu}\f{\p \mathcal{K}(q_2)}{\p q_2^\mu}+\ O(k^2)
	\Bigg). \label{eq:low_energy_action_1}
\end{eqnarray}
In the above expression we are not explicitly writing down the three dilaton interaction part, as this term involves four power of dilaton momenta and will not be important for the computation of $AB\rightarrow AB$ scattering amplitude up to the order we are interested in.
Using the definition \eqref{eq:K_fourier_transform} and the properties \eqref{eq:property_1} and \eqref{eq:property_2} the expression \eqref{eq:A2_eff_x} can be written in momentum space as
\begin{multline}
A_\text{eff}^2  =\f{1}{4f^2}\int \f{d^4q_1}{(2\pi)^4}\f{d^4q_2}{(2\pi)^4}\f{d^4 k_1}{(2\pi)^4}\f{d^4 k_2}{(2\pi)^4} (2\pi)^4 \delta^{(4)}(q_1 +q_2+k_1+k_2)\Phi(q_1)\Phi(q_2)\varphi(k_1)\varphi(k_2) \\
\Bigg( (2-\Delta)(3-2\Delta)\mathcal{K}(q_2)+(-3+2\Delta)q_{2}^{\mu}\f{\p \mathcal{K}(q_2)}{\p q_2^\mu} +\f{1}{2}q_{2}^{\mu}q_2^\nu\f{\p^2 \mathcal{K}(q_2)}{\p q_2^\mu \p q_{2}^{\nu}}+O(k_1, k_2)\Bigg).\label{eq:low_energy_action_2}
\end{multline}

\subsection{Low energy amplitude}
\label{secABtoAB_amplitude}
Having obtained the low energy effective action components given by \eqref{eq:low_energy_action_0} - \eqref{eq:low_energy_action_2} we can finally compute the scattering amplitude $AB\rightarrow AB$. We start by deriving the Feynman rules. 
The Feynman propagators are defined as
\begin{equation}
	(2\pi)^4 \delta^{(4)}(q_1 +q_2)\times D_F^{-1}(q_1)\equiv -i\frac{\delta^2 A_\text{eff}[\Phi,\varphi]}{\delta\Phi(q_1)\delta\Phi(q_2)}\Bigg{|}_{\Phi,\varphi=0},\qquad
	D_F(q_1) \equiv 
	\begin{tikzpicture}
			\begin{feynman}
				\vertex (i1);
				\vertex[right=of i1] (i2);
				\diagram* {
				(i1) -- [momentum=$q_1$] (i2)
				};
				\end{feynman}
			\end{tikzpicture}
	\end{equation}
	\begin{equation}
	(2\pi)^4 \delta^{(4)}(k_1 +k_2)\times \Delta_F^{-1}(k_1)\equiv -i\frac{\delta^2 A_\text{eff}[\Phi,\varphi]}{\delta\varphi(k_1)\delta\varphi(k_2)}\Bigg{|}_{\Phi,\varphi=0},\qquad
	\Delta_F(k_1) \equiv \begin{tikzpicture}
			\begin{feynman}
				\vertex (i1);
				\vertex[right=of i1] (i2);
				\diagram* {
				(i1) -- [dashed, momentum=$k_1$] (i2)
				};
				\end{feynman}
			\end{tikzpicture}
	\end{equation}
The cubic effective vertex $\Phi\Phi\varphi$ is defined as
\begin{equation}
	(2\pi)^4 \delta^{(4)}(q_1 +q_2+k)\times\Gamma^{(3)}(q_1,q_2;k) \equiv i\frac{\delta^3 A_\text{eff}[\Phi,\varphi]}{\delta\Phi(q_1)\delta\Phi(q_2)\delta \varphi(k)}\Bigg{|}_{\Phi,\varphi=0} \equiv
	\begin{tikzpicture}[baseline=(a.base)]
		\begin{feynman}
			\vertex (i1);
			\vertex [right=of i1, blob, pattern=none, anchor=center, minimum size=.9cm] (a) {$\Gamma^{(3)}$};
			\vertex [right=of a] (o1) ;
			\vertex [below=of a] (i2);
			\diagram* {
				(a) -- [rmomentum=$q_2$] (o1),
				(i1) --[momentum=$q_1$] (a),				
				(i2)-- [dashed, momentum=$k$] (a)
			};
		\end{feynman}
	\end{tikzpicture}
\end{equation}
The quartic $\Phi\Phi\varphi\varphi$ vertex is defined as 
\begin{equation}
	\begin{aligned}
		(2\pi)^4 \delta^{(4)}(q_1 +q_2&+k_1+k_2) \times
		\Gamma^{(4)}(q_1,q_2;k_1,k_2) \\
		&\equiv i\frac{\delta^4 A_\text{eff}[\Phi,\varphi]}{\delta\Phi(q_1)\delta\Phi(q_2)\delta \varphi(k_1)\delta \varphi(k_2)}\Bigg{|}_{\Phi,\varphi=0}
		\equiv
		\begin{tikzpicture}[baseline=(a.base)]
			\begin{feynman}
				\vertex (i1);
				\vertex [right=of i1, blob, pattern=none, anchor=center, minimum size=.9cm] (a) {$\Gamma^{(4)}$};
				\vertex [right=of a] (o1) ;
				\vertex [below right=of a] (i2);
				\vertex [below left=of a] (o2); 
				\diagram* {
					(i1) --[momentum=$q_1$] (a),
					(a)--[rmomentum=$q_2$] (o1),
					(i2)-- [dashed, momentum={[arrow distance=2mm]$k_2$}] (a),
					(a) --[dashed, rmomentum={[arrow distance=2mm]$k_1$}] (o2) 
				};
			\end{feynman}
		\end{tikzpicture}
	\end{aligned}
\end{equation}
The cubic $\varphi\varphi\varphi$ vertex is defined as 
\begin{align}
(2\pi)^4\delta^{(4)}(k_1+k_2+k_3)&\times V^{(3)}(k_1,k_2,k_3)\nonumber\\
&\equiv  i\frac{\delta^3 A_\text{eff}[\Phi,\varphi]}{\delta \varphi(k_1)\delta \varphi(k_2)\delta\varphi(k_3)}\Bigg{|}_{\Phi,\varphi=0}
\equiv   \begin{tikzpicture}[baseline=(a.base)]
	\begin{feynman}
		\vertex (i1);
		\vertex [right=of i1, blob, pattern=none, anchor=center, minimum size=.9cm] (a) {$V^{(3)}$};
		\vertex [right=of a] (o1) ;
		\vertex [below=of a] (i2);
		\diagram* {
			(a) -- [dashed, rmomentum=$k_2$] (o1),
			(i1) --[dashed, momentum=$k_1$] (a),				
			(i2)-- [dashed, momentum=$k_3$] (a)
		};
	\end{feynman}
\end{tikzpicture}
\end{align}

Applying these definitions to \eqref{eq:low_energy_action_0} - \eqref{eq:low_energy_action_2} we conclude that
the Feynman propagators read as
\begin{equation}
	\label{eq:propagator}
	D_F(q)=i\big(\mathcal{K}(q)\big)^{-1}\equiv \lbrace q^2+m^2 -i\epsilon \rbrace^{-1}\Xi(q), \hspace{1cm} \Delta_{F}(k)=-i(k^2-i\epsilon)^{-1}.
\end{equation}
where we have introduced for later convenience a new object $\Xi(q)$ which is the numerator of the scalar field propagator.
The cubic vertex $\Phi\Phi\varphi$ reads
\begin{equation}
	\begin{aligned}
		\Gamma^{(3)}(q_1,q_2;k) &= \f{i}{2\sqrt{2}f}\Bigg(
		-(4-2\Delta)\mathcal{K}(q_2)+q_{2}^{\mu}\f{\p \mathcal{K}(q_2)}{\p q_2^\mu}+\f{1}{2}\big\lbrace \delta_\mu^\nu k_\rho +\delta_\rho^\nu k_\mu -\eta_{\mu\rho}k^\nu\big\rbrace \\
	&\times q_{2\nu} \f{\p^2 \mathcal{K}(q_2)}{\p q_{2\mu}\p q_{2\rho}}
	+\Delta\ k^{\mu}\f{\p \mathcal{K}(q_2)}{\p q_2^\mu}+\ O(k^2)
		\Bigg) + (q_1 \leftrightarrow q_2),
	\end{aligned}
\end{equation}
where $q_1+q_2+k=0$. In the above expression of $\Gamma^{(3)}$ the terms linear in $k$ actually vanishes when we explicitly write the terms under $q_1 \leftrightarrow q_2$ exchange and substitute $q_2=-q_1-k$. 
The quartic vertex $\Phi\Phi\varphi\varphi$ reads as
\begin{equation}
	\begin{aligned}
		\Gamma^{(4)}(q_1,q_2;k_1,k_2) &= \f{i}{2f^2}
		\Bigg((2-\Delta)(3-2\Delta)\mathcal{K}(q_2)\\
		&-(3-2\Delta)q_{2}^{\mu}\f{\p \mathcal{K}(q_2)}{\p q_2^\mu}
		+\f{1}{2}q_{2}^{\mu}q_2^\nu\f{\p^2 \mathcal{K}(q_2)}{\p q_2^\mu \p q_{2}^{\nu}}+O(k_1, k_2)\Bigg)
		+ (q_1 \leftrightarrow q_2),
	\end{aligned}
\end{equation}
where $q_1+q_2+k_1+k_2=0$. Three dilaton field interaction vertex $\varphi\varphi\varphi$ reads as
\be
V^{(3)}(k_1,k_2,k_3)&=&\ \f{i}{f} \ \mathcal{O}(k_i^4).
\ee

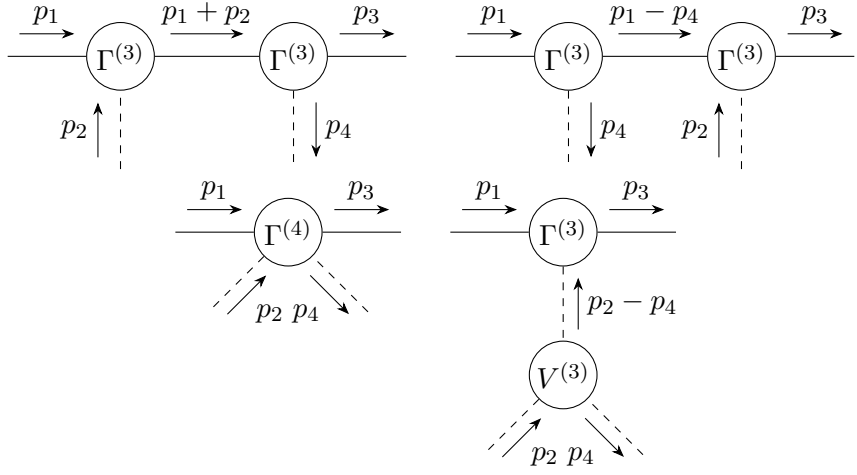
\begin{figure}[t!]
\begin{center}
\begin{tikzpicture}[baseline=(a.base)]
	\begin{feynman}
		\vertex (i1);
		\vertex [right=of i1, blob, pattern=none, anchor=center, minimum size=.9cm] (a) {$\Gamma^{(3)}$};
		\vertex [node distance = 2.3cm, right=of a, blob, pattern=none, anchor=center, minimum size=.9cm] (b) {$\Gamma^{(3)}$};
		\vertex [right=of b] (o1) ;
		\vertex [below=of a] (i2);
		\vertex [below=of b] (o2); 
		\diagram* {
			(i1) --[momentum=$p_1$] (a) --[momentum=$p_1+p_2$] (b) --[momentum=$p_3$] (o1),
			(i2)-- [dashed, momentum=$p_2$] (a),
			(b) --[dashed, momentum=$p_4$] (o2) 
		};
	\end{feynman}
\end{tikzpicture}
\hspace{1em}
\begin{tikzpicture}[baseline=(a.base)]
	\begin{feynman}
		\vertex (i1);
		\vertex [right=of i1, blob, pattern=none, anchor=center, minimum size=.9cm] (a) {$\Gamma^{(3)}$};
		\vertex [node distance = 2.3cm, right=of a, blob, pattern=none, anchor=center, minimum size=.9cm] (b) {$\Gamma^{(3)}$};
		\vertex [right=of b] (o1) ;
		\vertex [below=of a] (i2);
		\vertex [below=of b] (o2); 
		\diagram* {
			(i1) --[momentum=$p_1$] (a) --[momentum=$p_1-p_4$] (b) --[momentum=$p_3$] (o1),
			(a)-- [dashed, momentum=$p_4$] (i2),
			(o2) --[dashed, momentum=$p_2$] (b) 
		};
	\end{feynman}
\end{tikzpicture}
\\
\begin{tikzpicture}[baseline=(a.base)]
	\begin{feynman}
		\vertex (i1);
		\vertex [right=of i1, blob, pattern=none, anchor=center, minimum size=.9cm] (a) {$\Gamma^{(4)}$};
		\vertex [right=of a] (o1) ;
		\vertex [below right=of a] (i2);
		\vertex [below left=of a] (o2); 
		\diagram* {
			(i1) --[momentum=$p_1$] (a)--[momentum=$p_3$] (o1),
			(i2)-- [dashed, rmomentum={[arrow distance=2mm]$p_4$}] (a),
			(a) --[dashed, rmomentum={[arrow distance=2mm]$p_2$}] (o2) 
		};
	\end{feynman}
\end{tikzpicture}
\hspace{1em}
\begin{tikzpicture}[baseline=(a.base)]
	\begin{feynman}
		\vertex (i1);
		\vertex [right=of i1, blob, pattern=none, anchor=center, minimum size=.9cm] (a) {$\Gamma^{(3)}$};
		\vertex [node distance = 1.9cm, below=of a, blob, pattern=none, anchor=center, minimum size=.9cm] (b) {$V^{(3)}$};
		\vertex [right=of a] (o1) ;
		\vertex [below right=of b] (i2);
		\vertex [below left=of b] (o2); 
		\diagram* {
			(i1) --[momentum=$p_1$] (a)--[momentum=$p_3$] (o1),
			(a)-- [dashed, rmomentum={[arrow distance=2mm]$p_2-p_4$}] (b),
			(i2)-- [dashed, rmomentum={[arrow distance=2mm]$p_4$}] (b),
			(b) --[dashed, rmomentum={[arrow distance=2mm]$p_2$}] (o2) 
		};
	\end{feynman}
\end{tikzpicture}
\end{center}
\vspace{-1em}
	\caption{The set of diagrams contributing to the $\mathcal{T}_{AB\rightarrow AB}$ amplitudes. Here the solid lines represent scalar particles $A$ and dashed lines represent dilatons $B$.}
	\label{fig:feynman_diagrams_effective}
\end{figure}

We can now compute the $AB\rightarrow AB$ amplitude using the   Feynman diagrams depicted in figure \ref{fig:feynman_diagrams_effective}.
The total amplitude describing the scattering process $AB\rightarrow AB$ becomes
\begin{equation}
	\label{eq:amplitude_sum}
	\mathcal{T}_{AB\rightarrow AB} = \mathcal{T}_1+\mathcal{T}_2+\mathcal{T}_3+\mathcal{T}_4,
\end{equation}
where we have
\begin{equation}
	\begin{aligned}
		i\mathcal{T}_1&=\Gamma^{(3)}(p_1,-p_1-p_2;+p_2)D_F(p_1+p_2)\Gamma^{(3)}(p_1+p_2,-p_3;-p_4),\\
		i\mathcal{T}_2&=\Gamma^{(3)}(p_1,-p_1+p_4;-p_4)D_F(p_1-p_4)\Gamma^{(3)}(p_1-p_4,-p_3;+p_2),\\
		i\mathcal{T}_3&=\Gamma^{(4)}(p_1,-p_3;p_2,-p_4),\\
		i\mathcal{T}_4&=\Gamma^{(3)}(p_1,-p_3;p_2-p_4)\Delta_F(p_2-p_4)V^{(3)}(-p_2+p_4,p_2,-p_4).
	\end{aligned}
\end{equation}
Plugging the explicit expressions of the Feynman propagators and the effective vertices, expanding around $p^\mu_2=0$ and $p^\mu_4=0$ one obtains the following
\begin{equation}
	\label{eq:three_contributions}
	\begin{aligned}
        f^2\mathcal{T}_1 &=m^4(p_1 \cdot p_2)^{-1} -2m^2(\Delta -1) -2im^4 \Xi'(m^2)+O\big( (p_1\cdot p_2), (p_3\cdot p_4) \big),\\
        f^2\mathcal{T}_2 &=-m^4(p_1 \cdot p_4)^{-1} -2m^2(\Delta -1)-2im^4 \Xi'(m^2)+O\big( (p_1\cdot p_2), (p_3\cdot p_4) \big),\\
        f^2\mathcal{T}_3 &=(4\Delta -5)m^2\ +4im^4\Xi'(m^2)+O\big( (p_1\cdot p_2), (p_3\cdot p_4) \big),\\
        f^2\mathcal{T}_4 &=0\ +O\big( (p_1\cdot p_2), (p_3\cdot p_4) \big),
	\end{aligned}
\end{equation}
where $\Xi'(m^2)\equiv \f{\p\Xi(q)}{\p q^2}\Big{|}_{q^2=-m^2}$.

In deriving these we have used several straightforward relations which   follow from the fact that $\mathcal{K}$ is a scalar quantity and thus can depend only on $q^2$. As a consequence the same is true for $\Xi$ defined in \eqref{eq:propagator}. These relations read as
\begin{equation}
	\mathcal{K}(q)=\mathcal{K}(-q), \quad
	\Xi(q)=\Xi(-q), \quad
	 \f{\p \mathcal{K}(q)}{\p q^\mu}=2q_\mu \f{\p \mathcal{K}(q)}{\p q^2},\quad
	 \f{\p \Xi(q)}{\p q^\mu}=2q_\mu \f{\p \Xi(q)}{\p q^2},
\end{equation}
together with
\begin{equation}
	\begin{aligned}
		&\f{\p \mathcal{K}(q)}{\p q^\mu}\Xi(q)+\mathcal{K}(q)\f{\p \Xi(q)}{\p q^\mu} = 2iq_\mu,\\
		&\f{\p^2 \mathcal{K}(q)}{\p q^\mu \p q^\nu}\Xi(q)+\f{\p\mathcal{K}(q)}{\p q^\nu}\f{\p \Xi(q)}{\p q^\mu}+\f{\p\mathcal{K}(q)}{\p q^\mu}\f{\p \Xi(q)}{\p q^\nu}+\mathcal{K}(q)\f{\p^2 \Xi(q)}{\p q^\mu \p q^\nu} = 2i\eta_{\mu\nu}.
	\end{aligned}
\end{equation}
At the end of the evaluation of the Feynman diagrams we use the following on-shell conditions,
\be
&&\mathcal{K}(q)\Big{|}_{q^2=-m^2}=0\ ,\ \Xi(q)\Big{|}_{q^2=-m^2}=-i\ ,\ \f{\p \mathcal{K}(q)}{\p q^\mu}\Big{|}_{q^2=-m^2}=-2q_\mu,\\
&&\f{\p^2 \mathcal{K}(q)}{\p q^\mu \p q^\nu}\Big{|}_{q^2=-m^2}=-2\eta_{\mu\nu}+8iq_\mu q_\nu\ \Xi'(m^2).
\ee
Plugging \eqref{eq:three_contributions} into \eqref{eq:amplitude_sum} we obtain
\begin{equation}
	f^2\mathcal{T}_{AB\rightarrow AB} = \f{m^4}{(p_1\cdot p_2)}-\f{m^4}{(p_1\cdot p_4)}-m^2
	+O\big( (p_1\cdot p_2), (p_3\cdot p_4)\big).
\end{equation}
Using the definition of the Mandelsatm variables \eqref{eq:mandelstam} and the definition of the tilded amplitudes \eqref{eq:new_amplitudes} we obtain the final result
\begin{equation}
	\label{eq:soft_ABtoAB_final}
	\widetilde{T}_{AB\rightarrow AB}(s,t,u)=
    \f{-2m^4}{s-m^2}+\f{-2m^4}{u-m^2}-m^2 +\ O(u-m^2,s-m^2).
\end{equation}
The above result is independent of $\Delta$, which is not surprising since $\Delta$ appears in the definition of $\widehat{\Phi}(x)$ which is nothing but a field redefinition of $\Phi(x)$. Here we want to emphasize that up to the subleading order in the expansion parameters $s-m^2$ and $u-m^2$, our amplitude $\widetilde{T}_{AB\rightarrow AB}(s,t,u)$ is theory independent. We perturbatively verify the pole parts of the above result for a particular model in appendix \ref{app:perturbative_AABB}.
In addition, in appendix \ref{app:worldline} we argue for this universal soft behavior by analysing the worldline action of a massive particle in a dilaton background.

At the next order in $ O(u-m^2,s-m^2)$ the contribution to \eqref{eq:soft_ABtoAB_final} will depend on $\Xi'(m^2), \Xi''(m^2)$ and the non-minimal interaction strength of the scalar field with the dilaton e.g. flat space interaction follows from $\widehat{R}^{\mu\nu}\p_\mu \widehat{\Phi}\p_\nu \widehat{\Phi}$.\footnote{The theory dependence of sub-subleading soft graviton theorem along the same line of derivation has been worked out in \cite{Laddha:2017ygw}.} Our derivation is also motivated from the literatures \cite{DiVecchia:2015jaq,DiVecchia:2017uqn,Guerrieri:2017ujb}, where tree level single and double soft dilaton theorems have been studied.

\section{Numerical bounds}
\label{sec:num}
In order to put numerical bounds on the $a$-anomaly we use the numerical approach of \cite{Paulos:2017fhb,Homrich:2019cbt}. For the concise summary of this approach see section 1 and 4.1 of \cite{Hebbar:2020ukp}. In what follows we first explain very briefly our numerical setup and then present our numerical bounds. All the technical details of the numerical setup can be found in appendix \ref{app:details_numerical_setup}.

\subsection{Numerical setup}
\label{sec:setup}
We start by introducing two types of the rho-variables which automatically take care of the branch cuts discussed in \eqref{eq:disc1} and \eqref{eq:disc2}. They read
\begin{equation}
\label{eq:rho}
\begin{aligned}
\myRho_1(z;z_0) &\equiv \frac{\sqrt{4m^2-z_0}-\sqrt{4m^2-z}}{\sqrt{4m^2-z_0}+\sqrt{4m^2-z}},\\
\myRho_2(z;z_0) &\equiv \frac{\sqrt{9m^2-z_0}-\sqrt{9m^2-z}}{\sqrt{9m^2-z_0}+\sqrt{9m^2-z}}.
\end{aligned}
\end{equation}
Here $z_0$ is a free parameter and can be set to any convenient value for both variables $\myRho_1$ and $\myRho_2$ independently. Using these variables we can write the following ansatze
\begin{equation}
\label{eq:ansatze}
\begin{aligned}
	\mathcal{T}_{AA \rightarrow AA}(s,t,u) &=\sum_{a=0}^{\infty}\sum_{b=0}^{\infty}\sum_{c=0}^{\infty}\alpha_{abc}\left(\myRho_1(s;4m^2/3)\right)^a\left(\myRho_1(t;4m^2/3)\right)^b\left(\myRho_1(u;4m^2/3)\right)^c,\\
	\widetilde{\mathcal{T}}_{AB \rightarrow AB}(s,t,u) &=
	-\frac{2m^4}{s-m^2}-\frac{2m^4}{u-m^2}\\
	&+m^2
	\sum_{a=0}^{\infty}\sum_{b=0}^{\infty}\sum_{c=0}^{\infty}\beta_{abc}\left(\myRho_2(s;2m^2/3)\right)^a\left(\myRho_1(t;2m^2/3)\right)^b\left(\myRho_2(u;2m^2/3)\right)^c,\\
	\widetilde{\mathcal{T}}_{BB \rightarrow BB}(s,t,u) &=
	m^4\sum_{a=0}^{\infty}\sum_{b=0}^{\infty}\sum_{c=0}^{\infty}\gamma_{abc}\left(\myRho_1(s;0)\right)^a\left(\myRho_1(t;0)\right)^b\left(\myRho_1(u;0)\right)^c.
\end{aligned}
\end{equation}
Here $\alpha$, $\beta$ and $\gamma$ are the unknown real coefficients. Due to the crossing symmetry discussed in section \ref{sec:crossing} the coefficients $\alpha$ and $\gamma$ are fully symmetric in their indices. Instead the coefficients $\beta$ obey the following condition
\begin{equation}
\beta_{abc} = \beta_{cba}.
\end{equation}
The ansatz for the amplitude $\widetilde{\mathcal{T}}_{AA \rightarrow BB}(s,t,u)$ is obtained from \eqref{eq:ansatze} by exchanging  $s\leftrightarrow t$ according to the crossing equation \eqref{eq:crossin_3}. In order to simplify the ansatze and take into account that the variable $u$ is not independent we further demand that our coefficients obey
\begin{equation}
(abc)\neq 0:\qquad	\alpha_{abc}=	\beta_{abc} = 	\gamma_{abc} =0.
\end{equation}
We will be working in the units of the physical mass $m$. Effectively this allows us to set
\begin{equation}
	m =1
\end{equation}
in all the equations below.

Using the ansatze \eqref{eq:ansatze} one can define the following optimization problem: numerically find the set of coefficients $\alpha$, $\beta$ and $\gamma$ which maximizes some linear combination of these coefficients such that the unitarity conditions \eqref{eq:final_condition_1} and \eqref{eq:unitarity_2_final} are satisfied for all physical values of $s$ and spins $\ell$. Below we will explain how the coefficients $\alpha$, $\beta$ and $\gamma$ are related to the $a$-anomaly and to the physical observables.

In order to make the above optimization problem feasible in practice we first truncate the ansatz \eqref{eq:ansatze} to a finite sum with the parameter $N_{max}$ according to
\begin{equation}
	a+b+c \leq N_{max}.
\end{equation}
We impose unitarity conditions for a finite number of spins $\ell$ up to some fixed value $L_{max}$, namely
\begin{equation}
	\ell = 0, 1, 2, \ldots, L_{max}.
\end{equation}
Finally, we pick a finite grid of $s$ values where the unitarity conditions are imposed. The number of points in this grid is denoted by $N_\text{grid}$. After performing these ``truncations'' we use SDPB software \cite{Simmons-Duffin:2015qma,Landry:2019qug} to solve the optimization problem numerically.

\paragraph{Soft conditions}
At low energy the $AA\rightarrow BB$ and $BB\rightarrow BB$ amplitudes have a very particular (soft) behavior according to \eqref{eq:soft_AABB} and \eqref{eq:soft_BBBB} respectively. As a result we have additional constraints on the coefficients of the ansatze. Plugging here the definition of $(t,u)$ variables in terms of $(s,\cos\theta)$ according to \eqref{eq:definitionsTU} and expanding around small values of $s$ keeping $\cos\theta$ fixed we get expressions which should match \eqref{eq:soft_AABB} and \eqref{eq:soft_BBBB}. Requiring this matching one obtains the following constraints
\begin{equation}
	\label{eq:soft_conditions_main}
	\beta_{000} = -1 - (98-40\sqrt{6})\beta_{001} + \ldots,\quad
	\gamma_{000} = 0,\quad
	\gamma_{001} = \frac{1}{4} \left(512 a -2\gamma_{002}+\gamma_{011}\right),
\end{equation}
where $a$ is the $a$-anomaly. Plugging the solutions \eqref{eq:soft_conditions_main} into the anstatze \eqref{eq:ansatze} we effectively eliminate three coefficients $\beta_{000}$,  $\gamma_{000}$, $\gamma_{001}$ and introduce one additional coefficient $a$ (the $a$-anomaly).

\paragraph{Physical observables}
As explained in the introduction we will use either the pair of observables $(\lambda_0, \lambda_2)$ or $(\Lambda_0, \Lambda_2)$ in order to describe the scattering amplitude $AA \rightarrow AA$.\footnote{In the definitions \eqref{eq:lambdas} and \eqref{eq:Lambdas} one could take the derivative in $s$ is two different ways. First, one could plug the value of $u=4m^2-s-t$ and then treating $s$ and $t$ as independent variables evaluate the derivative in $s$ at the $s_0$ and $t_0$ point. Second, one could replace both $t$ and $u$ variables in terms of $s$ and $\cos\theta$, where $\theta$ is the scattering angle and only then take the derivative in $s$ at the $s_0$ and $x_0$ point. In this paper we will use the second option. At the crossing symmetric point $\lambda_2^\text{former} = 4/3 \ \lambda_2^\text{latter} $. At the forward point $\Lambda_2^\text{former} = \Lambda_2^\text{latter} $. } Applying either the definition \eqref{eq:lambdas} or \eqref{eq:Lambdas} to the ansatz \eqref{eq:ansatze} for the amplitude $AA \rightarrow AA$ we obtain a linear relation between $(\lambda_0, \lambda_2)$ or $(\Lambda_0, \Lambda_2)$ and the coefficients $\vec \alpha$. We solve these relations for
\begin{equation}
	\alpha_{000}\quad\text{and}\quad \alpha_{001}
\end{equation}
in terms of either $(\lambda_0, \lambda_2)$ or $(\Lambda_0, \Lambda_2)$ and the rest of $\vec \alpha$. We substitute this solution in the ansatz. This effectively removes two coefficients $\alpha_{000}$ and $\alpha_{001}$ in the ansatz and instead introduce the depends either on $(\lambda_0, \lambda_2)$ or on $(\Lambda_0, \Lambda_2)$.

\paragraph{Extension of the ansatz}
The ansatz  \eqref{eq:ansatze} is very powerful. Nevertheless since in practice we work at finite value of $N_{max}$ the bounds depend on $N_{max}$. And in order to get accurate bounds one needs to perform an extrapolation $N_{max}\rightarrow \infty$. The convergence with $N_{max}$ depends on a particular optimization problem. 

It is convenient to enlarge the ansatz for the  $AA \rightarrow AA$ by adding the following term
\begin{equation}
	\label{eq:singularity}
	\mathcal{T}_{AA\rightarrow AA}(s,t,u)  \supset \alpha'_{000} \times \left( \frac{1}{\myRho_1(s;4/3)-1}+ \frac{1}{\myRho_1(t;4/3)-1}+ \frac{1}{\myRho_1(u;4/3)-1}\right),
\end{equation}
which contains a new parameter $\alpha'_{000}$. 
When maximizing $\lambda_0$ this term gives a major contribution and improves the convergence with $N_{max}$ drastically \cite{Paulos:2017fhb}.
We also enlarge the ansatz  for $BB \rightarrow BB$ amplitude by adding the following term
\begin{equation}
	\label{eq:extra_term_ansatz}
	\widetilde{\mathcal{T}} _{BB \rightarrow BB}(s,t,u) \supset c\, \widetilde{\mathcal{T}}^\text{free} _{BB \rightarrow BB}(s,t,u),
\end{equation}
where $c$ is a free parameter and $\widetilde{\mathcal{T}}^\text{free} _{BB \rightarrow BB}$ is the dilaton scattering amplitude in the free scalar theory. It is hard to compute $\widetilde{\mathcal{T}}^\text{free} _{BB \rightarrow BB}$ exactly. Luckily we do not need its explicit expression,  in practice we will only need its imaginary part and the imaginary part of the associated partial amplitude, since only the latter enter in the unitarity constraint \eqref{eq:unitarity_2_final}.  These will be computed in equations \eqref{ImTbbbb}, \eqref{eq:projection_poles_free} and \eqref{eq:imBBtoBB_partial}.

\subsection{Results}
\label{sec:results}

We start in  section \ref{sec:absolute_minimum_a} by addressing the simplest possible question: what is the lowest value of the $a$-anomaly in the UV CFT which leads to a single $\ZZ$ odd asymptotic state given some relevant deformation. We will reconstruct the spin=0 partial amplitudes of our setup which lead to the absolute minimum of the $a$-anomaly. In section \ref{sec:consistency_check} we discuss several consistency checks of our numerical code. In section \ref{sec:a_bounds_lambdas} we construct a lower bound on the $a$-anomaly as a function of $\lambda_0$, $\lambda_2$, $\Lambda_0$ and $\Lambda_2$. We reconstruct the spin=0 partial amplitudes of our setup corresponding to the maximally allowed value of $\lambda_0$.

\subsubsection{Absolute minimum of the $a$-anomaly}
\label{sec:absolute_minimum_a}
Let us start by addressing the following question: what is the lowest value of the $a$-anomaly in the UV CFT which leads to a single $\ZZ$ odd asymptotic state given some relevant deformation.  For running the numerics we have found the optimal size of the grid to be $N_\text{grid}=300$. We have checked that $N_\text{grid}=350$ and $N_\text{grid}=400$ lead to the same solution.  For $N_\text{grid}=200$ the results differ significantly from the ones found with larger grids.

\begin{figure}[th]
	\centering
	\includegraphics[width=0.7\textwidth]{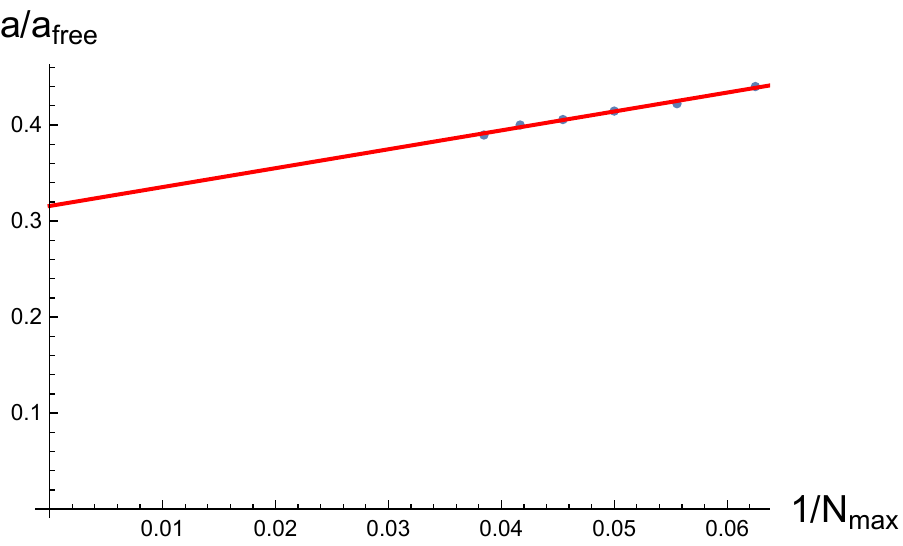}
	\caption{Minimum possible value of the a-anomaly without any further assumptions as a function of $1/N_{max}$ with $L_{max}=N_{max}+10$. The numerical results are depicted by blue points. Linear extrapolation to $N_{max} \rightarrow \infty$ depicted by the red line gives  $0.316\pm 0.015$ for the minimum of $a/a_\text{free}$. }
	\label{fig:plot_I}
\end{figure}

Let us fix the size of the ansatz to be $N_{max}=20$. The minimum of the $a$-anomaly for various values of $L_{max}$ is found to be
\begin{multline}
	\label{eq:values}
	\{L_{max}, a/a_\text{free}\} = \{  (16, 0.4015), (18, 0.4074), (20, 0.4100), (22, 0.4115), (24, 
			0.4125),\\ (26, 0.4133), (28, 0.4140), (30, 0.4146), (32, 
			0.4150), (34, 0.4154) \}.
\end{multline}
We see the the lower bound on the $a$-anomaly is stable under the change of $L_{max}$ and gets stronger when $L_{max}$  increases. In all the future numerical studies we make the following conservative choice for
\begin{equation}
	L_{max} = N_{max}+10.
\end{equation}
Let us now investigate the dependence of the numerical solution on $N_{max}$. We obtain the following values for the minimum of the $a$-anomaly for various values of $N_{max}$ 
\begin{multline}
	\label{eq:minimum_a_chane_Nmax}
	\{N_{max}, a/a_\text{free}\} = \{  (16, 0.4401), (18, 0.4223), (20, 0.4146), \\
	(22, 0.4058), (24, 0.4001), (26, 0.3897) \}.
\end{multline}
Using $N_{max}=$ $16$, $18$, $20$, $22$, $24$ and $26$ we can extrapolate our data to $N_{max}=\infty$ with the following linear function
\begin{equation}
	a/a_\text{free} = 0.316+ 1.969 /N_{max}.
\end{equation}
The dependence of the minimum of the $a$-anomaly on $N_{max}$ and its extrapolation is given in  figure \ref{fig:plot_I}.
We conclude that the absolute minimum of the $a$-anomaly in our setup is
\begin{equation}
	a/a_\text{free} \gtrsim 0.316 \pm 0.015.
\end{equation}
Here we have also included the estimated extrapolation error.

\begin{figure}[ht]
	\centering
	
	\includegraphics[width=0.45\textwidth]{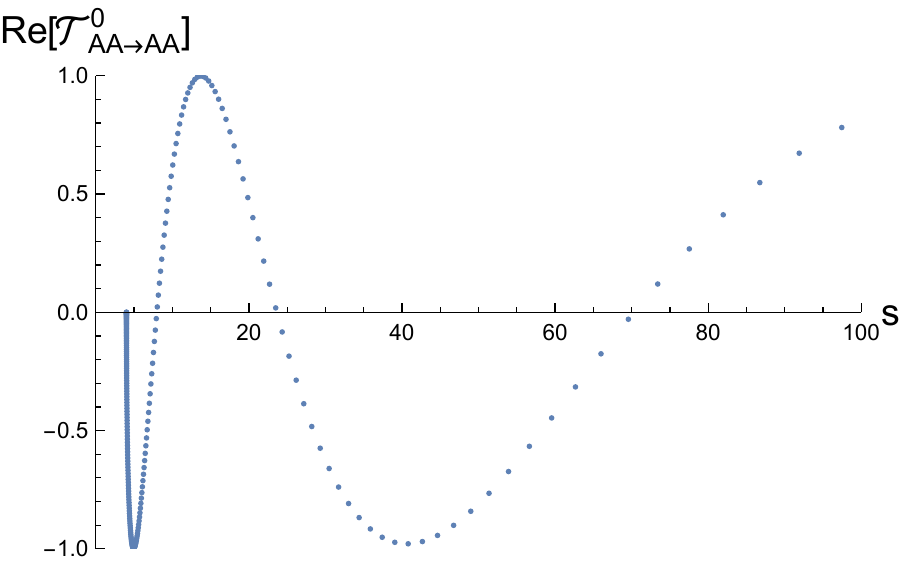}
	\includegraphics[width=0.45\textwidth]{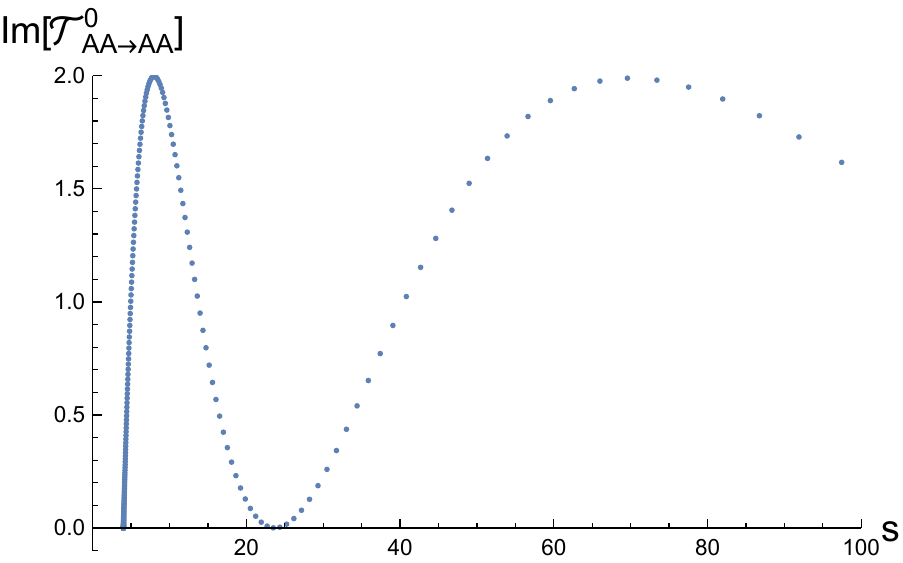}
	\caption{Real and imaginary parts of the spin 0 interacting part of the $AA\rightarrow AA$ partial amplitude leading to the absolute minimum of the $a$-anomaly. It is constructed at $N_{max}=20$ and $L_{max}=30$.}
	\label{fig:minimumAnomaly_1}
	\vspace{5mm}
	
	\includegraphics[width=0.45\textwidth]{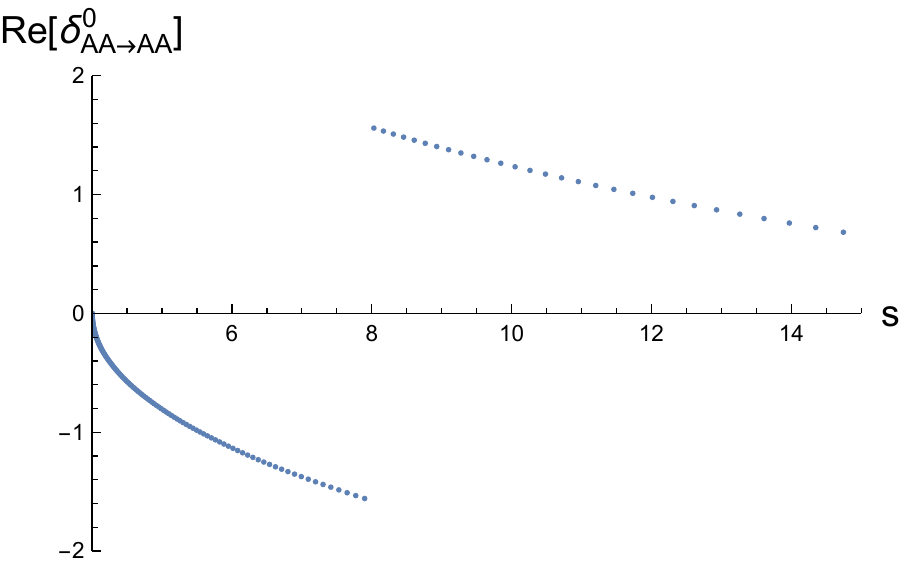}
	\includegraphics[width=0.45\textwidth]{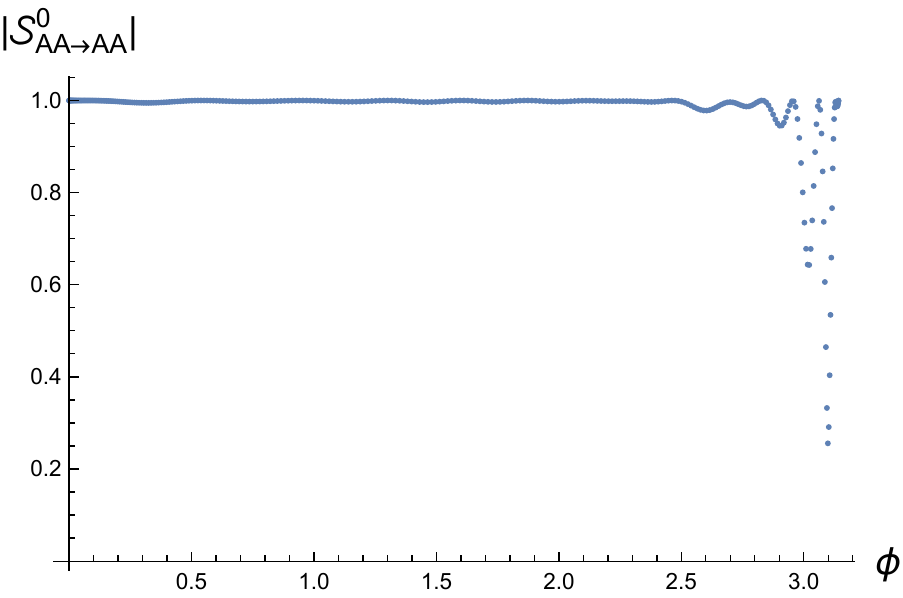}
	\caption{An alternative representation of the amplitude given in figure \ref{fig:minimumAnomaly_1}. Left plot represents the real part of the spin 0 phase shift of the $AA\rightarrow AA$ scattering defined in \eqref{eq:phase_shift}. The apparent jump around $s=8$ is due to the periodicity $\delta \simeq \delta+\pi$.
	 Right plot represents the absolute value of the spin 0 partial amplitude of the $AA\rightarrow AA$ scattering. On the real axis instead of the $s$ variable we use the $\phi$ variable defined in \eqref{eq:phi_variable}. The amplitude is fully ``elastic'' up to very high energies.}
	\label{fig:minimumAnomaly_2}
	\vspace{5mm}
	
	\includegraphics[width=0.45\textwidth]{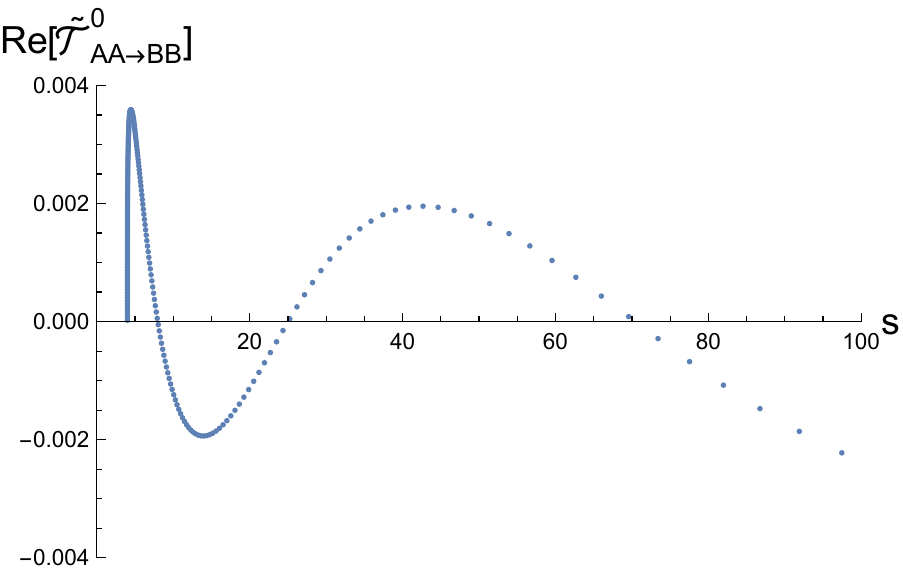}
	\includegraphics[width=0.45\textwidth]{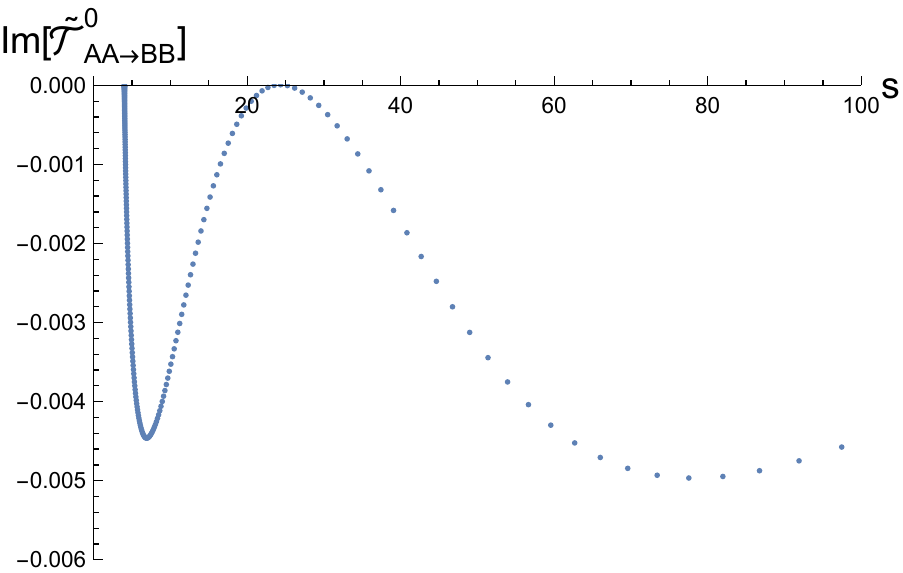}
	\caption{Real and imaginary parts of the spin 0 interacting part of the $AA\rightarrow BB$ partial amplitude leading to the absolute minimum of the $a$-anomaly. It is constructed at $N_{max}=20$ and $L_{max}=30$.}
	\label{fig:minimumAnomaly_3}
\end{figure}

At the absolute minimum of the $a$-anomaly we can actually reconstruct numerically scattering and partial amplitudes of all the process of our setup. In figures \ref{fig:minimumAnomaly_1} - \ref{fig:minimumAnomaly_3} we present the spin zero partial amplitudes of the $AA\rightarrow AA$ and $AA\rightarrow BB$ processes. In the left figure \ref{fig:minimumAnomaly_2} we have plotted the real part of the spin zero  phase shift of the $AA\rightarrow AA$ process. We recall that the spin $\ell$ phase shift $\delta^\ell$ of the $AA\rightarrow AA$ process is defined via
\begin{equation}
	\label{eq:phase_shift}
	\mathcal{S}^\ell_{AA\rightarrow AA}(s) = e^{2i\delta^\ell_{AA\rightarrow AA}(s)}.
\end{equation}
From the right figure \ref{fig:minimumAnomaly_2} we see that the amplitude is fully ``elastic'' up to very high energies.

\begin{figure}[ht]
	\centering
	\includegraphics[width=0.9\textwidth]{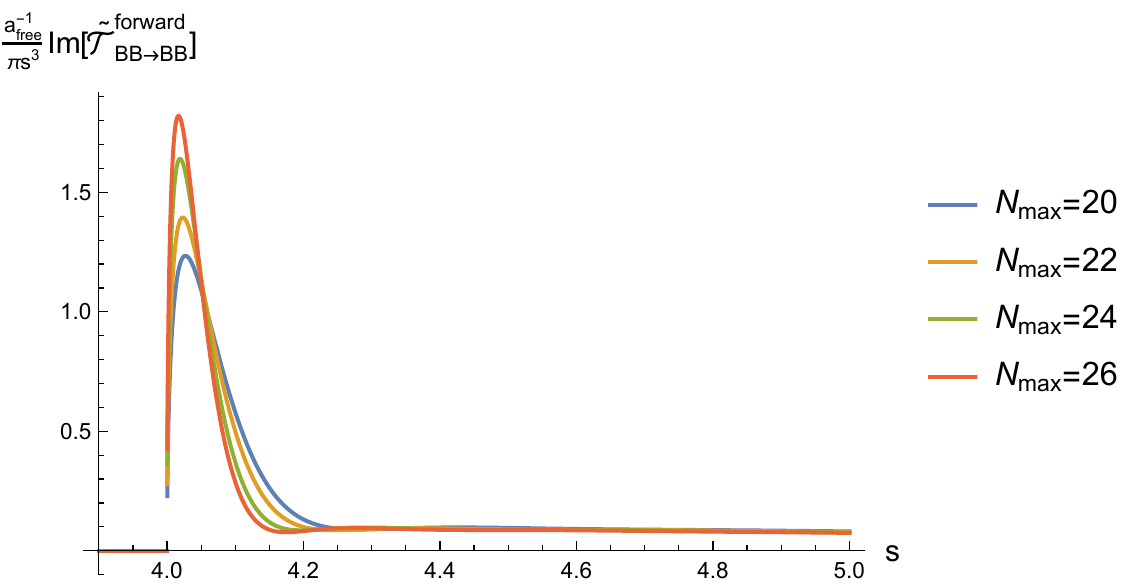}
	\caption{Integrand appearing in the sum-rule \eqref{eq:sum_rule_amplitude}  for the absolute minimum of the $a$-anomaly. Different colors indicate different values of $N_{max}$. Numerical integration of these function leads to $a/a_\text{free}=0.4146$, $0.4058$, $0.4001$ and $0.3896$ in a perfect agreement with \eqref{eq:minimum_a_chane_Nmax}.}
	\label{fig:minimumAnomaly_4}
	
	\vspace{10mm}
	
	\includegraphics[width=0.9\textwidth]{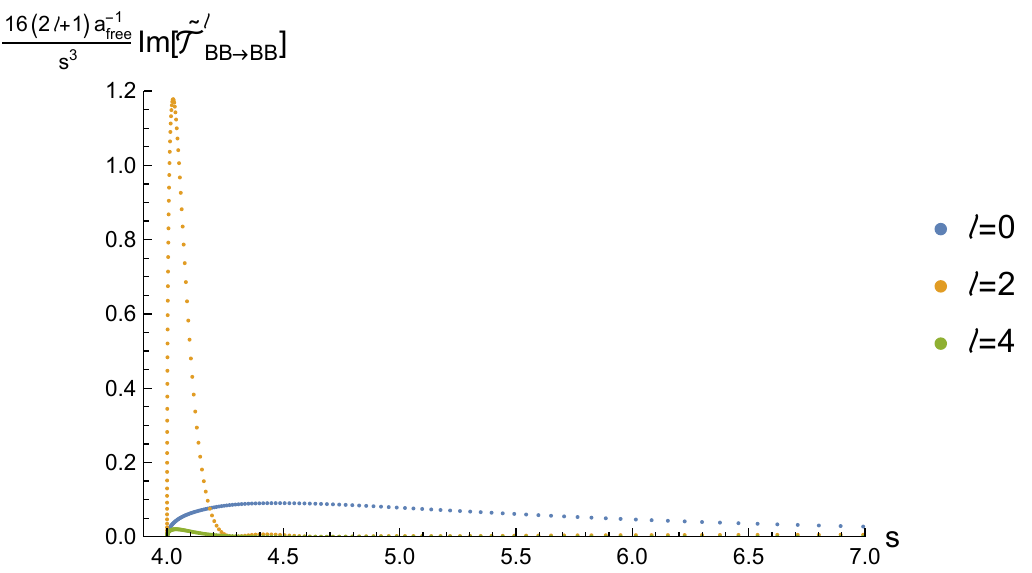}
	\caption{Integrand appearing in the sum-rule \eqref{eq:sum_rule_pamplitude}  for the absolute minimum of the $a$-anomaly for spin 0, 2 and 4. Numerical integration of this function leads to $a/a_\text{free}=0.2396+0.1605+0.0105+\ldots=0.4106+\ldots$, where the three entries correspond to spin 0, 2 and 4 respectively and the dots indicate the contribution due to higher spins. Plots are constructed at $N_{max}=20$ and $L_{max}=30$. Summing up these three contribution gives the $N_{max}=20$ curve of figure \ref{fig:minimumAnomaly_4}.}
	\label{fig:minimumAnomaly_5}
\end{figure}

In figure \ref{fig:minimumAnomaly_4} we plot the integrand of the sum-rule \eqref{eq:sum_rule_amplitude} for various values of $N_{max}$. Numerical integration of these functions gives the values $a/a_\text{free}=0.4146$, $0.4058$, $0.4001$ and $0.3897$  which are in a perfect agreement with \eqref{eq:minimum_a_chane_Nmax}. In figure  \ref{fig:minimumAnomaly_5} we plot the integrand of the sum-rule \eqref{eq:sum_rule_pamplitude} for spin 0, 2 and 4. Numerical integration of these functions gives the following value of the $a$-anomaly $a/a_\text{free}=0.2396+0.1605+0.0105+\ldots=0.4106+\ldots$, where the three entries correspond to spin 0, 2 and 4 respectively and the dots indicate the contribution due to higher spins..

\subsubsection{Consistency checks}
\label{sec:consistency_check}
Before presenting more bounds let us perform several consistency checks. 

First, we can set all the coefficients $\alpha$ and $\beta$ in the ansatz \eqref{eq:ansatze} to zero after imposing the first condition in \eqref{eq:soft_conditions_main}. This situation corresponds to particle $A$ being a free massive scalar, with the $AA\rightarrow BB$ scattering amplitude given by \eqref{eq:matter_compensator_free}. As explained below \eqref{eq:matter_compensator_free} this inevitably leads to $a/a_\text{free} = 1$. We successfully reproduce this theoretical outcome numerically.
It is important to notice that the ansatz \eqref{eq:ansatze} requires very large numbers $N_{max}$ in order to reproduce the free theory accurately. In practice it is very non-economical to work with such a big value of $N_{max}$. This was the motivation behind the introduction of the additional term \eqref{eq:extra_term_ansatz} in the $BB\rightarrow BB$ ansatz which improves the convergence with $N_{max}$ drastically in this particular situation.

Second, let us set all the $\alpha$ coefficients to zero and leave $\beta$ coefficients to be completely free. One might expect the same outcome as before, however solving the optimization problem we obtain
\begin{equation}
	\label{eq:semi_free_solution}
	 a/a_\text{free} \approx 0.9071.
 \end{equation}
This result was obtained with $N_{max}=L_{max}=20$. For comparison we get $0.9107$ for $N_{max}=L_{max}=10$.
This result is very stable under the change of $N_{max}$ and $L_{max}$. The solution of the optimization problem gives the coefficients of the ansatz such that only
\begin{equation}
	\label{eq:solution_semi_free}
	\beta_{00n} \neq 0.
\end{equation}
Under closer inspection of the unitarity conditions one observes that setting $\alpha$ to zero forces the imaginary part of the $\widetilde{\mathcal{T}}_{AA\rightarrow BB}^{\ell}(s)$ partial amplitude to be zero due to the last line in \eqref{eq:conditions}. However, the real part can still be non-zero. The solution \eqref{eq:solution_semi_free} exactly reproduces such a situation. We, thus, conclude that the unitarity conditions we use do not fully constraint the behavior of the matter - dilaton scattering given the form of the matter scattering.

It is possible to guess analytically the $AA\rightarrow BB$ scattering amplitude which gives the solution \eqref{eq:semi_free_solution}, \eqref{eq:solution_semi_free}.  Let us assume no discontinuity in the s-channel of the amplitude $AA\rightarrow BB$ (as follows from absence of $AA\to AA$ scattering), we can then write the general ansatz
\begin{equation}
\widetilde{\mathcal{T}}_{AA\rightarrow BB}= -m^2-\f{2m^4}{t-m^2}-\f{2m^4}{u-m^2} 
-\int_{9m^2}^\infty dw q(w) \left(\f{m^2}{t-w}+\f{m^2}{u-w}-\f{2m^2}{m^2-w}\right),
\end{equation}
where $q(w)\ge 0$ by unitarity. For the special case when $q(w)=Q m^2\delta(w-9m^2)$ we obtain
\be
\label{eq:amplitude_guess}
\widetilde{\mathcal{T}}_{AA\rightarrow BB}= -m^2-\f{2m^4}{t-m^2}-\f{2m^4}{u-m^2}
-Q m^2 \left(\f{m^2}{t-9m^2}+\f{m^2}{u-9m^2}+\f{1}{4}\right).
\ee
Plugging it into \eqref{eq:sum_rule_a_better} we obtain $
a^\text{UV}/a_\text{free}\ge 0.905527$ for $Q=2.07869$. The amplitude \eqref{eq:amplitude_guess} matches precisely the one obtained numerically.

Finally, analogously to \eqref{eq:sum_rule_amplitude} one can write the following dispersion relation
\begin{equation}
\label{eq:dispersion_Lambda2}
\Lambda_2 =\f{m^4}{8\pi^2}
\int_{4m^2}^\infty\f{ds}{s^3}\ \text{Im}\mathcal{T}_{AA \rightarrow AA}(s,0,4m^2-s).
\end{equation}
Setting $\Lambda_2$ to zero will force the imaginary part of the $AA\rightarrow AA$ process to be zero since the integrand is non-zero due to the first inequality in the second line in \eqref{eq:conditions}. Due to the same inequality we see that the imaginary part in turn forces the whole amplitude to be zero. In practice we indeed observe that by setting $\Lambda_2$ to zero the numerical solution leads to all the coefficients $\alpha$ being zero. In other words our numerics leads to \eqref{eq:semi_free_solution} if $\Lambda_2=0$.

\subsubsection{Lower bound on the $a$-anomaly as a function of $\lambda_0$, $\lambda_2$, $\Lambda_0$ and $\Lambda_2$}
\label{sec:a_bounds_lambdas}

Let us now construct a lower bound on the $a$-anomaly as a function of the coupling constants $\lambda_0$, $\lambda_2$, $\Lambda_0$ and $\Lambda_2$. They were precisely defined in \eqref{eq:lambdas} and \eqref{eq:Lambdas}.

We begin by noticing that these couplings are bounded themselves. We can use our setup of section \ref{sec:setup} to obtain upper and lower bounds which read as
\begin{equation}
	\label{eq:lambdas_ranges}
	\begin{aligned}
		-6.0253\leq&\lambda_0\leq + 2.6613,\qquad
		0\leq \lambda_2\leq +2.2568,\\
		-2.8145\leq&\Lambda_0\leq + 2.8086,\qquad
		0\leq \Lambda_2\leq +0.6550.
	\end{aligned}
\end{equation}
These are obtained by setting $a=5a_\text{free}$ in the setup and using $N_{max}=20$ and $L_{max}=30$. Increasing the value of $a$ does not change the result. Around $a/a_\text{free}\sim 1$ we get a non-trivial dependence of the bounds on $a$ which will be better represented in the later plots.
The dependence of these bounds on $L_{max}$ is negligible. The dependence on $N_{max }$ is non-trivial and should be taken into account. The correct bounds are obtained using the extrapolation to $N_{max} \rightarrow \infty$. The exception is the upper bound on $\lambda_0$, it is independent of $N_{max}$ due to the presence of the singularity term \eqref{eq:singularity} which drastically improves the convergence of this particular bound. We do not perform $N_{max} \rightarrow \infty$ extrapolations in this section and instead we will always work at $N_{max}=20$ and $L_{max}=30$.

\begin{figure}[ht]
	\centering
	
	\includegraphics[width=0.45\textwidth]{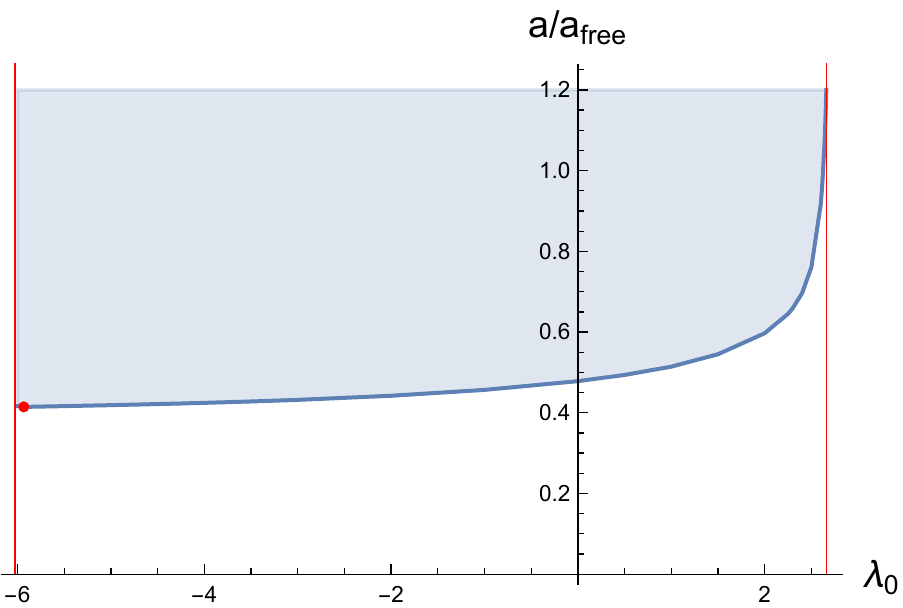}
	\includegraphics[width=0.45\textwidth]{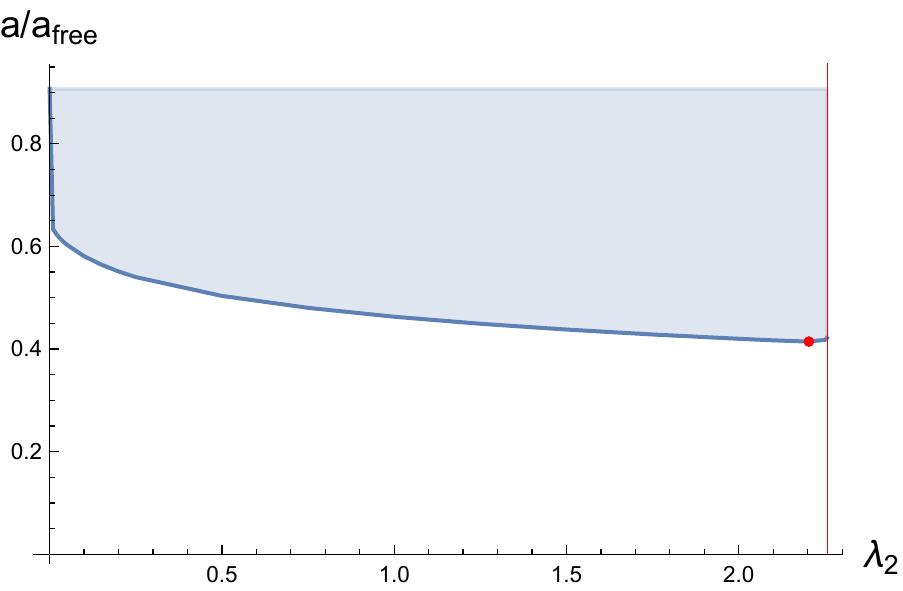}
	\caption{Lower bound on the $a$-anomaly as a function of $\lambda_0$ in the left plot and as a function of the $\lambda_2$ in the right plot. The allowed region is depicted in blue. Red dot represents the point with the lowest value of the $a$-anomaly. The red vertical lines indicate the boundaries of the allowed regions for $\lambda_0$ and $\lambda_2$ given in \eqref{eq:lambdas_ranges}. The plots are built with $N_{max}=20$ and $L_{max}=30$.
}
	\label{fig:plot_bounds_lambda_mainText}
	\vspace{8mm}
	
	\includegraphics[width=0.45\textwidth]{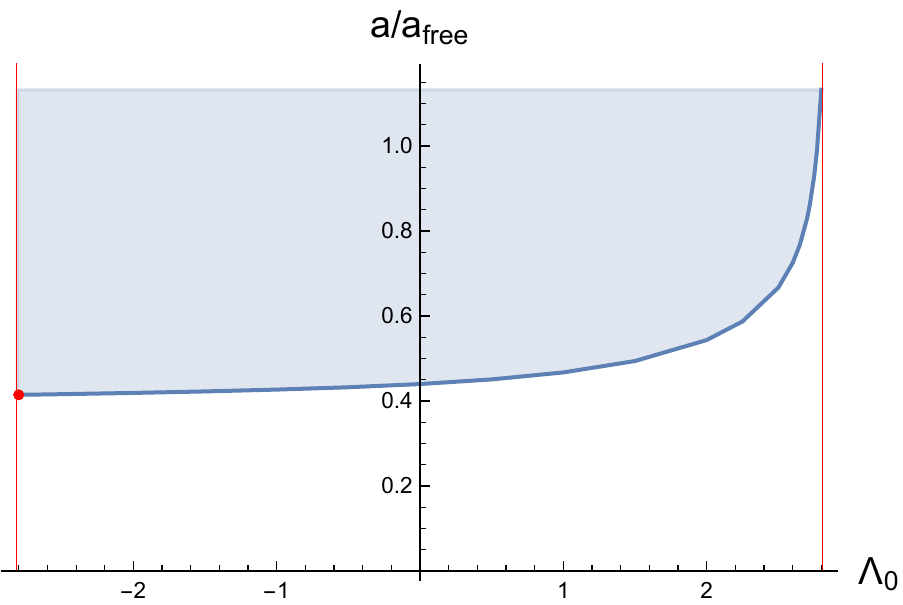}
	\includegraphics[width=0.45\textwidth]{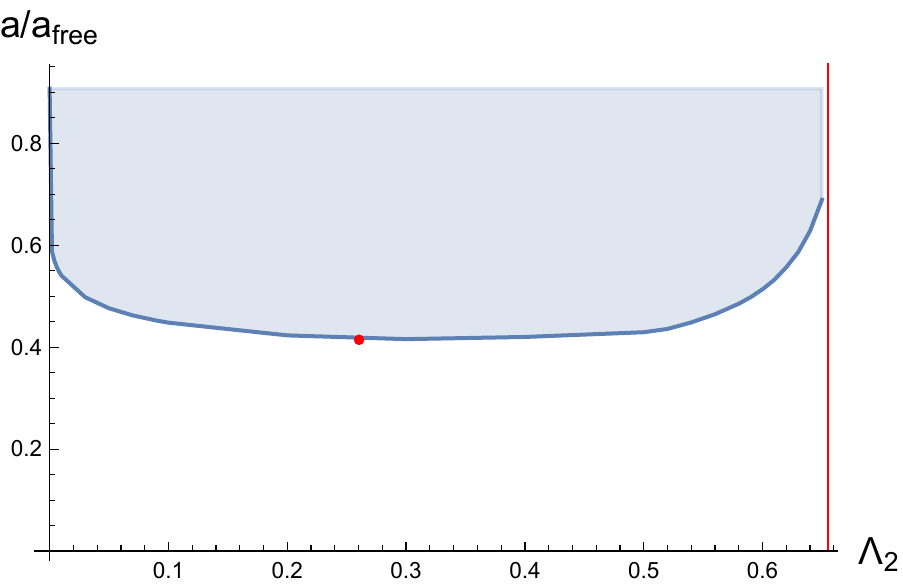}
	\caption{Lower bound on the $a$-anomaly as a function of $\Lambda_0$ in the left plot and as a function of the $\Lambda_2$ in the right plot. The allowed region is depicted in blue. Red dot represents the point with the lowest value of the $a$-anomaly. The red vertical lines indicate the boundaries of the allowed regions for $\Lambda_0$ and $\Lambda_2$ given in \eqref{eq:lambdas_ranges}.  The plots are built with $N_{max}=20$ and $L_{max}=30$.}
	\label{fig:plot_bounds_Lambda_mainText}
\end{figure}

We can now pick value of $\lambda_0$, $\lambda_2$, $\Lambda_0$ and $\Lambda_2$ from the allowed ranges \eqref{eq:lambdas_ranges} and minimize the $a$-anomaly. The result is presented in figures \ref{fig:plot_bounds_lambda_mainText} and \ref{fig:plot_bounds_Lambda_mainText}. The allowed area is shaded in blue. The red dot represents the solution with the absolute minimum of the $a$-anomaly found in section \ref{sec:absolute_minimum_a}. The right plots in figures \ref{fig:plot_bounds_lambda_mainText} and \ref{fig:plot_bounds_Lambda_mainText} have a sharp peak around $\lambda_2=0$ and $\Lambda_2=0$. This point corresponds to a freely propagating particle $A$ and gives the value of the $a$-anomaly quoted in \eqref{eq:semi_free_solution}. See the explanation below \eqref{eq:semi_free_solution} why this value is not exactly one.

\begin{figure}[ht]
	\centering
	
	\includegraphics[width=0.45\textwidth]{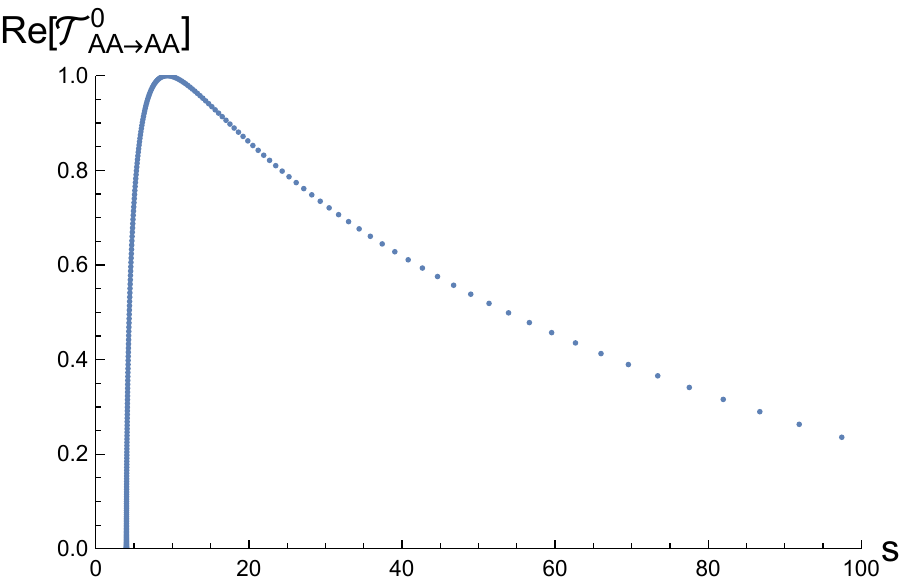}
	\includegraphics[width=0.45\textwidth]{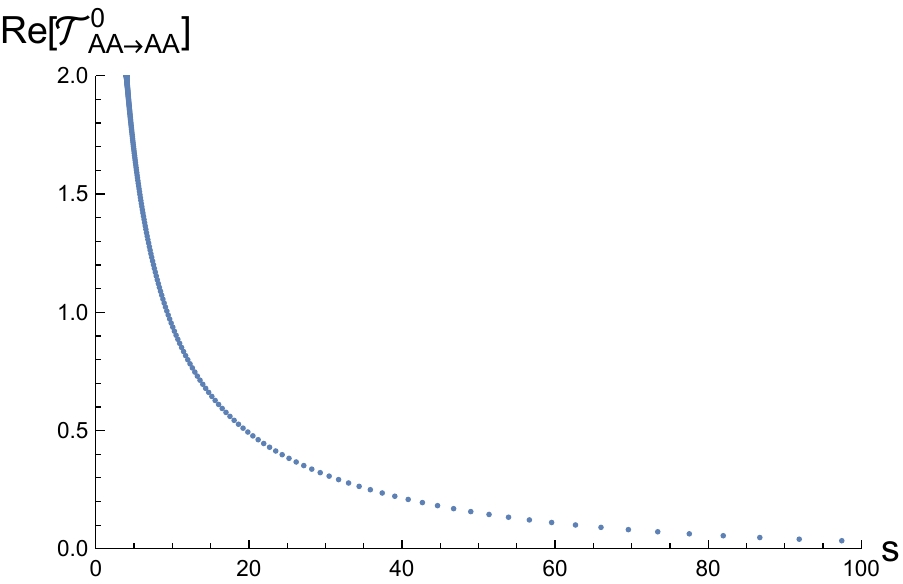}
	\caption{Real and imaginary parts of the spin 0 interacting part of the $AA\rightarrow AA$ partial amplitude for $\lambda_0=2.66$ which leads to $a/a_\text{free} = 1.2002$. It is constructed for $N_{max}=20$ and $L_{max}=30$.}
	\label{fig:Maxlambda0_1}
	\vspace{7mm}
	
	\includegraphics[width=0.45\textwidth]{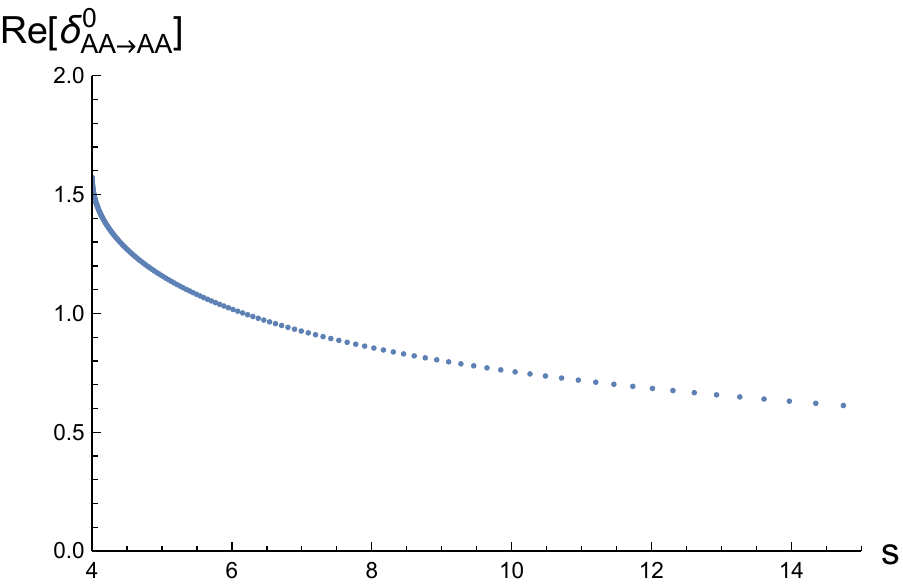}
	\includegraphics[width=0.45\textwidth]{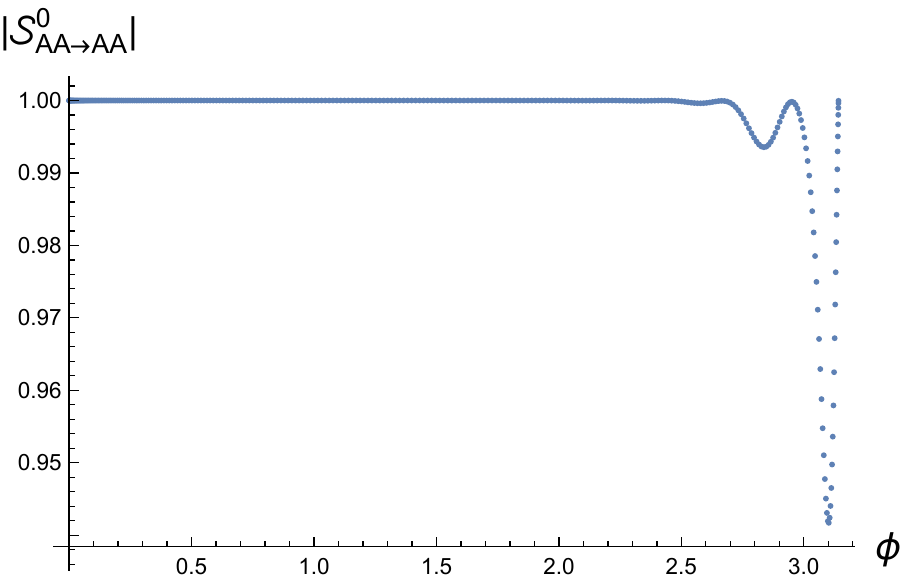}
	\caption{Left plot represents the real part of the spin 0 phase shift of the $AA\rightarrow AA$ scattering defined in \eqref{eq:phase_shift}. No resonances are present. Right plot represents the absolute value of the spin 0 partial amplitude of the $AA\rightarrow AA$ scattering. On the real axis instead of $s$ variable we use $\phi$ variable defined in \eqref{eq:phi_variable}. The amplitude is fully ``elastic'' up to very high energies.}
	\label{fig:Maxlambda0_2}
	\vspace{7mm}
	
	\includegraphics[width=0.45\textwidth]{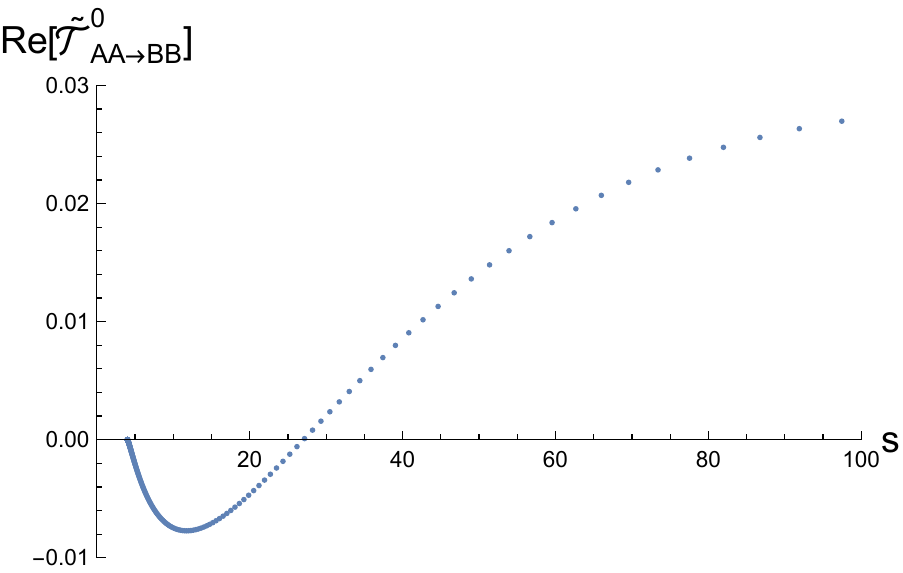}
	\includegraphics[width=0.45\textwidth]{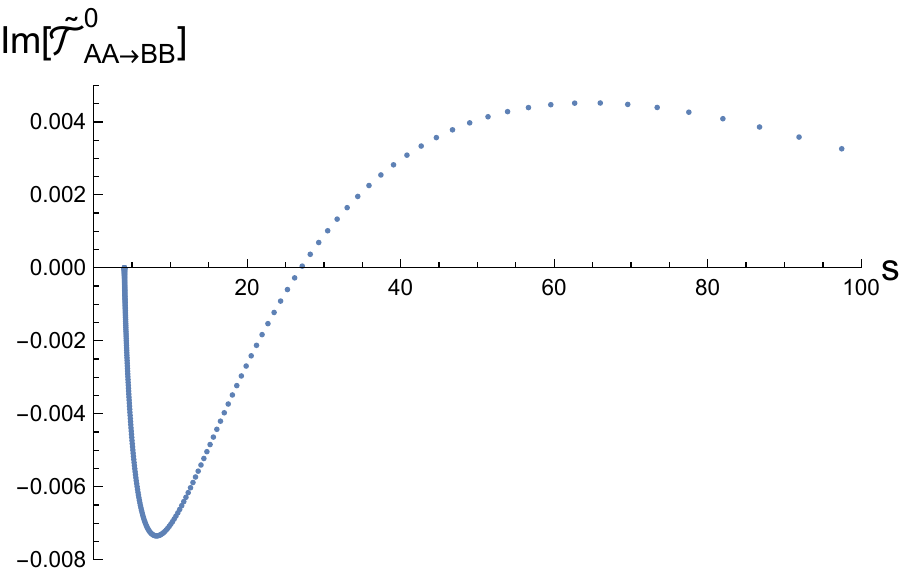}
	\caption{Real and imaginary parts of the spin 0 interacting part of the $AA\rightarrow BB$ partial amplitude for $\lambda_0=2.66$ which leads to $a/a_\text{free} = 1.2002$. It is constructed for $N_{max}=20$ and $L_{max}=30$.}
	\label{fig:Maxlambda0_3}
\end{figure}

\begin{figure}[ht]
	\centering
	\includegraphics[width=0.42\textwidth]{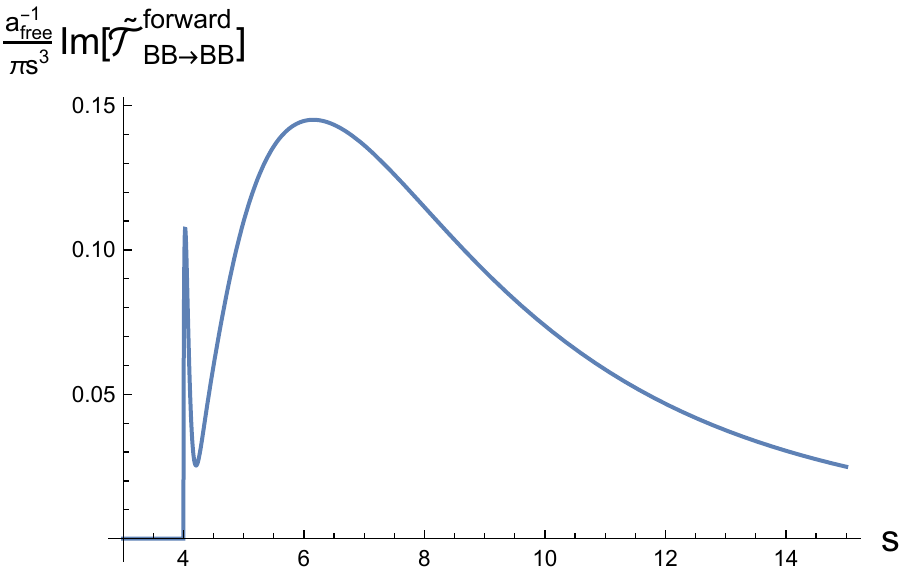}
	\includegraphics[width=0.49\textwidth]{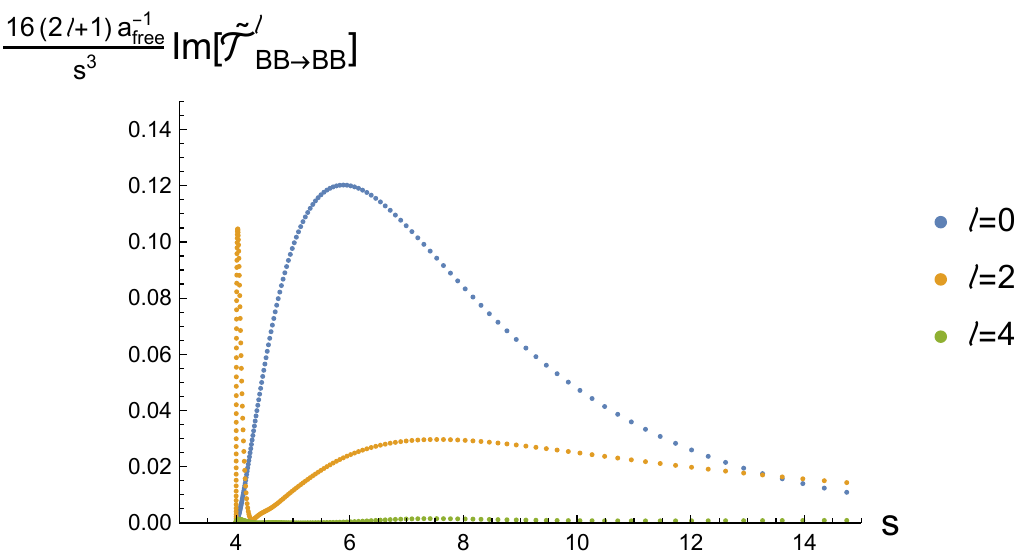}
	\caption{
		Left plot: integrand appearing in the sum-rule \eqref{eq:sum_rule_amplitude}  for the absolute minimum of the $a$-anomaly. Numerical integration of this function leads to $a/a_\text{free}=1.2002$. Right plot: integrand appearing in the sum-rule \eqref{eq:sum_rule_pamplitude}  for the absolute minimum of the $a$-anomaly. Numerical integration of this function leads to $a/a_\text{free}=0.71146 + 0.45094 + 0.03238 + \ldots = 1.19477 + \ldots$, where the three entries correspond to spin 0, 2 and 4 respectively and the dots indicate higher spin contributions. The plots are constructed at $N_{max}=20$ and $L_{max}=30$.
	}
	\label{fig:Maxlambda0_4}
\end{figure}

Another interesting point in these plots is the one with $\lambda_0\approx 2.66$ and $a/a_\text{free}=1.2$. As was done in section \ref{sec:absolute_minimum_a} we can plot spin zero partial amplitudes in our setup for this solution. They are given in figures \ref{fig:Maxlambda0_1} - \ref{fig:Maxlambda0_4}.

\section{Discussion}
\label{sec:conc}

We proposed a framework to study gapped QFTs defined as relevant deformations of an UV CFT. For simplicity we assumed the existence of a $\ZZ$ symmetry and a single stable $\ZZ$-odd particle 
 at low energies denoted by $A$. By studying the unitarity constraints on the system of scattering amplitudes between the particle $A$ and the dilaton particle $B$, as depicted  in figure \ref{fig:set_amplitudes}, we related the physical scattering data for the particle $A$ to the $a$-anomaly   of the UV CFT.
This allowed us to derive lower bounds on the $a$-anomaly as a function of several parameters defined in \eqref{eq:lambdas} and \eqref{eq:Lambdas} in terms of the amplitude $\mathcal{T}_{AA\to AA}$ (see figures \ref{fig:plot_bounds_lambda} and \ref{fig:plot_bounds_Lambda}).

In practice, we used the primal numerical S-matrix bootstrap method for multiple amplitudes.\footnote{For an example of the primal numerical setup with multiple amplitudes in 2d see \cite{Homrich:2019cbt}.} This method requires extrapolation in the number of parameters of the ansatz ($N_{max} \to \infty$) and the number of unitarity constraints ($L_{max}\to \infty$).
As usual in S-matrix bootstrap problems with a mass gap, the extrapolation $L_{max}\to \infty$ is easy. On the other hand, the extrapolation $N_{max} \to \infty$ is more subtle and deserves further study. In this context, it would be extremely useful to have a dual S-matrix bootstrap method to approach the minimal $a$-anomaly from below. It should be possible to generalize the methods of \cite{Lopez:1976zs, He:2021eqn, Guerrieri:2021tak} to our setup.

Our main result is the derivation of a universal lower bound on the $a$-anomaly. We estimated the bound to be $a/a_\text{free} > 0.3 $.
Combining this result with the conformal collider bound $\frac{31}{18}\ge \frac{a}{c} \ge \frac{1}{3}$ from \cite{Hofman:2008ar}, we obtain a lower bound $c > 0.17\  a_\text{free}$.

It would be interesting to generalize our analysis by dropping some of the simplifying assumptions about the presence of the $\ZZ$ symmetry and the low energy spectrum of particles. For example, we can start by allowing more stable particles either even or odd under  $\ZZ$. More stable particles lead to extra poles in the ansatz \eqref{eq:ansatze} and change the branch point in $\myRho_2$ associated with the beginning of the continuum in the $\ZZ$ even sector (in practice $9m^2 \to (m+m')^2$ where $m'$ is the mass of the lightest $\ZZ$ even particle). Dropping the assumption of $\ZZ$ symmetry leads to similar effects: now particle $A$ gives rise to a pole in every channel of every amplitude and all cuts start at $4m^2$. More ambitiously, we could include in our setup also the processes $\mathcal{T}_{AA\to AB}$ and $\mathcal{T}_{AB\to BB}$. This would give rise to a stronger set of unitarity constraints.
Any of these more general scenarios must give rise to a weaker lower bound than $a/a_\text{free} > 0.3 $.\footnote{It may be possible to derive a stronger lower bound for QFTs with more than one stable particle by considering  $2 \to 2$ scattering amplitudes with other external particles besides the dilaton $B$ and the lightest particle $A$. This is similar to the discussion below about promoting particle $A$ to a vector multiplet of $O(3)$.} This is intriguing because we do not know of any 4d QFT with $a<a_\text{free}$. On the one hand, one may interpret our results as pointing to the existence of an unknown QFT with $a<a_\text{free}$. On the other hand, it is also possible that our setup is just not constraining enough to approach the true minimal value of $a$ in the space of QFTs.

Another interesting generalization would be to study QFTs with a continuous global symmetry. 
In practice, this is achieved by promoting particle $A$ to an irreducible representation of the global symmetry group. For example, we could study pions as a triplet of $O(3)$.
 Building on \cite{Guerrieri:2018uew}, we could impose the known value of the $a$-anomaly in QCD and study several observables like scattering lengths, chiral zeros and resonances.
 Alternatively, we could inject some experimental data about $\pi-\pi$ scattering and then minimize the $a$-anomaly. 
 
It would also be interesting to apply our approach to QFTs with fermionic asymptotic states. For example, we can ask the question: what is the lowest $a_{UV}$ that can give rise to massive   fermions in the IR?\footnote{We thank H. Osborn for suggesting this question.}
 In practice, this would require a combination of the techniques of \cite{Hebbar:2020ukp} with the present paper.
 
Our setup can easily accommodate a massless physical particle described by some Effective Field Theory (EFT) at low energies.
This would have several interesting applications. The simplest one is for particle $A$ to be a Goldstone boson of spontaneous symmetry breaking. For example, particle $A$ could describe massless pions like in \cite{Guerrieri:2020bto}.
More challenging, would be to take particle $A$ to be a photon. This would allows to ask the question: what is the minimum $a$-anomaly of the UV CFT that can give rise to a photon in the IR?
Our method creates a non-perturbative bridge between low-energy  EFT and their UV completions (within the realm of QFT). Unfortunately, such applications will be very challenging numerically for the current methods. The main challenge is the extrapolation $L_{max} \to \infty$ in gapless theories (see for instance \cite{Guerrieri:2021ivu}). 
 
In this paper, the dilaton particle $B$ was an external probe. However, one can also study physical dilatons, which are the Goldstone bosons from spontaneous conformal symmetry breaking. In this case, the dilaton decay constant $f$ would be finite and generically of order $m$. For this reason, dilatons can now appear as internal massless particles in the amplitudes. Such non-perturbative setup would be more challenging numerically.

 So far, we restricted ourselves to 4 spacetime dimensions.
 However, our setup can be easily adapted to 6d. The main difference with 4d is the low energy behavior of the dilaton scattering amplitude which reads in 6d as \cite{Elvang:2012st}
\begin{equation}
 \widetilde{\mathcal{T}}_{BB\rightarrow BB} (s,t,u) = 
\frac{b}{8} \times (s^2+t^2+u^2) +
\frac{9}{8}(a^\text{UV}-a^\text{IR}) \times stu + O(s^4).
\end{equation}
Here $b$ is some parameter of the theory.  It is an outstanding open question to prove the $a$-theorem in 6d.
The paper \cite{Kundu:2019zsl} made an interesting proposal: consider the RG flow on a fixed Anti-de Sitter  background and formulate the $a$-theorem as a conformal bootstrap problem for its boundary correlation functions. Unfortunately, this approach has not yet succeeded. 
In flat space,  there seems to be no positive sum rule for $ (a^\text{UV}-a^\text{IR})$ in 6d  \cite{Elvang:2012st}. 
Nevertheless, we can try to run our numerical approach and minimize $ a^\text{UV}$ in QFTs with a mass gap. We leave this exploration for the future.

\section*{Acknowledgements}
The authors thank Hongbin Chen, Liam Fitzpatrick,  Andrea Guerrieri, Brian Henning,  Riccardo Rattazzi, Sébastien Reymond, Matt Walters and Pedro Vieira for useful discussions.
We also thank Zohar Komargodski for   comments on the draft.

The work of DK is supported by the SNSF Ambizione grant PZ00P2\_193411.
The work of JM, BS and JP is supported by the Simons Foundation grant 488649 (Simons Collaboration on the Nonperturbative Bootstrap) and by the Swiss National Science Foundation through the project 
200020\_197160  
and through the National Centre of Competence in Research SwissMAP.  

\appendix

\section{Correlation functions of the stress tensor}
\label{app:tensor_structures}
In this appendix we summarize results for the two- and three-point correlation functions of the stress-tensor in four spacetime dimension derived in a seminal paper \cite{Osborn:1993cr} by Osborn and Petkou. Let us start by defining the short-hand notation
\begin{equation}
x_{ij}^\mu \equiv x_i^\mu -x_j^\nu.
\end{equation}
All the tensor structures for the stress-tensor correlators are built out of these elementary tensors
\begin{equation}
	\begin{aligned}
		I^{\mu\nu}(x) &\equiv \delta^{\mu\nu} - 2\,\frac{x^\mu x^\nu}{x^2},\\
		\mathcal{I}^{\mu\nu,\rho\sigma}(x)\ &\equiv \f{1}{2}I^{\mu\rho}(x)I^{\nu\sigma}(x)+\f{1}{2}I^{\mu\sigma}(x)I^{\nu\rho}(x)-\f{1}{4}\eta^{\mu\nu}\eta^{\rho\sigma},\\
		X_{3,12}^{\mu}&\equiv \f{x_{13}^{\mu}}{x_{13}^2}-\f{x_{23}^{\mu}}{x_{23}^2}.
	\end{aligned}
\end{equation}
According to \cite{Osborn:1993cr} we have
\begin{align}
	\< T^{\mu\nu}(x_1)T^{\rho\sigma}(x_2)\> &=C_{T}\, \f{1}{x_{12}^8}\ \mathbf{T}^{\mu\nu;\rho\sigma}_0,\\
	\< T^{\mu\nu}(x_1)T^{\rho\sigma}(x_2)T^{\alpha\beta}(x_3)\> &=  \f{1}{x_{12}^{4}x_{23}^{4}x_{31}^{4}}\Big[\mathbb{A}\ \mathbf{T}^{\mu\nu;\rho\sigma;\alpha\beta}_1+\mathbb{B}\ \mathbf{T}^{\mu\nu;\rho\sigma;\alpha\beta}_2 +\mathbb{C}\ \mathbf{T}^{\mu\nu;\rho\sigma;\alpha\beta}_3\Big].
\end{align}
Here $C_{T}$ is the central charge and $\mathbb{A}$, $\mathbb{B}$ and $\mathbb{C}$ are the stress-tensor OPE coefficients with itself. These obey the following relation
\be
C_{T}\ =\ \f{\pi^2}{3}\Big(14\mathbb{A}-2\mathbb{B}-5\mathbb{C}\Big).
\ee
The four tensor structures written above are defined as
\begin{equation}
\begin{aligned}
\mathbf{T}^{\mu\nu;\rho\sigma}_0 &\equiv
\mathcal{I}_{\mu\nu,\rho\sigma}(x_{12}),\\
\mathbf{T}^{\mu\nu;\rho\sigma;\alpha\beta}_1 &\equiv \mathcal{I}^{\mu\nu,\mu'\nu'}(x_{13}) \mathcal{I}^{\rho\sigma,\rho'\sigma'}(x_{23}) t^1_{\mu'\nu',\rho'\sigma'}\ ^{
\alpha\beta}(X_{3,12}), \\
\mathbf{T}^{\mu\nu;\rho\sigma;\alpha\beta}_2 &\equiv \mathcal{I}^{\mu\nu,\mu'\nu'}(x_{13}) \mathcal{I}^{\rho\sigma,\rho'\sigma'}(x_{23}) t^2_{\mu'\nu',\rho'\sigma'}\ ^{
\alpha\beta}(X_{3,12}), \\
\mathbf{T}^{\mu\nu;\rho\sigma;\alpha\beta}_3 &\equiv \mathcal{I}^{\mu\nu,\mu'\nu'}(x_{13}) \mathcal{I}^{\rho\sigma,\rho'\sigma'}(x_{23}) t^3_{\mu'\nu',\rho'\sigma'}\ ^{
\alpha\beta}(X_{3,12}),
\end{aligned}
\end{equation}
where
\be
 t^1_{\mu\nu,\rho\sigma,\alpha\beta}(X)& \equiv &
 	h^5_{\mu\nu,\rho\sigma,\alpha\beta}(X)
  - 2h^4_{\mu\nu,\rho\sigma,\alpha\beta}(X)- 2h^4_{\rho\sigma,\mu\nu ,\alpha\beta}(X)
  + 24 h^2_{\mu\nu,\rho\sigma}(X) h^1_{\alpha\beta}(X)\nn\\
  &&- 16  h^1_{\rho\sigma}(X) h^2_{\mu\nu,\alpha\beta}(X)- 16  h^1_{\mu\nu}(X) h^2_{\rho\sigma,\alpha\beta}(X)
  + 64 h^1_{\mu\nu}(X) h^1_{\rho\sigma}(X) h^1_{\alpha\beta}(X),\nn\\
  t^2_{\mu\nu,\rho\sigma,\alpha\beta}(X) &\equiv& 
  h^4_{\alpha\beta,\mu\nu,\rho\sigma}(X) 
  - h^4_{\mu\nu,\rho\sigma,\alpha\beta}(X) - h^4_{\rho\sigma,\mu\nu ,\alpha\beta}(X)+
  6 h^2_{\mu\nu,\rho\sigma}(X) h^1_{\alpha\beta}(X)\nn  \\
 && - 2 h^1_{\rho\sigma}(X) h^2_{\mu\nu,\alpha\beta}(X)- 2 h^1_{\mu\nu}(X) h^2_{\rho\sigma ,\alpha\beta}(X)
  + 32 h^1_{\mu\nu}(X) h^1_{\rho\sigma}(X) h^1_{\alpha\beta}(X),\nn\\
  t^3_{\mu\nu,\rho\sigma,\alpha\beta}(X) &\equiv& 
  	h^3_{\mu\nu, \rho\sigma}(X) h^1_{\alpha\beta}(X) + 
  	 h^1_{\mu\nu}(X) h^3_{\rho\sigma,\alpha\beta}(X)+h^1_{\rho\sigma}(X) h^3_{\mu\nu,\alpha\beta}(X)\nn\\
  	 && -
  	6 h^2_{\mu\nu,\rho\sigma}(X) h^1_{\alpha\beta}(X)
  	+4 h^1_{\rho\sigma}(X) h^2_{\mu\nu,\alpha\beta}(X)+4 h^1_{\mu\nu}(X) h^2_{\rho\sigma,\alpha\beta}(X)\nn\\
  	&& -
  	16 h^1_{\mu\nu}(X) h^1_{\rho\sigma}(X) h^1_{\alpha\beta}(X).
\ee
and
\be
h^1_{\mu\nu}(X) &\equiv& \f{X_\mu X_\nu}{X^2} - \frac{1}{4} \eta_{\mu\nu},\nn\\
h^2_{\mu\nu ,\rho\sigma}(X)&\equiv&\f{1}{X^2}\Big[X_\mu X_\rho \eta_{\nu\sigma}+X_{\mu}X_{\sigma}\eta_{\nu\rho}+X_{\nu}X_{\rho}\eta_{\mu\sigma}+X_{\nu}X_{\sigma}\eta_{\mu\rho}-X_\mu X_\nu \eta_{\rho\sigma}\nn\\
&&\ -X_\rho X_\sigma \eta_{\mu\nu}\Big]+\f{1}{4}\eta_{\mu\nu}\eta_{\rho\sigma},\nn\\
h^3_{\mu\nu ,\rho\sigma}(X)&\equiv&\ \eta_{\mu\rho}\eta_{\nu\sigma}+\eta_{\mu\sigma}\eta_{\nu\rho}-\f{1}{2}\eta_{\mu\nu}\eta_{\rho\sigma},\nn\\
h^4_{\mu\nu ,\rho\sigma,\alpha\beta}(X)&\equiv&\f{1}{X^2}\Big[h^3_{\mu\nu,\rho\alpha}X_\sigma X_\beta +h^3_{\mu\nu,\sigma\alpha}X_\rho X_\beta +h^3_{\mu\nu,\rho\beta}X_\sigma X_\alpha +h^3_{\mu\nu,\sigma\beta}X_\rho X_\alpha \Big]\nn\\
&&-\f{1}{2}\eta_{\rho\sigma}h^2_{\mu\nu ,\alpha\beta}(X)-\f{1}{2}\eta_{\alpha\beta}h^2_{\mu\nu ,\rho\sigma}(X)-\f{1}{2}\eta_{\rho\sigma}\eta_{\alpha\beta}h^1_{\mu\nu}(X),\nn\\
h^5_{\mu\nu ,\rho\sigma,\alpha\beta}(X)&\equiv& \ \eta_{\mu\rho}\eta_{\nu\alpha}\eta_{\sigma\beta}+\eta_{\nu\rho}\eta_{\mu\alpha}\eta_{\sigma\beta}+\eta_{\mu\sigma}\eta_{\nu\alpha}\eta_{\rho\beta}+\eta_{\nu\sigma}\eta_{\mu\alpha}\eta_{\rho\beta}+ \eta_{\mu\rho}\eta_{\nu\beta}\eta_{\sigma\alpha}\nn\\
&&+\eta_{\nu\rho}\eta_{\mu\beta}\eta_{\sigma\alpha}+\eta_{\mu\sigma}\eta_{\nu\beta}\eta_{\rho\alpha}+\eta_{\nu\sigma}\eta_{\mu\beta}\eta_{\rho\alpha}-\f{1}{2}\eta_{\mu\nu}\eta_{\rho\sigma}\eta_{\alpha\beta}\nn\\
&&-\eta_{\mu\nu}h^3_{\rho\sigma ,\alpha\beta}(X)-\eta_{\rho\sigma}h^3_{\mu\nu ,\alpha\beta}(X)-\eta_{\alpha\beta}h^3_{\mu\nu ,\rho\sigma}(X).
\ee

In curved spacetime one-point function of the trace trace of the stress-tensor is non-zero. It has the following form
\be
\<T^\mu_{\mu}\>_{g} \ &=&\  -a\times\ E_{4}\ +\times\  c\ W^2,
\ee
where $E_{4}$ is the Eular density and $W^2$ is the square of the Weyl tensor. They have the following expressions in $4d$,
\be
\label{eq:euler}
E_4\ &=&\ R^{\alpha\beta\gamma\delta}R_{\alpha\beta\gamma\delta}-4R^{\alpha\beta}R_{\alpha\beta}+R^2,\\
\label{eq:weyl}
W^2\ &=&\ R^{\alpha\beta\gamma\delta}R_{\alpha\beta\gamma\delta}-2R^{\alpha\beta}R_{\alpha\beta}+\f{1}{3}R^2.
\ee
The coefficients $a$ and $c$ are called the Weyl anomalies. They are related to the OPE coefficients $\mathbb{A}$, $\mathbb B$ and $\mathbb C$ as
\be
a\ =\ \f{\pi^4}{64\times 90}\Big(9\mathbb{A}-2\mathbb{B}-10\mathbb{C}\Big),\quad
c\ =\ \f{\pi^4}{64\times 30}\Big(14\mathbb{A}-2\mathbb{B}-5\mathbb{C}\Big).
\ee
As an example let us consider a theory of a free massless (conformally coupled) scalar $\Phi(x)$ which is described by the free CFT. According to \cite{CALLAN197042} it has the following stress-tensor
\be
T_{\mu\nu}=\p_{\mu}\Phi\p_\nu \Phi -\f{1}{12}\Big[2\p_\mu \p_\nu \Phi^2 +\eta_{\mu\nu}\p^2 \Phi^2\Big].
\ee
Such a CFT has the following parameters
\begin{equation}
\mathbb{A}=\f{1}{27\pi^6},\quad
\mathbb{B}=-\f{4}{27\pi^6},\quad
\mathbb{C}=-\f{1}{27\pi^6},\quad
C_{T}=\f{1}{3\pi^4},
\end{equation}
\begin{equation}
	\label{eq:ac_free_scalar}
a =\f{1}{5760\pi^2},\quad
c=\f{1}{1920\pi^2}.
\end{equation}

\section{Example of the free scalar theory}
\label{sec:example}
In this appendix we consider the theory of a free massive scalar field $\Phi(x)$ which has the following action
\begin{equation}
\label{QFTaction_free}
A_\text{free}(m)=\ \int d^{4}x\ \Big[-\f{1}{2}\p_{\mu}\Phi\p^{\mu}\Phi -\f{1}{2}m^2\Phi^2\Big].
\end{equation}
It can be interpreted as a free massless CFT in the UV deformed by the mass term.
As reviewed in sections \ref{KS_setup} and \ref{S:KS_a_theorem_review} one can define the following modified action
\begin{equation}
	\label{QFTaction}
	A'_\text{free}(m)=\int d^{4}x \Big[-\f{1}{2}\p_{\mu}\Phi\p^{\mu}\Phi -\f{1}{2}m^2\Phi^2-\f{1}{2}  \p_{\mu}\varphi \p^{\mu}\varphi+\f{m^2}{\sqrt{2}f}\ \varphi \Phi^2 - \f{m^2}{4f^2}\varphi^2\Phi^2\Big].
\end{equation}
In what follows using this action we will compute the $BB\rightarrow BB$ scattering amplitude, where $B$ is the dilaton particle created by the dilaton field $\varphi(x)$ from the vacuum. We will show that in this particular model this amplitude at low energy is given by equation \eqref{eq:dilaton_example}. The particle created by the field $\Phi(x)$ from the vacuum is referred to as the particle $A$. We will do the computation in two different ways.

All the Feynman rules needed for the computation of scattering amplitudes in the model \eqref{QFTaction} read as
\begin{align}
	\feynmandiagram[horizontal=a to b, inline=(a.base)] {
		a -- b
	};\  &= \frac{-i}{p^2 + m^2 -i\epsilon}\\
	\feynmandiagram[horizontal=a to b, inline=(a.base)] {
		a --[dashed] b
	}; \ &= \frac{-i}{ p^2 -i\epsilon}\\
	\begin{tikzpicture}[baseline=(a)]
		\begin{feynman}
			\vertex (a);
			\vertex [above left = of a] (i1);
			\vertex [below left = of a] (i2);
			\vertex [right = of a] (o1);
			\diagram* {
				(i1)--(a)--(i2), 
				(o1)--[dashed](a),
			};
		\end{feynman}
	\end{tikzpicture}\  &= \frac{i\sqrt{2}m^2}{f}\\
	\begin{tikzpicture}[baseline=(a)]
		\begin{feynman}
			\vertex (a);
			\vertex [above left = of a] (i1);
			\vertex [below left = of a] (i2);
			\vertex [above right = of a] (o1);
			\vertex [below right = of a] (o2);
			\diagram* {
				(i1)--(a)--(i2), 
				(o1)--[dashed](a)--[dashed](o2),
			};
		\end{feynman}
	\end{tikzpicture} &= - \f{im^2}{f^2}
\end{align} 
Here solid lines represent the field $\Phi(x)$ and dashed lines represent the dilaton field $\varphi(x)$.

\begin{figure}[h!]
\centering
\tikzfeynmanset{momentum/arrow distance=1mm}
\tikzfeynmanset{momentum/label distance=-0.15em}
\begin{tikzpicture}[baseline=(a.base)]
	\begin{feynman}
		\vertex (a);
		\vertex [right=1.5 of a] (b);
		\vertex [below left=of a] (i1);
		\vertex [above left=of a] (i2);
		\vertex [below right=of b] (o1);
		\vertex [above right=of b] (o2);
		\diagram* {
			(a) --[half left, momentum={$\ell$}] (b) -- [half left, rmomentum=$k_1+k_2-\ell$] (a),
			(i1) --[dashed, momentum={$k_2$}] (a) --[dashed, rmomentum'={$k_1$}] (i2),
			(o1) --[dashed, rmomentum'={$k'_2$}] (b) --[dashed, momentum={$k'_1$}] (o2),
		};
	\end{feynman}
\end{tikzpicture}
\qquad
\begin{tikzpicture}[baseline=(a.base)]
	\begin{feynman}
		\vertex (a);
		\vertex [below right=1 and 2 of a] (b);
		\vertex [above right=1 and 2 of a] (c);
		\vertex [below left=of a] (i1);
		\vertex [above left=of a] (i2);
		\vertex [right=of b] (o1);
		\vertex [right=of c] (o2);
		\diagram* {
			(a) --[momentum'={[label distance=0.5em, style={inner xsep= -1.3em}]$k_1+k_2-\ell$}] (b) --[rmomentum'=$\ell-k_1'$] (c) --[rmomentum={$\ell$}] (a),
			(i1) --[dashed, momentum={$k_2$}] (a) --[dashed, rmomentum'={$k_1$}] (i2),
			(o1) --[dashed, rmomentum={$k'_2$}] (b), (c) --[dashed, momentum=$k_1'$] (o2),
		};
	\end{feynman}
\end{tikzpicture}
\qquad
\begin{tikzpicture}[baseline=($(a.base)!0.5!(b.base)$)]
	\begin{feynman}
		\vertex (a);
		\vertex [below=of a] (b);
		\vertex [right=of b] (c);
		\vertex [above=of c] (d);
		\vertex [left=of a] (i1);
		\vertex [left=of b] (i2);
		\vertex [right=of c] (o1);
		\vertex [right=of d] (o2);
		\diagram* {
			(a) -- (b) -- (c) -- (d) -- (a),
			(i1) --[dashed, momentum=$k_1$] (a),
			(i2) --[dashed, momentum'=$k_2$] (b),
			(c) --[dashed, momentum'=$k'_2$] (o1),
			(d) --[dashed, momentum=$k'_1$] (o2),
		};
	\end{feynman}
\end{tikzpicture}
	\caption{We consider the momenta of incoming dilatons be $k_1$ and $k_2$ and the momenta of outgoing dilatons be $k'_1$ and $k'_2$. In the total amplitude contribution we also need to add the contributions of two topologically in-equivalent diagrams for the above individuals which we can easily read off using crossing symmetry.}\label{1loop}
\end{figure}
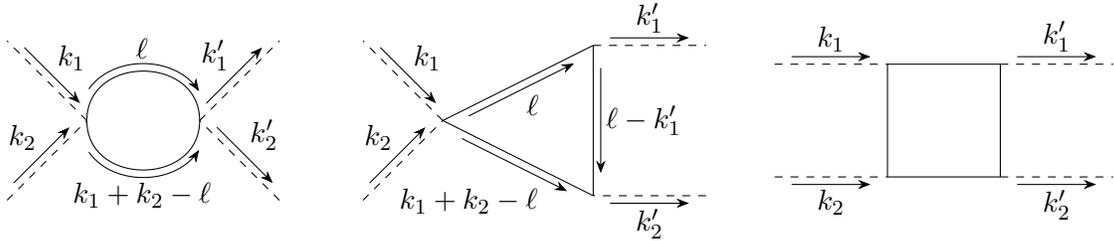

\paragraph{Direct computation}
The $BB\rightarrow BB$ dilaton scattering at the order $\mathcal{O}(f^{-4})$ is described by the Feynman diagram depicted in figure \ref{1loop}. We compute these diagrams one by one using the standard Feynman parametrization. We will then expand these expression at the leading order in energy and the perform the Feynman integrals.

The amplitude described by the first diagram in figure \ref{1loop} together with the ones obtained from it by using crossing symmetry has the following form  
\be
\label{AI}
&&i\mathcal{A}_{I}(s,t)\\
 &=&\ \f{m^4}{2f^4}\int \f{d^4 \ell}{(2\pi)^4}\f{1}{\ell^2+m^2-i\epsilon}\ \f{1}{(k_1+k_2-\ell)^2+m^2-i\epsilon} + \text{(cross-sym)}\nn\\
&=& -\f{im^4}{32\pi^2 f^4}\int_{0}^{1}dx\ \Bigg[\ln\Bigg(\f{m^2-sx(1-x)}{\Lambda^2}\Bigg)+\ln\Bigg(\f{m^2-tx(1-x)}{\Lambda^2}\Bigg)+\ln\Bigg(\f{m^2-ux(1-x)}{\Lambda^2}\Bigg)\Bigg]\nn. 
\ee
Here $\Lambda$ is the Pauli-Villars regularisation parameter representing the $UV$ cut-off. Contribution from the second diagram in figure \ref{1loop} together the ones obtained from it by corssing symmetry has the form
\be
 \label{AII}
&&i\mathcal{A}_{II}(s,t)\\
 &=&\ -\f{2m^6}{f^4}\int\f{d^4\ell}{(2\pi)^4}\f{1}{\ell^2+m^2-i\epsilon}\f{1}{(\ell-k'_1)^2+m^2-i\epsilon}\f{1}{(k_1+k_2-\ell)^2+m^2-i\epsilon}+\text{(cross-sym)}\nn\\
&=&\ -\f{im^6}{4\pi^2f^4}\int_{0}^{1}dx\ dy\ dz\ \delta(x+y+z-1)\Bigg[\f{1}{m^2-sxy}+\f{1}{m^2-txy}+\f{1}{m^2-uxy}\Bigg].\nn
\ee
Contribution from the third diagram in figure \ref{1loop} together the ones obtained from it by corssing symmetry has the form
\be
\label{AIII}
&&i\mathcal{A}_{III}(s,t)\\
&=&+\f{im^8}{4\pi^2f^4}\int_{0}^1 dxdydzdw\ \delta(x+y+z+w-1)\ \Bigg[\f{1}{\big[m^2-\lbrace sy(z+w)+tyz+uz(1-z-w)\rbrace\big]^2}\nn\\
&&\ + \f{1}{\big[m^2-\lbrace ty(z+w)+uyz+sz(1-z-w)\rbrace\big]^2}+\f{1}{\big[m^2-\lbrace uy(z+w)+syz+tz(1-z-w)\rbrace\big]^2}\Bigg].\nn
\ee
Above the Mandelstam variables are defined as $s=-(k_1 +k_2)^2\ ,\ t=-(k_1-k'_1)^2 ,\ u=-(k_1-k'_2)^2$ with $s+t+u=0$. Summing all the above contributions we get the four dilaton scattering amplitude
\be
\mathcal{T}_{BB\rightarrow BB}(s,t,u)\ &=&\ \mathcal{A}_{I}(s,t)+\mathcal{A}_{II}(s,t)+\mathcal{A}_{III}(s,t)+O(f^{-5})\label{A}
\ee
Let us for simplicity work in the forward limit and focus on low energies when $s<< m^2$. Up to the order $O(s^2)$ we obtain
\be
\mathcal{T}_{BB\rightarrow BB}(s,0,-s)\ &=& \f{3m^4}{16\pi^2 f^4}\ln\Big(\f{\Lambda}{m}\Big)+\f{s^2}{32\pi^2 f^4}\int_{0}^1 dx \ x^2(1-x)^2\ -\ \f{s^2}{4\pi^2 f^4}\int_{0}^1 dx\int_{0}^{1-x}dy\ 2x^2y^2\nn\\
&&\ +\f{3s^2}{4\pi^2f^4}\int_{0}^1 dy\int_0^{1-y}dz\int_{0}^{1-y-z}dw\ \Big[(yz+yw-z+z^2+zw)^2\nn\\
&&\ +(-yz+z-z^2-wz)^2+(yw)^2\Big]+O(s^3)\nn\\
&=&\ \f{3m^4}{16\pi^2 f^4}\ln\Big(\f{\Lambda}{m}\Big)+\f{s^2}{960\pi^2 f^4}-\f{s^2}{360\pi^2 f^4}+\f{s^2}{480\pi^2 f^4}+O(s^3)\nn\\
&=&\f{3m^4}{16\pi^2 f^4}\ln\Big(\f{\Lambda}{m}\Big)+\f{s^2}{2880\pi^2 f^4}+O(s^3).
\label{eq:result_forward}
\ee
The amplitude away from the forward limit at the order $O(s^2)$ can be obtained from \eqref{eq:result_forward} by using crossing symmetry, then it reads
\begin{equation}
	\label{eq:result_nonforward}
	\mathcal{T}_{BB\rightarrow BB}(s,t,u) = \f{3m^4}{16\pi^2 f^4}\ln\Big(\f{\Lambda}{m}\Big)+ \frac{1}{5760\pi^2f^4}\times(s^2+t^2+u^2) + O(s^3).
\end{equation}
It is obvious that \eqref{eq:result_nonforward} reduces to \eqref{eq:result_forward} in the forward limit. The equation \eqref{eq:result_nonforward} is precisely \eqref{eq:dilaton_example} quoted in the main text once we set $\Lambda=m$ to make the cosmological constant equals to zero.

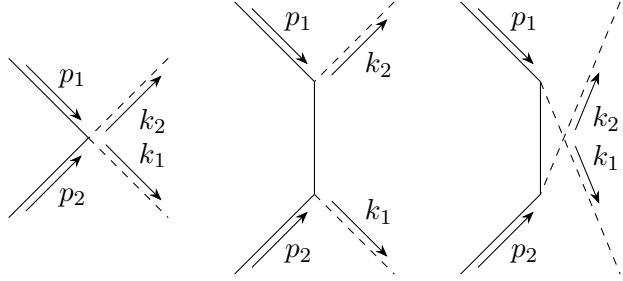
\begin{figure}[h!]
	\tikzfeynmanset{momentum/arrow distance=1mm}
	\tikzfeynmanset{momentum/label distance=-0.15em}
	\centering
	\begin{tikzpicture}[baseline=(a)]
		\begin{feynman}
			\vertex (a);
			\vertex [above left = of a] (i1);
			\vertex [below left = of a] (i2);
			\vertex [above right = of a] (o1);
			\vertex [below right = of a] (o2);
			\diagram* {
				(i1)--[momentum=$p_1$](a)--[rmomentum=$p_2$](i2), 
				(o1)--[rmomentum=$k_2$,dashed](a)--[momentum=$k_1$,dashed](o2),
			};
		\end{feynman}
	\end{tikzpicture}\qquad
	\begin{tikzpicture}[baseline=($(a.base)!.5!(b.base)$)]
		\begin{feynman}
			\vertex (a);
			\vertex [below = of a] (b);
			\vertex [above left = of a] (i1);
			\vertex [below left = of b] (i2);
			\vertex [above right = of a] (o1);
			\vertex [below right = of b] (o2);
			\diagram* {
				(i1)--[momentum=$p_1$](a)--(b)--[rmomentum=$p_2$](i2), 
				(o1)--[rmomentum=$k_2$,dashed](a),
				(b)--[momentum=$k_1$,dashed](o2),
			};
		\end{feynman}
	\end{tikzpicture}\qquad
	\begin{tikzpicture}[baseline=($(a.base)!.5!(b.base)$)]
		\begin{feynman}
			\vertex (a);
			\vertex [below = of a] (b);
			\vertex [above left = of a] (i1);
			\vertex [below left = of b] (i2);
			\vertex [above right = of a] (o1);
			\vertex [below right = of b] (o2);
			\diagram* {
				(i1)--[momentum=$p_1$](a)--(b)--[rmomentum=$p_2$](i2), 
				(o1)--[rmomentum={[arrow shorten = 0.35]$k_2$},dashed](b),
				(a)--[momentum={[arrow shorten = 0.35]$k_1$},dashed](o2),
			};
		\end{feynman}
	\end{tikzpicture}

\caption{Tree level Feynman diagrams describing the $AA\rightarrow BB$ scattering amplitude.}
\label{fig:FDtreeAAtoBB}
\end{figure}

\paragraph{Indirect computation}
The imaginary part of the $BB\rightarrow BB$ scattering amplitude at one loop is related to the tree level scattering amplitude $BB\rightarrow AA$ via the optical theorem which can be written as
\begin{multline}
	\text{Im}\mathcal{T}_{BB\rightarrow BB}(s,0,-s)=\\
	 \f{1}{2}\Bigg[\f{1}{2}\int \f{d^3\vec{p}_1}{(2\pi)^3 2E_{\vec{p}_1}}\f{d^3\vec{p}_2}{(2\pi)^3 2E_{\vec{p}_2}}\Bigg]\ (2\pi)^4 \delta^{(4)}(k_1+k_2-p_1-p_2) \times \Big{|}\mathcal{T}_{BB\rightarrow AA}(s,t,u)\Big{|}^2 \label{ImA}.
\end{multline}
The terms within the square bracket is the two identical particle phase space integral. To derive the above relation we considered only two massive particle exchange in the unitarity cut, which is the leading order contribution in large $f$. Above we can use crossing symmetry to write $\mathcal{T}_{BB\rightarrow AA}(s,t,u)=\mathcal{T}_{AA\rightarrow BB}(s,t,u)$.
The tree level Feynman diagrams describing the $AA\rightarrow BB$ scattering process are depicted in figure \ref{fig:FDtreeAAtoBB}.
This leads to the following explicit expression for the amplitude
\be
\label{TAABB}
i\mathcal{T}_{AA\rightarrow BB}(s,t,u)&=&\ -\f{im^2}{f^2}\Bigg[1\ +\ \f{2m^2}{t-m^2}\ +\ \f{2m^2}{u-m^2}\Bigg],
\ee
where $s=-(p_1+p_2)^2\ ,\ t=-(p_1-k_1)^2\ ,\ u=-(p_1-k_2)^2$ with $s+t+u=2m^2$. We recall that in the center of mass frame the $t$ and $u$ variables can be expressed in terms of total energy squared $s$ and the scattering angle $\theta$ according to the second entry in \eqref{eq:definitionsTU}. We write this relation here again for convenience
\begin{equation}
	\begin{aligned}
		t&= m^2-\f{s}{2}+\f{1}{2}\sqrt{s(s-4m^2)}\cos\theta,\nn\\
		u&=m^2-\f{s}{2}-\f{1}{2}\sqrt{s(s-4m^2)}\cos\theta.
	\end{aligned}
\end{equation}

Plugging \eqref{TAABB} into \eqref{ImA} we obtain
\be
&&\text{Im}\mathcal{T}_{BB\rightarrow BB}(s,0,-s)\nn\\
&=& \f{1}{64\pi}\ \f{\sqrt{s-4m^2}}{\sqrt{s}}\ \f{m^4}{f^4}\int_{-1}^{1}d(\cos\theta)\ \Bigg[1\ +\ \f{2m^2}{t-m^2}\ +\ \f{2m^2}{u-m^2}\Bigg]^2\nn\\
&=&\ \f{1}{64\pi}\f{\sqrt{s-4m^2}}{\sqrt{s}}\ \f{m^4}{f^4}\int_{-1}^{1}dx\ \Bigg[1-\f{8sm^2}{s^2-s(s-4m^2)x^2}\Bigg]^2\nn\\
&=&\ \f{1}{64\pi}\f{\sqrt{s-4m^2}}{\sqrt{s}}\ \f{m^4}{f^4} \Bigg[2+\f{16m^2}{s}-\f{16m^2(s-2m^2)}{s\sqrt{s(s-4m^2)}}\ln\Bigg(\f{s+\sqrt{s(s-4m^2)}}{s-\sqrt{s(s-4m^2)}}\Bigg)\Bigg].
\label{eq:imBBtoBB}
\ee 
At low energy the $BB\rightarrow BB$ amplitude will have the form \eqref{eq:sum_rule}. We remind that for the QFT under consideration $a^\text{IR} =0$. The $a^\text{UV}$ is given by the sum rule \eqref{eq:dispersion_relation_general} which is completely determined by the imaginary part \eqref{eq:imBBtoBB}. Plugging \eqref{eq:imBBtoBB} into \eqref{eq:dispersion_relation_general} we conclude that
\be
a^\text{UV}&=&\ \f{m^4}{64\pi^2}\ \int_{4m^2}^{\infty}\f{ds}{s^3}\ \f{\sqrt{s-4m^2}}{\sqrt{s}}\Bigg[2+\f{16m^2}{s}-\f{16m^2(s-2m^2)}{s\sqrt{s(s-4m^2)}}\ln\Bigg(\f{s+\sqrt{s(s-4m^2)}}{s-\sqrt{s(s-4m^2)}}\Bigg)\Bigg]\nn\\
&=&\ \f{m^4}{64\pi^2}\Big[\f{1}{30m^4}+\f{4}{105m^4}-\f{19}{315m^4}\Big]\nn\\
&=&\ \f{1}{(64\times 90)\pi^2}\ =\ \f{1}{5760\pi^2}.
\ee
This together with \eqref{eq:sum_rule} is in a perfect agreement with \eqref{eq:result_nonforward}.

One can obtain the imaginary part of the $BB\rightarrow BB$ amplitude away from the forward limit away using  \eqref{A}. It reads
\begin{equation}
	\label{ImTbbbb}
	\begin{aligned}
		\text{Im}[\widetilde{\mathcal{T}}_{BB\rightarrow BB}(s,t)]&=\ \frac{1}{32\pi}\sqrt{1-\frac{4}{s}} -\frac{1}{4\pi s}\ln\Bigg(\frac{1+\sqrt{1-\frac{4}{s}}}{1-\sqrt{1-\frac{4}{s}}}\Bigg)\\
		&-\frac{1}{4\pi}\ \frac{1}{su}\frac{1}{\sqrt{1+\frac{4t}{su}}}\ln\Bigg(\frac{\frac{1}{s}-\frac{u}{st}+\frac{u}{2t}\Big[1+\sqrt{1-\frac{4}{s}}\sqrt{1+\frac{4t}{us}}\Big]}{\frac{1}{s}-\frac{u}{st}+\frac{u}{2t}\Big[1-\sqrt{1-\frac{4}{s}}\sqrt{1+\frac{4t}{us}}\Big]}\Bigg)\\
		&-\frac{1}{4\pi}\ \frac{1}{st}\frac{1}{\sqrt{1+\frac{4u}{st}}}\ln\Bigg(\frac{\frac{1}{s}-\frac{t}{su}+\frac{t}{2u}\Big[1+\sqrt{1-\frac{4}{s}}\sqrt{1+\frac{4u}{ts}}\Big]}{\frac{1}{s}-\frac{t}{su}+\frac{t}{2u}\Big[1-\sqrt{1-\frac{4}{s}}\sqrt{1+\frac{4u}{ts}}\Big]}\Bigg).
	\end{aligned}
\end{equation}
We can  check that this expression in the forward limit $t=0$ reproduces \eqref{eq:imBBtoBB}.

\paragraph{Partial amplitudes and unitarity}
Using the definitions \eqref{eq:Tell} and the explicit expressions \eqref{TAABB} and \eqref{ImTbbbb} in free theory we obtain the following spin 0 and 2 partial amplitudes
\begin{eqnarray}
	\widetilde{\mathcal{T}}^{0}_{AA\rightarrow BB}(s)&=& -\frac{1}{32\pi}\Big(1-4/s\Big)^{1/4}\Bigg[2-\frac{8}{s\sqrt{1-4/s}}\ln\Bigg(\frac{1+\sqrt{1-4/s}}{1-\sqrt{1-4/s}}\Bigg)\Bigg]\label{Taabb0},\\
	\widetilde{\mathcal{T}}^{2}_{AA\rightarrow BB}(s)&=& -\frac{1}{4\pi}\Big(1-4/s\Big)^{1/4}\Bigg[\frac{3}{s-4}-\frac{1+2/s}{s\Big(1-4/s\Big)^{3/2}}\ln\Bigg(\frac{1+\sqrt{1-4/s}}{1-\sqrt{1-4/s}}\Bigg)\Bigg].
\end{eqnarray}
\begin{eqnarray}
	\label{eq:imBBtoBB_0}
	\text{Im}\Big[\widetilde{\mathcal{T}}^{0}_{BB\rightarrow BB}(s)\Big]&=& \frac{1}{2(16\pi)^2}\sqrt{1-4/s}\ -\ \frac{1}{64\pi^2 s}\ln\Bigg(\frac{1+\sqrt{1-4/s}}{1-\sqrt{1-4/s}}\Bigg)\nonumber\\
	\label{eq:imBBtoBB_2}
	&&\ +\frac{1}{32\pi^2 s^2}\frac{1}{\sqrt{1-4/s}}\Bigg[\ln\Bigg(\frac{1+\sqrt{1-4/s}}{1-\sqrt{1-4/s}}\Bigg)\Bigg]^2\label{Tbbbb0},\\
	\text{Im}\Big[\widetilde{\mathcal{T}}^{2}_{BB\rightarrow BB}(s)\Big]&=& \frac{9}{32\pi^2}\ \frac{1}{s^2 \Big(1-4/s\Big)^{3/2}} -\frac{3}{16\pi^2}\frac{s+2}{s^3 \Big(1-4/s\Big)^2}\ln\Bigg(\frac{1+\sqrt{1-4/s}}{1-\sqrt{1-4/s}}\Bigg)\nonumber\\
	&& +\frac{1}{32\pi^2}\frac{(s+2)^2}{s^4 \Big(1-4/s\Big)^{5/2}}\Bigg[\ln\Bigg(\frac{1+\sqrt{1-4/s}}{1-\sqrt{1-4/s}}\Bigg)\Bigg]^2.
\end{eqnarray}

For the free scalar theory the unitarity condition \eqref{eq:unitarity_2_final} simplifies to the following expression
\begin{equation}
\label{eq:unitarity_free_scalar}
\begin{aligned}
\forall \ell=0,2,4,\ldots\\
\forall s\in[4m^2,\infty)
\end{aligned}
:\qquad
\begin{pmatrix}
1   & \widetilde{\mathcal{T}}_{AA\rightarrow BB}^{*\ell}(s) \\
 \widetilde{\mathcal{T}}_{AA\rightarrow BB}^{\ell}(s)    & 2\text{Im}\widetilde{\mathcal{T}}_{BB\rightarrow BB}^{\ell}(s)
\end{pmatrix} \succeq 0.
\end{equation}
One explicitly check that the expressions obtained for spin 0 and 2 partial amplitude saturate this matrix inequality as expected.

\section{Derivation of poles}
\label{app:poles}
Let us consider the scattering amplitude $AB\rightarrow AB$ defined in section \ref{sec:amplitudes}. Unitarity allows to determine part of this amplitude non-perturbatively. This is explained for example in section 2.5.1 in \cite{Karateev:2019ymz}. One can argue that the amplitude $AB\rightarrow AB$ has a pole in the s-channel due to the presence of one-particle states $A$, namely
\begin{equation}
	\label{eq:pole_app}
	\mathcal{T}_{AB\rightarrow AB}(s,t,u) = -\frac{\left|g\right|^2}{s-m^2}+\ldots,
\end{equation}
where the residue $g$ is given as the limit
\begin{equation}
	\label{eq:limit}
	g\equiv \lim_{s\rightarrow m^2} g(s).
\end{equation}
The function $g(s)$ is defined as the following matrix element
\begin{equation}
	\label{eq:definition}
	g(s)  \times (2\pi^4)\delta^4(p-p_1-p_2) = \<p^0,\vec p\,|T|m_A,\vec p_1; m_B,\vec p_2\>,
\end{equation}
where $T$ is the interacting part of the scattering operator and  the total energy squared $s$ reads as
\begin{equation}
	\label{eq:s_def}
	s\equiv - p^2=-(p_1+p_2)^2.
\end{equation}
The $\ldots$ in \eqref{eq:pole_app} denote al the finite contributions at $s=m^2$.
The physical range of energies in \eqref{eq:s_def} is $s\in [m^2,\infty)$. The masses of particles A and B are given by \eqref{eq:masses}, we remind here for the readers convenience that $m_A=m$ and $m_B=0$. Due to the presence of the $\ZZ$ symmetry, the bra-state in the right-hand side of \eqref{eq:definition} is $\ZZ$ odd.

From the explicit expression of the modified action \eqref{eq:action_modified_equivalent} one can conclude that the interacting part of the scattering operator has the form
\begin{equation}
	T =  - \frac{i}{\sqrt{2}f}\int d^4 x\; \Theta(x) \varphi(x)+ O\left(f^{-2}\right),
\end{equation}
where $\Theta(x)$ is the trace of the stress-tensor. Plugging this expression into \eqref{eq:definition} we obtain
\begin{equation}
	\label{eq:g(s)}
	g(s)  \times (2\pi^4)\delta^4(p-p_1-p_2) = -\frac{i}{\sqrt{2}f}\,
	\int d^4x\, e^{ip_2\cdot x}\<p^0,\vec p\,|\Theta(x)|m_A,\vec p_1\>+O(f^{-2}).
\end{equation}
Here we have used the contraction between the dilaton field $\varphi(x)$ and the dilaton state $|m_B,\vec p_2\,\rangle$. The translation symmetry allows us to write
\begin{equation}
	\label{eq:operator_transformation_Theta}
	\Theta(x) = e^{-iP\cdot x} \Theta(0) e^{+iP\cdot x}.
\end{equation}
Here $P^\mu$ are the generators of translation. Using \eqref{eq:operator_transformation_Theta} and taking into account the fact that the states in \eqref{eq:g(s)} are eigenstates of $P^\mu$, writing the integral over $x$ as a $\delta$-function
we get the final expression for the function $g(s)$ which reads
\begin{equation}
	g(s) = -\frac{i}{\sqrt{2}f}\,\<p^0,\vec p\,|\Theta(0)|m_A,\vec p_1\>+O(f^{-2}).
\end{equation}
We remind that the total energy squared $s$ was defined in \eqref{eq:s_def}, as a result we have
\begin{equation}
	p^0 = |\vec p_2| + \sqrt{m_A^2+\vec p_1^{\,2}},\quad
	\vec p = \vec p_1 + \vec p_2.
\end{equation}
Let us now take the limit \eqref{eq:limit}. This limit is achieved by setting  $\vec p_2 \rightarrow 0$. Hence we get,
\begin{equation}
	\label{eq:g_value}
	g =  -\frac{i}{\sqrt{2}f}\,\<m,\vec{p}_1 \,|\Theta(0)|m,\vec{p}_1\>+O(f^{-2})
\end{equation}

As derived in \cite{Karateev:2019ymz,Karateev:2020axc}, in particular see appendix G of \cite{Karateev:2020axc}, the following normalization condition holds
\begin{equation}
	\label{eq:trace_condition}
	\lim_{\vec p_2 \rightarrow \vec p_1} 
	\< m, \vec p_1| \Theta(0)|m,\vec p_2\> = -2 m^2.  
\end{equation}
Plugging it into \eqref{eq:g_value} we conclude that
\begin{equation}\label{eq:g2}
	|g|^2 = \frac{2m^4}{f^2}.
\end{equation}
In turn, plugging this into \eqref{eq:pole_app}, using crossing symmetry \eqref{eq:crossin_4} and the definitions \eqref{eq:new_amplitudes} we finally obtain
\begin{equation}
	\widetilde{\mathcal{T}}_{AB\rightarrow AB}(s,t,u) = -\frac{2m^4}{s-m^2}-\frac{2m^4}{u-m^2}+\ldots \label{eq:residue_ABAB}
\end{equation}

\section{Useful identities}
\label{S:Fourier_transforms}
In this appendix we derive a set of identities used in section \ref{sec:ABtoAB_EFT}.

Let us start with the following Fourier transform
\begin{equation}
\sum_{n=0}^\infty c_n^{\mu_1 \mu_2\ldots \mu_n} \; n\p_{\mu_1}\p_{\mu_2}\cdots \p_{\mu_n}\Phi(x)  \longrightarrow
H_1(q)\equiv\sum_{n=0}^\infty (i)^n c_n^{\mu_1\ldots \mu_n} \; n\; q_{\mu_1}q_{\mu_2}\cdots q_{\mu_n}\Phi(q).
\end{equation}
Recall that in section \ref{sec:ABtoAB_EFT} we introduced the object $\mathcal{K}(q)$, it was defined in \eqref{eq:K_fourier_transform}. Let us reproduce this definition here for the readers convenience
\begin{equation}
	\mathcal{K}(q)\equiv\sum_{n=0}^\infty (i)^n\ c_n^{a_1 a_2\ldots a_n}  q_{a_1}q_{a_2}\ldots q_{a_n}.
\end{equation}
Using the obvious fact that the object $\mathcal{K}(q)$ is homogeneous in $q_\mu$ we conclude that
\begin{equation}
\label{eq:property_1}
H_1(q) = q^\mu \frac{\partial \mathcal{K}(q)}{\partial q^\mu}\Phi(q).
\end{equation}

Let us denote arbitrary tensors of rank 1 and 3  by $\mathcal{E}$ and $\mathcal{F}$ respectively. In section \ref{sec:ABtoAB_EFT} we had another three Fourier transforms which are
\begin{multline}
\sum_{n=0}^\infty c_n^{\mu_1 \mu_2\ldots \mu_n} \; \f{n(n+1)}{2}\p_{\mu_1}\p_{\mu_2}\cdots \p_{\mu_n}\Phi(x)  \longrightarrow\\
H_2(q) \equiv \sum_{n=0}^\infty (i)^n c_n^{\mu_1\ldots \mu_n} \; \f{n(n+1)}{2}\ q_{\mu_1}q_{\mu_2}\cdots q_{\mu_n}\Phi(q),
\end{multline}
\begin{multline}
\sum_{n=1}^{\infty}c_n^{\mu_1\cdots \mu_n}\sum_{i=1}^{n}\mathcal{E}_{\mu_i}\ \p_{\mu_1}\cdots \p_{\mu_{i-1}}\p_{\mu_{i+1}}\cdots \p_{\mu_n}\Phi(x)
\longrightarrow\\
H_3(q) \equiv \sum_{n=1}^{\infty}(i)^{n-1}c_n^{\mu_1\cdots \mu_n}\sum_{i=1}^{n}\mathcal{E}_{\mu_i}\ q_{\mu_1}\cdots q_{\mu_{i-1}}q_{\mu_{i+1}}\cdots q_{\mu_n}\Phi(q),
\end{multline}
\begin{multline}
\sum_{n=2}^{\infty}c_n^{\mu_1 \ldots \mu_n}\sum_{\substack{i, j=1\\ i<j}}^{n}\mathcal{F}_{\mu_i \mu_j}^\nu \p_{\mu_1}\cdots \p_{\mu_{i-1}}\p_{\mu_{i+1}}\cdots \p_{\mu_{j-1}}\p_{\mu_{j+1}}\cdots \p_{\mu_n}\p_\nu \Phi(x)
\longrightarrow\\
H_4(q) \equiv\f{1}{2}\sum_{n=2}^{\infty}(i)^{n-1}c_n^{\mu_1 \ldots \mu_n}\sum_{\substack{i, j=1\\ i\neq j}}^{n}\mathcal{F}_{\mu_i \mu_j}^\nu q_{\mu_1}\cdots q_{\mu_{i-1}}q_{\mu_{i+1}}\cdots q_{\mu_{j-1}}q_{\mu_{j+1}}\cdots q_{\mu_n} q_\nu \Phi(q).
\end{multline}
Analogously to \eqref{eq:property_1} we can write
\begin{align}
	\label{eq:property_2}
H_2(q) &= \f{1}{2}q^\mu q^\nu \frac{\partial^2 \mathcal{K}(q)}{\partial q^\mu \p q^\nu}\Phi(q)+q^\mu  \f{\p \mathcal{K}(q)}{\partial q^\mu}\Phi(q),\\
\label{eq:property_3}
H_3(q) &= -i\ \mathcal{E}_{\mu}\f{\p \mathcal{K}(q)}{\p q_\mu}\Phi(q),\\
\label{eq:property_4}
H_4(q) &= -\f{i}{2}\mathcal{F}_{\mu\rho}^\nu q_\nu \f{\p^2 \mathcal{K}(q)}{\p q_\mu \p q_\rho}\Phi(q).
\end{align}

\section{Matter - dilaton scattering: perturbative example}
\label{app:perturbative_AABB}
In this appendix we consider the $\Phi^3$ perturbative model defined by the following action
\begin{equation}
\label{eq:Phi3}
A(\lambda_0)=
\int d^{4}x \Big[-\f{1}{2}\p_{\mu}\Phi\p^{\mu}\Phi -\f{1}{2}m_0^2\Phi^2-\frac{\lambda_0}{3!}\Phi_0^3\Big],
\end{equation}
where $m_0$ is the bare mass parameter and $\lambda_0$ is the bare cubic coupling constant of mass dimension one. In order to simplify the computations of this section we will restrict our attention to the case when
\begin{equation}
	m_0 = 0.
\end{equation}
The model \eqref{eq:Phi3} can then be interpreted as the $\lambda_0\Phi^3$ deformation of the free massless CFT in the UV.
According to the discussion of sections \ref{KS_setup} and \ref{S:KS_a_theorem_review} one can define the following modified action
\begin{equation}
\label{lambda_deformed_QFTaction}
A'(\lambda_0)=
\int d^{4}x\ \Big[-\f{1}{2}\p_{\mu}\Phi_0\p^{\mu}\Phi_0 -\f{1}{2}  \p_{\mu}\varphi \p^{\mu}\varphi-\frac{\lambda_0}{3!}\Phi_0^3 + \f{\lambda_0}{3! \sqrt{2}f}\Phi_0^3 \varphi\Big]+O(f^{-2}).
\end{equation}
Here as usual the field $\Phi(x)$ creates the particle $A$ and the dilaton field $\varphi(x)$ creates the dilaton particle $B$ from the vacuum. Throughout this paper we assumed that the particle $A$ is $\ZZ$-odd. Thus, the associated field must be $\ZZ$-odd. In this appendix we relax this requirement which makes our statements here even more general. 

Using the action \eqref{lambda_deformed_QFTaction} one could compute the $BB \rightarrow BB$ scattering amplitude at low energies and show that it is given by \eqref{eq:result_nonforward} exactly as in the case of the free massive field theory discussed in appendix \ref{sec:example}. This is because the two models in appendices \ref{sec:example} and \ref{app:perturbative_AABB} have the same UV fixed point. We will not do this computation here. Instead we will focus on the $AB \rightarrow AB$ scattering process. We will show that the residues of the $s$- and $u$- channel poles matches with the result in \eqref{eq:residue_ABAB}. This confirms the general result of appendix \ref{app:poles}.

In order to proceed with the computation let us defined the renormalized field $\Phi(x)$ defined as $\Phi(x)=Z^{- 1/2}\Phi_0(x)$ with the renormalized mass $m$. Here $Z$ is the field strength redefinition constant. In terms of the renormalized field the action \eqref{lambda_deformed_QFTaction} takes the form
\be
A'(\lambda)&=&\ -\f{1}{2}\int d^{4}x\ \Big[\p_{\mu}\Phi \p^{\mu}\Phi +m^2\Phi^2 +\  \p_{\mu}\varphi \p^{\mu}\varphi\Big]\ -\int d^{4}x\Big[\frac{\lambda}{3!}\Phi^3 -\f{\lambda}{3! \sqrt{2}f}\Phi^3 \varphi\Big]\nonumber\\
&&\ -\f{1}{2}\int d^{4}x\ \Big[\delta_Z\ \p_{\mu}\Phi \p^{\mu}\Phi +\ \delta_m\ \Phi^2 \Big]\ -\int d^{4}x\Big[\frac{\delta_\lambda}{3!}\Phi^3 -\f{\delta_\lambda}{3! \sqrt{2}f}\Phi^3 \varphi\Big],
\ee
where we have defined the counter terms as
\begin{equation}
	\delta_Z =Z-1,\qquad
	\delta_m =-m^2,\qquad
	\delta_\lambda =\lambda_0 Z^{3/2}-\lambda.
\end{equation}

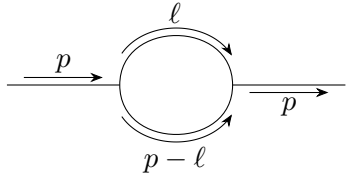
\begin{figure}[t!]
	\centering
\tikzfeynmanset{momentum/arrow distance=1mm}
\tikzfeynmanset{momentum/label distance=-0.15em}
\begin{tikzpicture}
	\begin{feynman}
		\vertex (a);
		\vertex [right=of a] (b);
		\vertex [left =of a] (i1);
		\vertex [right =of b] (o1);
		\diagram* {
			(a) --[half left, momentum={$\ell$}] (b) -- [half left, rmomentum=$p-\ell$] (a),
			(i1) --[ momentum={$p$}] (a),
			(o1) --[rmomentum={$p$}] (b),
		}; 
	\end{feynman}
\end{tikzpicture}
\caption{One loop contribution to the scalar propagator of the particle $A$.}\label{oneloop_propagator}
\end{figure}\ 
Up to one loop order and linear in counter terms the scalar propagator for the particle $A$ has the following form
\be
D_F(p)= \f{-i}{p^2+m^2 -i\epsilon} +\f{-i}{p^2+m^2 -i\epsilon}\ \Big[i\Sigma_2(p) -i\delta_m  -i\delta_Z\ p^2 \Big]\ \f{-i}{p^2 +m^2 -i\epsilon}, 
\ee
where
\be
i\Sigma_2(p)\ &=&\ \f{\lambda^2}{2}\int \f{d^4 \ell}{(2\pi)^4}\ \f{1}{\ell^2 +m^2 -i\epsilon}\ \f{1}{(p-\ell)^2 +m^2-i\epsilon}.
\ee
and the corresponding Feynman diagram contributing to it is drawn in figure \ref{oneloop_propagator}. We can in principle evaluate the above loop integral using dimensional regularization, but we do not need to evaluate it here. Now if we impose the condition that the remormalized scalar propagator has a pole at $p^2=-m^2$ with the residue $(-i)$, we get
\be
\delta_m=-m^2 =\Sigma_2(p)\Big{|}_{p^2=-m^2},\quad
\delta_{Z}=\f{\p \Sigma_2(p)}{\p p^2}\Big{|}_{p^2=-m^2}.
\ee
So the non-vanishing contribution to the mass square for the scalar particle $A$ appears at order $\lambda^2$. Similarly we can find out the counter term $\delta_\lambda$ at order $\lambda^3$ analyzing three point scalar correlation function up to one loop order. 
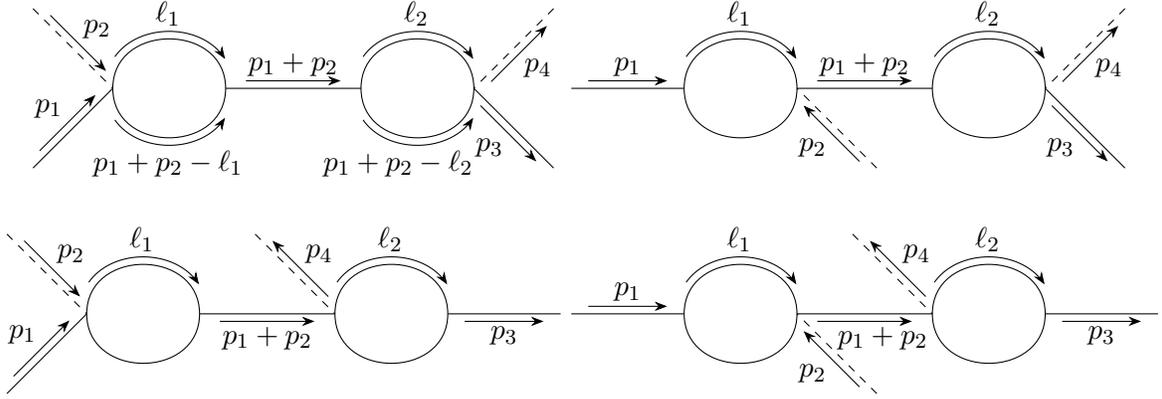
\begin{figure}[h!]
\centering
\tikzfeynmanset{momentum/arrow distance=1mm}
\tikzfeynmanset{momentum/label distance=-0.15em}
\begin{tikzpicture}[baseline=(a)]
	\begin{feynman}
		\vertex (a);
		\vertex [right=of a] (b);
		\vertex [right=1.8 of b] (c);
		\vertex [right=of c] (d);
		\vertex [below left =of a] (i1);
		\vertex [above left =of a] (i2);
		\vertex [above right=of d] (o1);
		\vertex [below right=of d] (o2);
		\diagram* {
			(a) --[half left, momentum={$\ell_1$}] (b) -- [half left, rmomentum=$p_1+p_2-\ell_1$] (a),
			(i1) --[ momentum={$p_1$}] (a),
			(i2) --[dashed, momentum={$p_2$}] (a),
			(b) --[momentum={$p_1+p_2$}] (c),
			(c) --[half left, momentum={$\ell_2$}] (d) -- [half left, rmomentum=$\hspace{-1.25em}p_1+p_2-\ell_2$] (c),
			(o1) --[dashed, rmomentum={$p_4$}] (d),
			(o2) --[rmomentum={$p_3$}] (d),
		}; 
	\end{feynman}
\end{tikzpicture}
\begin{tikzpicture}[baseline=(a)]
	\begin{feynman}
		\vertex (a);
		\vertex [right=of a] (b);
		\vertex [right=1.8 of b] (c);
		\vertex [right=of c] (d);
		\vertex [left =of a] (i1);
		\vertex [below right =of b] (i2);
		\vertex [above right=of d] (o1);
		\vertex [below right=of d] (o2);
		\diagram* {
			(a) --[half left, momentum={$\ell_1$}] (b) -- [half left] (a),
			(i1) --[ momentum={$p_1$}] (a),
			(i2) --[dashed, momentum={$p_2$}] (b),
			(b) --[momentum={$p_1+p_2$}] (c),
			(c) --[half left, momentum={$\ell_2$}] (d) -- [half left] (c),
			(o1) --[dashed, rmomentum={$p_4$}] (d),
			(o2) --[rmomentum={$p_3$}] (d),
		}; 
	\end{feynman}
\end{tikzpicture}\\\vspace{1em}
\begin{tikzpicture}[baseline=(a)]
	\begin{feynman}
		\vertex (a);
		\vertex [right=of a] (b);
		\vertex [right=1.8 of b] (c);
		\vertex [right=of c] (d);
		\vertex [below left =of a] (i1);
		\vertex [above left =of a] (i2);
		\vertex [above left=of c] (o1);
		\vertex [right=of d] (o2);
		\diagram* {
			(a) --[half left, momentum={$\ell_1$}] (b) -- [half left] (a),
			(i1) --[ momentum={$p_1$}] (a),
			(i2) --[dashed, momentum={$p_2$}] (a),
			(b) --[momentum'={$p_1+p_2$}] (c),
			(c) --[half left, momentum={$\ell_2$}] (d) -- [half left] (c),
			(o1) --[dashed, rmomentum={$p_4$}] (c),
			(o2) --[rmomentum={$p_3$}] (d),
		}; 
	\end{feynman}
\end{tikzpicture}
\begin{tikzpicture}[baseline=(a)]
	\begin{feynman}
		\vertex (a);
		\vertex [right=of a] (b);
		\vertex [right=1.8 of b] (c);
		\vertex [right=of c] (d);
		\vertex [left =of a] (i1);
		\vertex [below right =of b] (i2);
		\vertex [right =of b] (p);
		\vertex [above left=of c] (o1);
		\vertex [right=of d] (o2);
		\diagram* {
			(a) --[half left, momentum={$\ell_1$}] (b) -- [half left] (a),
			(i1) --[ momentum={$p_1$}] (a),
			(i2) --[dashed, momentum={$p_2$}] (b),
			(b) --[momentum'={$\hspace{1.25em}p_1+p_2$}] (c),
			(c) --[half left, momentum={$\ell_2$}] (d) -- [half left] (c),
			(o1) --[dashed, rmomentum={$p_4$}] (c),
			(o2) --[rmomentum={$p_3$}] (d),
		}; 
	\end{feynman}
\end{tikzpicture}
\caption{Loop diagrams with $s$-channel pole for the scattering process $AB\rightarrow AB$. The solid lines represent scalar particles (A) and dashed line represent dilatons (B). Using crossing symmetry $p_2\leftrightarrow -p_4$ we can get the Feynman diagrams with $u$-channel pole.}\label{Feynman_loop}
\end{figure}\ 
The amplitude for the scattering process $A(p_1)+B(p_2)\rightarrow A(p_3)+B(p_4)$ from the Feynman diagrams in figure \ref{Feynman_loop} reads,
\begin{equation}
\mathcal{T}_{AB\rightarrow AB}(s,t,u) = -\frac{g(s)^2}{s-m^2}-\frac{g(u)^2}{u-m^2}+\ldots	
\end{equation}
where $s=-(p_1+p_2)^2$ , $u=-(p_1-p_4)^2$ and ``$\ldots$" represents the contributions coming from the Feynman diagrams which do not contain any propagator with momenta $(p_1+p_2)$ or $(p_1-p_4)$. At order $\mathcal{O}(\lambda^4)$ the contribution of $g(s)^2$ takes the following form, from the loop diagrams in figure \ref{Feynman_loop}
\be
g(s)^2 &=& \Big[\f{1}{\sqrt{2}f}\Sigma_2(p_1+p_2)+\f{1}{\sqrt{2}f}\Sigma_2(p_1)\Big]\times \Big[\f{1}{\sqrt{2}f}\Sigma_2(p_1+p_2)+\f{1}{\sqrt{2}f}\Sigma_2(p_3)\Big]
\ee
Now to read off the residue of the pole at $s=m^2$, we need to evaluate the above expression at $s=m^2$ with all the external particles being on-shell. This reduces to substituting $p_2=p_4=0$ and evaluating $g(s)^2$ at $p_1^2=p_3^2=-m^2$. We get,
\be
g(s)^2\Bigg{|}_{\substack{p_2=p_4=0\\ p_1^2=p_3^2=-m^2}}\ &=&\ \f{2}{f^2}\Bigg[\Sigma_2(p_1)\Big{|}_{p_1^2 =-m^2}\Bigg]^2\ =\ \f{2m^4}{f^2}
\ee
This verifies the general result in \eqref{eq:residue_ABAB}. Though the above verification has been done at one-loop order, the proof can be generalized to all orders in perturbation theory.

\section{Worldline action in dilaton background}
\label{app:worldline}

In this appendix, we consider the effective worldline action for a massive particle moving in a background geometry with metric   $g_{\mu\nu} = e^{-2\tau(x)} \eta_{\mu\nu}$. Writing $e^{- \tau(x)} = 1 -\frac{1}{\sqrt{2} f} \varphi(x) $, we shall show that the worldline action is universal up to two derivatives and quadratic order in the dilaton field $\varphi(x)$.

The most general coordinate  invariant worldline action is 
\begin{equation}
\label{worldlineaction}
	S = -m\int dt \left[ 1 + c_1 \ddot{x}^\mu \ddot{x}^\nu g_{\mu\nu}+c_2 R + c_3 \dot{x}^\mu \dot{x}^\nu R_{\mu\nu} +\dots \right]
\end{equation}
where $m$ is the mass of the particle and $c_i$ are non-universal Wilson coefficients.  
 The   4-vector $\dot{x}^\mu$ is equal to $\frac{d x^\mu }{ d t}$ with $t$ the proper time defined by
\begin{equation}
dt^2=   - g_{\mu\nu} dx^\mu  dx^\nu    \,.
\end{equation} 
 $R$   ($R_{\mu\nu}$) stands for the Ricci scalar   (tensor) of the background metric evaluated on the worldline, and the dots represent higher derivative terms.
Notice that the extrinsic curvature of a worldline is simply given in terms of  $ \dot{x}^\mu$ and $ \ddot{x}^\mu$.

For the conformally flat metric $g_{\mu\nu} = e^{-2\tau } \eta_{\mu\nu}$, the Riemann curvature tensor is \begin{equation}  
	R_{\alpha \beta \gamma \delta } = e^{-2\tau}(\eta_{\alpha \gamma}T_{\beta \delta} + \eta_{\beta \delta }T_{\alpha \gamma} - \eta_{\alpha \delta }T_{\beta \gamma} - \eta_{\beta \gamma}T_{\alpha \delta}) \,,\\
\end{equation}
with
\begin{equation}
		\quad T_{\alpha \beta } = \partial _\alpha \partial _\beta \tau + \partial _\alpha \tau \partial _\beta \tau - \frac{1}{2}(\partial  \tau)^2 \eta_{\alpha \beta}\,.
\end{equation}
Therefore, up to quadratic order in the dilaton field, both $R$ and $R_{\mu \nu}$ are of order $  O(\partial ^2 \varphi,(\partial \varphi)^2)$. Clearly, higher derivative terms will contain  more derivatives (and more powers of $\varphi$ in some cases). Notice that $\ddot{x}^\mu=0$ is the leading order equation of motion, thus we can neglect the second term in \eqref{worldlineaction}.
We conclude that non-universal terms contribute to  the scattering amplitude $\widetilde{T}_{AB\to AB}$   at order at least $p^2$ where $p$ is dilaton 4-momentum.
This confirms the universality of the result \eqref{eq:soft_ABtoAB_final}.

\addtocontents{toc}{\protect\enlargethispage{\baselineskip}}
\section{Details of the numerical setup}
\label{app:details_numerical_setup}
In this appendix we provide further technical details of the numerical setup described in section \ref{sec:setup}. These details will be useful to someone who wants to reproduce our numerical results.

Recall that the full ansatze describing the scattering amplitudes $AA\rightarrow AA$, $AB\rightarrow AB$, $AA\rightarrow BB$ and $BB\rightarrow BB$ is given by \eqref{eq:ansatze} together with \eqref{eq:singularity} and \eqref{eq:extra_term_ansatz}.
The unknown coefficients entering the anstatze  are
\begin{equation}
	\label{eq:variables_to_optimize}
	\overrightarrow{\text{coefficients}} = \{\alpha'_{000}, \vec\alpha,c, \vec\beta, \vec\gamma \},
\end{equation}
where we have defined
\begin{equation}
	\vec \alpha \equiv \{\alpha_{000}, \alpha_{001}, \ldots\},\quad 
	\vec \beta \equiv \{\beta_{000},  \beta_{001}, \ldots\},\quad
	\vec \gamma \equiv \{\gamma_{000}, \gamma_{001},  \ldots\}.
\end{equation}
These coefficients should be determined numerically by solving some optimization problem. In order to make any concrete computations we need to make the vector of coefficients \eqref{eq:variables_to_optimize} to be finite. This is done by keepping only a finite number of terms in the ansatze \eqref{eq:ansatze}. This truncation is governed by  the parameter $N_{max}$ defined as 
\begin{equation}
	a+b+c \leq N_{max}.
\end{equation}

\paragraph{Soft behavior}
At low energy the $AA\rightarrow BB$ and $BB\rightarrow BB$ amplitudes have a very particular (soft) behavior according to \eqref{eq:soft_AABB} and \eqref{eq:soft_BBBB} respectively. As a result we have additional constraints on the coefficients of the ansatze. Plugging here the definition of $(t,u)$ variables in terms of $(s,\cos\theta)$ according to \eqref{eq:definitionsTU} and expanding around small values of $s$ keeping $\cos\theta$ fixed we get expressions which match \eqref{eq:soft_AABB} and \eqref{eq:soft_BBBB}. Doing this one obtains the following constraints
\begin{equation}
	\label{eq:soft_conditions}
	\beta_{000} = -1 - (98-40\sqrt{6})\beta_{001} + \ldots,\quad
	\gamma_{000} = 0,\quad
	\gamma_{001} = \frac{1}{4} \left(512 a -2\gamma_{002}+\gamma_{011}\right),
\end{equation}
where $a$ is the $a$-anomaly. Plugging the solutions \eqref{eq:soft_conditions} into the anstatze \eqref{eq:ansatze} we effectively eliminate 3 coefficients $\beta_{000}$,  $\gamma_{000}$, $\gamma_{001}$ and introduce one additional coefficient $a$ (the $a$-anomaly).  As a result the list of unknown variables \eqref{eq:variables_to_optimize} gets modified. 
Let us denote the final list of variables by
\begin{equation}
	\label{eq:variables_to_optimize_new}
	\overrightarrow{\text{coefficients}}'.
\end{equation}

\paragraph{Partial amplitudes}
In order to impose unitarity we need to compute partial amplitudes defined in \eqref{eq:Tell}. For that let us define the following integrals
\begin{align}
	\label{eq:int11AAAA}
	\text{int11}_{AA\rightarrow AA}^{bc;\,\ell} &\equiv 
	\int_{-1}^{+1}d\cos\theta P_\ell(\cos\theta)
	\left(\myRho_1(t;4/3)\right)^b\left(\myRho_1(u;4/3)\right)^c, \\
	\text{int12}_{AB\rightarrow AB}^{bc;\,\ell} &\equiv 
	\int_{-1}^{+1}d\cos\theta P_\ell(\cos\theta)
	\left(\myRho_1(t;2/3)\right)^b\left(\myRho_2(u;2/3)\right)^c, \\
	\text{int22}_{AA\rightarrow BB}^{bc;\,\ell} &\equiv 
	\int_{-1}^{+1}d\cos\theta P_\ell(\cos\theta)
	\left(\myRho_2(t;2/3)\right)^b\left(\myRho_2(u;2/3)\right)^c, \\
	\label{eq:int11BBBB}
	\text{int11}_{BB\rightarrow BB}^{bc;\,\ell} &\equiv 
	\int_{-1}^{+1}d\cos\theta P_\ell(\cos\theta)
	\left(\myRho_1(t;0)\right)^b\left(\myRho_1(u;0)\right)^c
\end{align}
together with
\begin{equation}
	\label{eq:projection_poles_AAtoBB}
	\begin{aligned}
		\text{poleT}_{AA\rightarrow BB}^{\ell} &\equiv \int_{-1}^{+1}d\cos\theta P_\ell(\cos\theta) \frac{1}{t-1} ,\\
		\text{poleU}_{AA\rightarrow BB}^{\ell} &\equiv \int_{-1}^{+1}d\cos\theta P_\ell(\cos\theta) \frac{1}{u-1}.
	\end{aligned}
\end{equation}
\begin{equation}
	\label{eq:projection_poles_ABtoAB}
	\begin{aligned}
		\text{poleS}_{AB\rightarrow AB}^{\ell} &\equiv \int_{-1}^{+1}d\cos\theta P_\ell(\cos\theta) \frac{1}{s-1} ,\\
		\text{poleU}_{AB\rightarrow AB}^{\ell} &\equiv \int_{-1}^{+1}d\cos\theta P_\ell(\cos\theta) \frac{1}{u-1}.
	\end{aligned}
\end{equation}
Recall that the definitions of the $(t,u)$ variables in terms of $(s,\cos\theta)$ depend on the process, see \eqref{eq:definitionsTU}. We compute the integrals \eqref{eq:int11AAAA} - \eqref{eq:int11BBBB} numerically  in Mathematica for 
\begin{equation}
	\label{eq:Ngrid}
	N_\text{grid}
\end{equation}
different values of $s\geq 4m^2$ for a finite number of spins
\begin{equation}
	\label{eq:Lmax}
	\ell \in [0,\; L_{max}].
\end{equation}
We will explain how the grid of $s$ values is chosen in the very end of this appendix.
We demand 30 significant digits of precision in the numerical evaluation of integrals. The poles appearing in the $AB\rightarrow AB$ process do not enter in our unitarity conditions and thus there is no need to compute the associated contributions into partial amplitudes \eqref{eq:projection_poles_ABtoAB}.  
The integrals \eqref{eq:projection_poles_AAtoBB} are computed analytically. Their explicit expressions will be given in appendix \ref{app:analytic_integrals}.

We also have to define partial amplitudes associated to the additional terms \eqref{eq:singularity} and \eqref{eq:extra_term_ansatz}. They are given by
\begin{multline}
	\label{eq:singularity_PA}
\text{sing} \mathcal{T}^\ell_{AA\rightarrow AA}(s) \equiv \frac{1}{32\pi}\left(1-4/s\right)^{1/2}\times\\
\int_{-1}^{+1}d\cos\theta P_\ell(\cos\theta)
\left( \frac{1}{\myRho_1(s;4/3)-1}+ \frac{1}{\myRho_1(t;4/3)-1}+ \frac{1}{\myRho_1(u;4/3)-1}\right),
\end{multline}
and
\begin{equation}
	\label{eq:free_PA}
\widetilde{\mathcal{T}} ^{\text{free}\;\ell}_{BB \rightarrow BB}(s) \equiv \frac{1}{32\pi}\times
\int_{-1}^{+1}d\cos\theta P_\ell(\cos\theta) \widetilde{\mathcal{T}}^\text{free} _{BB \rightarrow BB}(s,t,u).
\end{equation}
Their explicit expressions will also be computed analytically in appendix \ref{app:analytic_integrals}.

Using the definitions \eqref{eq:int11AAAA} - \eqref{eq:projection_poles_AAtoBB}, \eqref{eq:singularity_PA} and  \eqref{eq:free_PA} we convert our ansatze for the scattering amplitudes \eqref{eq:ansatze} into partial amplitudes as
\begin{multline}
	\label{eq:partial_1}
	\mathcal{T}^\ell_{AA \rightarrow AA}(s) = \alpha'_{000}\;\text{sing} \mathcal{T}^\ell_{AA\rightarrow AA}(s)+\\
	\frac{1}{32\pi}\left(1-4/s\right)^{1/2}\times
	\sum_{a=0}^{\infty}\sum_{b=0}^{\infty}\sum_{c=0}^{\infty}\alpha_{abc}\left(\myRho_1(s;4/3)\right)^a
	\text{int11}_{AA\rightarrow AA}^{bc;\,\ell},
\end{multline}
\begin{multline}
	\label{eq:partial_2}
	\widetilde{\mathcal{T}}^\ell_{AB \rightarrow AB}(s) = \frac{1}{16\pi}(1-1/s)\times
	\Big(-2\text{poleS}_{AB\rightarrow AB}^{\ell}-2\text{poleU}_{AB\rightarrow AB}^{\ell}\\
	+\sum_{a=0}^{\infty}\sum_{b=0}^{\infty}\sum_{c=0}^{\infty}\beta_{abc}\left(\myRho_2(s;2/3)\right)^a
	\text{int12}_{AB\rightarrow AB}^{bc;\,\ell}\Big),
\end{multline}
\begin{multline}
	\label{eq:partial_3}
	\widetilde{\mathcal{T}}^\ell_{AA \rightarrow BB}(s) = \frac{1}{32\pi}\left(1-4/s\right)^{1/4}\times
	\Big(-2\text{poleT}_{AA\rightarrow BB}^{\ell}-2\text{poleU}_{AA\rightarrow BB}^{\ell}\\
	\sum_{a=0}^{\infty}\sum_{b=0}^{\infty}\sum_{c=0}^{\infty}\beta_{abc}\left(\myRho_1(s;2/3)\right)^a
	\text{int22}_{AA\rightarrow BB}^{bc;\,\ell}\Big),
\end{multline}
\begin{equation}
	\label{eq:partial_4}
	\widetilde{\mathcal{T}}^\ell_{BB \rightarrow BB}(s) =  c\;\widetilde{\mathcal{T}} ^{\text{free}\;\ell}_{BB \rightarrow BB}(s)+
	\frac{1}{32\pi}\times
	\sum_{a=0}^{\infty}\sum_{b=0}^{\infty}\sum_{c=0}^{\infty}\gamma_{abc}\left(\myRho_1(s;0)\right)^a
	\text{int11}_{BB\rightarrow BB}^{bc;\,\ell}.
\end{equation}
Finally, we need to plug the solutions \eqref{eq:soft_conditions} into the expressions of partial amplitudes \eqref{eq:partial_1} - \eqref{eq:partial_4}. For each value $s$ the partial amplitudes \eqref{eq:partial_1} - \eqref{eq:partial_4} are simply numerical linear combinations of  unknown coefficients \eqref{eq:variables_to_optimize_new} which can be used to impose unitarity. 

\paragraph{Unitarity}
Let us now address the unitarity constraints \eqref{eq:final_condition_1} and \eqref{eq:unitarity_2_final}. Plugging \eqref{eq:partial_2} into \eqref{eq:final_condition_1} we get the unitarity constraint in the form
\begin{equation}
	\label{eq:unit1_exp}
	0 + \overrightarrow{\text{coefficients}}' \cdot \overrightarrow{M}^\ell_\text{1x1}(s) \geq 0.
\end{equation}
Plugging \eqref{eq:partial_1} together with \eqref{eq:partial_3} and \eqref{eq:partial_4} into \eqref{eq:unitarity_2_final} we obtain
\begin{equation}
	\label{eq:unit2_exp}
	M_{0,\,\text{3x3}}(s) + \overrightarrow{\text{coefficients}}' \cdot \overrightarrow{M}^\ell_\text{3x3}(s) \succeq 0,
\end{equation}
where we have defined
\begin{equation}
	M_{0,\,\text{3x3}}(s) \equiv
	\begin{pmatrix}
		1 & 0 & \text{e}(s)\\
		0 & 0 & 0 \\
		\text{e}(s) & 0 & 0
	\end{pmatrix}
\end{equation} 
together with
\begin{equation}
	\text{e}(s) \equiv  \frac{1}{32\pi}\left(1-4/s\right)^{1/4}\times
	\Big(-2\text{poleT}_{AA\rightarrow BB}^{\ell}-2\text{poleU}_{AA\rightarrow BB}^{\ell}\Big).
\end{equation}

Following the logic described in this appendix so far it is straightforward to obtain the explicit expressions $\overrightarrow{M}^\ell_\text{1x1}(s)$ and $ \overrightarrow{M}^\ell_\text{3x3}(s)$ in Mathematica. These are purely numerical 1x1 and 3x3 matrices for any particular value of $s$ and spin $\ell$. In other words we can obtain $N_\text{grid}$ numerical matrices $\overrightarrow{M}^\ell_\text{1x1}$ and $ \overrightarrow{M}^\ell_\text{3x3}$  for a given value of spin $\ell$.\footnote{To be precise we notice that the matrices $\overrightarrow{M}^\ell_\text{1x1}(s)$ do not exist for each value of $s$ from the grid since they are defined only at $s\geq 9$.} In practice we will then work with a finite number of spins $\ell$, namely
\begin{equation}
	\label{eq:spins}
	\ell = 0, 1, \ldots, L_{max},
\end{equation}
where $L_{max}$ is another truncation parameter analogous to $N_{max}$.
We export the numerical matrices $\overrightarrow{M}^\ell_\text{1x1}$ and $ \overrightarrow{M}^\ell_\text{3x3}$ to SDPB software \cite{Simmons-Duffin:2015qma,Landry:2019qug} which allows to determine numerically the coefficients \eqref{eq:variables_to_optimize_new} such that the unitarity constraints \eqref{eq:unit1_exp} and \eqref{eq:unit2_exp} are satisfied for the selected grid of $s$ values and spins \eqref{eq:spins} given one extra condition. This extra condition can be the minimization of the $a$-anomaly, which is simply one of the parameters in the list of coefficients \eqref{eq:variables_to_optimize_new}.

\paragraph{Grid of $s$ values}
Let us explain how we chose a grid of $s$ values. We first make the following change of variables
\begin{equation}
	\label{eq:phi_variable}
	\myRho_1(s;4/3) = e^{i \phi}
	\quad\Leftrightarrow\quad
	s = \frac{4}{3}\times \frac{5+\cos\phi}{1+\cos\phi} 
\end{equation}
which compactifies the ray $s\geq 4$ into a finite interval $\phi\in[0,\pi]$. We then distribute $N_\text{grid}$ points in $\phi$ in this interval using the Chebyshev grid. These values of $\phi$ are then used to obtain the $s$ values via  \eqref{eq:phi_variable}.

\subsection{Analytic integrals}
\label{app:analytic_integrals}
Let us provide here analytic expressions for some integrals used in appendix \ref{app:details_numerical_setup} above. Let us introduce the Legendre function of the second kind
\begin{align}
	\label{eq:LegendreQ_definition}
	Q_\ell (x) &\equiv -\frac{1}{2} \int_{-1}^{+1} dx'\; \frac{P_\ell(x')}{x'-x}\\
	&= \frac{\sqrt{\pi}\Gamma(\ell+1)}{2^{\ell+1}(x-1)^{\ell+1}\Gamma(3/2+\ell)}
	{}_2F_{1}\left(\ell+1,\ell+1,2(\ell+1), \frac{2}{1-x}\right).
	\label{eq:LegendreQ}
\end{align}
Mathematica has this function built in, it is called $\text{LegendreQ}[\ell,0,3,x]$. Numerically Mathematica has troubles evaluating it when $s$ is close to 4. One can efficiently evaluate the Lengendre Q function however in Mathematica using the form \eqref{eq:LegendreQ}.
From the definition \eqref{eq:LegendreQ_definition} one can see that $Q_\ell(x)$ has a branch cut in the complex plane $x$ on the real axis in the interval $[-1,+1]$. We define the discontinuity of $Q_\ell(x)$ as
\begin{equation}
	\label{eq:disc}
\text{disc} Q_\ell(x) \equiv \f{1}{2i}\lim_{\epsilon\rightarrow 0}\big(Q_\ell(x+i\epsilon)-Q_\ell(x-i\epsilon)\big).
\end{equation}
Plugging here the definition \eqref{eq:LegendreQ_definition} we conclude that
\begin{equation}
	\label{eq:discQ}
	\begin{aligned}
		\text{disc} Q_\ell(x) &= -\frac{\pi }{2} \int_{-1}^{+1} dx'\;P_\ell(x') \lim_{\epsilon\rightarrow 0} \frac{1}{\pi}\frac{\epsilon}{(x'-x)^2+\epsilon^2}\\
		&= -\frac{\pi }{2} \int_{-1}^{+1} dx'\;P_\ell(x')\delta(x'-x)\\
		&=-\frac{\pi }{2} P_\ell(x).
	\end{aligned}
\end{equation}
Here in the second line we have used one of the definitions of the Dirac $\delta$-functions, see for example section ``primary definition'' in \cite{WR}.

Plugging \eqref{eq:discQ} into \eqref{eq:projection_poles_AAtoBB} and using the definition \eqref{eq:LegendreQ_definition} it is straightforward to show that
\begin{equation}
	\label{eq:projection_poles_explicit}
	\text{poleT}_{AA\rightarrow BB}^{\ell} = 
	\text{poleU}_{AA\rightarrow BB}^{\ell} = 
	- \frac{4}{\sqrt{s(s-4)}}Q_\ell \left(\sqrt{\frac{s}{s-4}}\right).
\end{equation}
Here $\ell$ is assumed to be even. We remind that $t$ is related to $\cos\theta$ according to the second relation in \eqref{eq:definitionsTU} and $x\equiv \cos\theta$. 
Using this expression we can compute partial amplitudes of the process $AA\rightarrow BB$ in the free theory. Applying \eqref{eq:Tell} to  \eqref{eq:matter_compensator_free} and using \eqref{eq:projection_poles_explicit} we obtain
\begin{equation}
	\label{eq:projection_poles_free}
	\mathcal{T}_{AA\rightarrow BB}^{\text{free}\;\ell}(s)  \equiv \frac{1}{32\pi}\left(1-4/s\right)^{1/4}
	\left(-2\delta_{\ell, 0} + \frac{16}{\sqrt{s(s-4)}}Q_\ell \left(\sqrt{\frac{s}{s-4}} \right)\right).
\end{equation}
Using unitarity we get the imaginary part of the $BB\rightarrow BB$ process which reads in terms of the above equation as
\begin{equation}
	\label{eq:imBBtoBB_partial}
	\text{Im}
	\mathcal{T}_{BB\rightarrow BB}^{\text{free}\;\ell}(s) = \frac{1}{2}  \left(	\mathcal{T}_{AA\rightarrow BB}^{\text{free}\;\ell}(s)\right)^2.
\end{equation}
Finally, we quote the final result for the projection of the singularity \eqref{eq:singularity_PA} which reads as
\begin{multline}
\text{sing} \mathcal{T}^\ell_{AA\rightarrow AA}(s) = \left( 
	-\frac{\sqrt{\frac{s-4}{s}} \left(\sqrt{s}-2\right)^{\ell} \left(\sqrt{s}+2\right)^{-\ell-1}}{2 \sqrt{6} \pi  (2 \ell+1)}+\frac{\left(-9 \sqrt{s-4}-2 i \sqrt{6}\right) \delta _{0,\ell}}{96 \pi  \sqrt{s}}
	\right).
\end{multline}

\bibliographystyle{JHEP}
\bibliography{refs}

\end{document}